\pdfoutput=1
\documentclass[twoside]{report}
\usepackage{longtable,tabularx,array,url,ifthen,multicol}
\usepackage{fancyhdr,desyhead}
\usepackage{graphpap}
\usepackage[fit]{truncate}

\usepackage{makeidx}
\makeindex
\def\indexname{List of Authors}

\newcommand{\BLKP}{
  \ifthenelse{\isodd{\value{page}}}{\relax}{\mbox{}\thispagestyle{empty}\newpage}}
\newcommand{\CLDP}{\newpage 
  \ifthenelse{\isodd{\value{page}}}{\relax}{\mbox{}\thispagestyle{empty}\newpage}}
\usepackage{geometry}
\usepackage{fancyhdr}
\newcommand{\ARTauthor}{~}
\newcommand{\ARTtitle}{~}
\usepackage{pdfpages}
\newenvironment{papers}{\clearpage}{\clearpage}
\newcommand*\coltoctitle[1]{\def\CTIT{#1}}
\newcommand*\coltocauthor[1]{\def\CAUT{#1}}
\makeatletter
\renewcommand{\l@section}{\@dottedtocline{1}{2em}{0em}}
\renewcommand{\@dotsep}{1000}
\makeatother
\newcommand{\Includeart}[4][]{%
\def\AAA{#2}\def\TTT{#3}%
\renewcommand{\ARTauthor}{~}
\renewcommand{\ARTtitle}{~}
   \includepdf[
               pages=1,
               noautoscale,
               pagecommand={\pagestyle{fancy}},
               offset=0mm 0mm,
               addtotoc={1, subsubsection, 3, ~,  S#4},
               trim=19mm 21mm 19mm 27mm, clip]
               {#4.pdf}%
\addtocontents{toc}{\protect\contentsline{chapter}{\textbf{\TTT}}{\textbf{\pageref{S#4}}}}
\addtocontents{toc}{\protect\par}
\addtocontents{toc}{\protect\contentsline{section}{\AAA}{~}}
\ifthenelse{\equal{#1}{OnePage}}{
                                }{
  \renewcommand{\ARTauthor}{\truncate{.9\linewidth}{\AAA}}
  \renewcommand{\ARTtitle}{\truncate{.9\linewidth}{\TTT}}
    \includepdf[
                pages=2-,
                noautoscale,
                pagecommand={\pagestyle{fancy}},
                offset=0mm 0mm,
                trim=19mm 21mm 19mm 27mm, clip]
                {#4.pdf}}
}
\begin{document}
\pagestyle{empty}
\setlength{\fboxsep}{0pt}
\setlength{\fboxrule}{0.04pt}
\setlength{\unitlength}{1mm}
\pagestyle{plain}
\pagenumbering{roman}
\setcounter{page}{1}
\CLDP
\parindent 0mm
\hspace*{12 cm} {\small DESY--PROC--2009--02}\\
\vspace*{-0.5cm}
\vspace*{3cm}\\
{\sffamily\LARGE
Proceedings of the workshop\vspace*{1cm}\\
{\sffamily\Huge
HERA and the LHC\vspace*{1cm}}\\ 
\vspace*{1cm}
{\sffamily\LARGE workshop series on the implications of HERA for LHC physics\vspace*{2cm}\\}
\vspace*{1cm}
{\sffamily\LARGE 2006 - 2008, Hamburg - Geneva}
\vspace{7mm}
}
\vfill

{\Large
Editors:  
Hannes Jung,
Albert De Roeck

\vspace{1cm}
Verlag Deutsches Elektronen-Synchrotron}

\newpage

{\bfseries\Large
Impressum
\vspace{10mm}

Proceedings of the workshop \\
HERA and the LHC 
\\[1ex]
2nd workshop on the implications of HERA for LHC physics\\
2006 - 2008, Hamburg - Geneva}
\vspace{10mm}

Conference homepage\\
\url{http://www.desy.de/~heralhc}

\vspace{5mm}


\vspace{5mm}
Online proceedings at\\
\url{http://www.desy.de/~heralhc/proceedings-2008/proceedings.html}

\vfill

The copyright is governed by the Creative Commons agreement, which allows
for free use and distribution of the articels for non-commercial activity,
as long as the title, the authors' names and the place of the original are
referenced.

\vspace{10mm}

Editors:\\
Hannes Jung (DESY, U. Antwerp),
Albert De Roeck (CERN, U. Antwerp)
\\
March 2009\\
DESY-PROC-2009-02  \\
ISBN 978-3-935702-32-4\\
ISSN 1435-8077

\vspace{5mm}

Published by\\
Verlag Deutsches Elektronen-Synchrotron\\
Notkestra{\ss}e 85\\
22607 Hamburg\\
Germany
\newpage

\begin{flushleft} 
\mbox{}\\[1cm]
{\bfseries\large Organizing Committee:}\\[3mm]
H. Jung (DESY and U. Antwerp, chair), 
A. De Roeck (CERN and U. Antwerp, chair) \\
G.~Altarelli (CERN),
J.~Bl\"umlein (DESY),
M.~Botje (Nikhef), 
J.~Butterworth (UCL), 
K.~Eggert (CERN), 
E.~Gallo (INFN), 
T.~Haas (DESY),
R.~Heuer (CERN), 
M.~Klein (Liverpool),
M.~Mangano (CERN),
A.~Morsch (CERN), 
G.~Polesello (INFN/CERN), 
O.~Schneider (EPFL),
C.~Vallee (CPPM)
\\[6mm]
{\bfseries\large Conveners:}\\[3mm]
{\bfseries Parton Density Functions:} \\
M. Dittmar (ETH, Z\"urich, CMS), S. Forte (U. Milan),  A. Glazov (DESY, H1), S. Moch (DESY)

{\bfseries Multi-jet final states and energy flows:}\\
C. Gwenlan (UCL, ZEUS), L. L\"onnblad (Lund),  E. Rodrigues (LHCb), G. Zanderighi (CERN)  \\
Contact persons:  S. Banerjee (CMS), D. Traynor (H1)

{\bfseries Heavy quarks (charm and beauty):}\\
M. Cacciari (Paris VI \& VII.),  A. Dainese (INFN, ALICE), A. Geiser (DESY, ZEUS), H. Spiesberger (U. Mainz)\\
Contact persons: K. Lipka (U. Hamburg, H1), Ulrich Uwer (CERN)

{\bfseries Diffraction:}\\
M. Arneodo (U. Piemonte Orientale, Novara, INFN, CMS, ZEUS), M. Diehl (DESY), P. Newman (U. Birmingham, H1), V. A. Khoze (U. Durham)\\
Contact persons: A. Bruni (INFN, ZEUS), B. Cox (U. Manchester, ATLAS), R. Orava (U. Helsinki)

{\bfseries Cosmic Rays:}\\
C. Diaconu (DESY, CPPM, H1), Ch. Kiesling (MPI Munich, H1), T. Pierog (FZ Karlsruhe),

{\bfseries Monte Carlos and Tools:}\\
P. Bartalini (Taiwan, CMS),  S.Chekanov (Argonne, ZEUS), F. Krauss (IPP Durham), S. Gieseke (U. Karlsruhe) 
\\[6mm]

{\bfseries\large Advisory Committee:}\\[3mm]
G. Altarelli (CERN),
J.~Bartels~(Hamburg),
M.~Della~Negra~(CERN),
J.~Ellis~(CERN), J.~Engelen~(CERN),
G.~Gustafson~(Lund), G.~Ingelman~(Uppsala),
P.~Jenni~(CERN), R.~Klanner~(DESY),
M.~Klein~(DESY), L.~McLerran~(BNL),
T.~Nakada~(CERN), D.~Schlatter~(CERN),
F.~Schrempp~(DESY), J.~Schukraft~(CERN),
J.~Stirling~(Durham), W.K.~Tung~(Michigan~State),
A.~Wagner~(DESY), R.~Yoshida~(ANL) \\[6mm]

{\bfseries\large Supported by:}\\[3mm]
Deutsches Elektronen-Synchroton DESY (Hamburg)\\
CERN (Geneva) 
\end{flushleft}

\newpage

\begin{center}
\mbox{}\\[5mm]
{\bfseries\Large Preface }\\[1cm]
\end{center}
In April 2004 the first meeting of what would become a series of HERA-LHC 
meetings took place at CERN. Over 250 participants joined and helped to shape the 
goals and objectives of this workshop. 
These are:

\begin{itemize}
\item To identify and prioritize those measurements to be made at HERA which
have an impact on the physics reach of the LHC;
\item to encourage and stimulate transfer of knowledge between the HERA and 
LHC communities and establish an ongoing interaction;
\item to encourage and stimulate theory and phenomenology efforts;
\item to examine and improve theoretical and experimental tools;
\item to increase the quantitative understanding of the implication of 
HERA measurements on LHC physics;
\end{itemize}

That HERA deep inelastic scattering and photoproduction 
data and knowledge acquired will  have an impact on the analysis of 
LHC data is a priori obvious. 
First and foremost there is the question on the structure of 
the proton. 
HERA is the first and so far only collider for 
lepton-proton scattering to date. The data from the 27.5 GeV electron beams 
scattered on the 820 (920) GeV protons have delivered an accurate picture of the 
Structure of the proton in a wide kinematic range.
Precise predictions of cross sections at the LHC critically depend on 
the knowledge of the parton density functions (PDFs)
in the proton. It can be the 
largest uncertainty in  measurements, as the detector systematics will get
under control to the anticipated level. The precision 
measurements at HERA in the last 15 years have boosted our knowledge on the 
parton distributions by several orders in magnitude in kinematic reach and 
by specific measurements of heavy flavors, such as bottom and charm quark
PDFs. Final states allow to study multi-jet production, 
complementing the impressive LEP results by measurements in an 
environment with an additional important complication, namely in the 
presence of an object containing color: the proton. The understanding of these
data will be a key to the study of  LHC jet data especially at medium jet 
energies. Measurements of this type, together with the PDF data, allow for
precision tests of QCD dynamics, e.g. to test classical approaches such as 
DGLAP evolution, or more sophisticated ideas, including e.g. special 
$\log(1/x)$ terms, angular ordering etc. The HERA data are also instrumental in
understanding double parton scattering, a phenomenon which is expected 
to be very important at the LHC. HERA has elevated the studies of diffraction 
to precision physics, and the LHC is expected to carry on that program.
Finally, many tools have been developed over the last years for the analysis 
and understanding of HERA data, which can be adapted for future studies at 
the LHC.

In view of this anticipated synergy between HERA and the LHC the workshop
has defined six working groups
\begin{itemize}
\item Parton density functions and related questions
\item Multi-Jet final states and energy flows
\item Heavy quarks (charm and beauty)
\item Diffraction
\item Cosmic Rays, HERA and the LHC
\item Monte Carlo generators and tools
\end{itemize}

The {\it Parton Density Functions}  working group had the most obvious task, namely getting to understand
what the present precision - both from data and from theory-- is to determine 
parton distributions, and what are the consequences of these uncertainties
on LHC measurements. 
At an early stage in the workshop it became obvious that the combination of 
the H1 and ZEUS experiments would be very beneficial. Such lessons had been
learned from LEP and are now applied at HERA. It turns out that the gain
of a common analysis of the data of the two experiments on the precision of
the PDFs is substantially larger than when these data-sets 
are used individually in fits. 
Benchmark test have been performed to check the systematics of the different 
assumptions in the QCD fit procedures, keeping certain assumptions and data
sets in the fits fixed. 
At the start of the workshop there was some controversy on the NLO gluons at
low-x, being very different between different PDF fit groups. This could be 
resolved by measuring $F_L$, requiring lower energy running at HERA. 
The workshop has strongly supported that proposal and the last months
of HERA have been used to measure $F_L$. First results are now being released 
by the experiments.
Steps on getting towards common procedures to be used in the PDF fitting 
community and to get the most optimal PDFS 
are being defined and followed up in a special PDF forum called
PDF4LHC, which is a spin-off of this workshop

The {\it Multi-Jet final states and energy flows} working group has studied in detail the novel jet 
algorithms, designed to be infrared and collinear safe, such as the 
SISCone and the (anti)-$k_T$ algorithms. Jet algorithms and performances 
as used in the experiments are discussed. A jet quality measure has been 
defined.
The perturbative calculation of higher order corrections has been studied in detail and a
comparison of all order analytical resummation with Monte Carlo parton shower approaches
has been performed.
An important issue to understand better the details of the final states
in experimental data is the concept of $k_T$ factorization. A formalism
for extracting e.g. the needed unintegrated gluon distributions from 
fits to data is proposed. Implications for the LHC are studied e.g. on the 
case of gauge boson production, and boson production in association with
heavy quarks. Forward so called "Mueller-Navalet" jets predictions have been
made for the LHC. Very forward jet measurements opportunities e.g. using 
forward CASTOR detector in CMS look promising. Finally prompt 
photon production, high density systems and handles to multi-parton event discoveries have been
discussed.

In the {\it Heavy Quarks (Charm and Beauty)} working group a summary of experimental results 
on fragmentation functions, gluon densities and charm/beauty masses from HERA has been collected. Prospects for heavy quark measurement at the LHC are discussed. In the theory area important and significant progress has been made in the understanding of heavy quark mass effects in the evolution of parton density functions. In a common contribution from members of CTEQ and MRST the progress in understanding of mass effects and its impact on the global analysis of parton density functions is reviewed and documented. In addition also progress in the calculation of fragmentation functions including mass effects is discussed. Finally the progress in calculation of higher order corrections to $t\bar{t}$ production at the LHC is summarized.

The working group {\it Diffraction} brought about an important information transfer 
between HERA (and Tevatron) and the LHC on the experience with near beam 
detectors operation and calibration issues. Since the start of the workshop,
there are several near beam detector projects that have been launched in the 
experiments. Diffractive and forward physics is now in the blood of the LHC experiments.
CMS, TOTEM and ALICE present their physics program, also what can be achieved
without near beam detectors by using rapidity gaps instead.
Major progress has been achieved in understanding central exclusive production 
at $pp$ colliders, with a tight re-evaluation of the theoretical calculations, 
and foremost with the exclusive measurements made at the Tevatron.
Factorization in diffractive processes remains mysterious. It is known not 
to work between $ep$ and $pp$ data. Now also within $ep$ data at HERA 
it is found not to work for diffractive di-jet photoproduction events. The deployment of 
diffractive PDFs from $ep$ to $pp$ data has therefore to be done with care.

In the working group {\it  Cosmic Rays, HERA and the LHC} the impact of laboratory measurements for the understanding of the source and propagation of high energy cosmic rays has been discussed. These cosmic rays are measured mainly via air-showers and for their simulation measurements at high energy lepton hadron and hadron hadron colliders are important. The main sources of uncertainties come from cross sections (elastic and inelastic), secondary particle production and multiplicity distributions. Hadron production in the forward region especially from HERA and also the LHC can provide important constraints. On the theory side the application of perturbative QCD for the calculations, hadron production in the forward region, the relation to multi-parton interaction and non-linear effects arising at highest energies (i.e. at small $x$) have been discussed.

The goals of working group {\it  Monte Carlo and Tools} was to examine and improve the Monte Carlo event generators for the use at LHC, to provide a framework for tuning and to develop new tools and libraries for the analysis of data. The available Monte Carlo generators are reviewed and tools like HZtool and RIVET, tools for fitting like Professor and Proffit (in  {\it Multi-Jet Þnal states and energy flows} WG) are discussed. Multiparton interaction and underlying event structures was a major issue, also in close connection with {\it Multi-Jet Þnal states and energy flows}.

The special character of this workshop was -- apart from its clear charge
on the connection between HERA and LHC -- that it was alternative held at
CERN and at DESY. Note that Tevatron was always an invited guest at the 
table, and its data and interpretation of the results have always been
part of the input in the discussions.
  
The last workshop in this series was held at CERN, where the series started, 
and over 200 participants attended. This clearly shows that the 
workshop has been established to be a beacon and forum for discussions
of QCD for the preparation of the LHC. With the termination of 
the HERA accelerator in 2007 and the turn on of the LHC,  the series was terminated and the results are written up in 
these extensive proceedings. But clearly there is a need and community for 
targeted forum on LHC QCD questions, and no doubt a workshop of this kind
will emerge in the near future, as soon as the first data arrive.

Finally we wish to thank all the participants of the HERA an the LHC workshops for making this series so interesting and lively. 
We thank especially the conveners for their enormous work in the preparation of the many meetings and finally the proceedings

Last but not least we wish to thank 
A. Grabowksy, D. Denise, S. Platz and L. Schmidt for their continuous help and support during
all the meetings. We thank B. Liebaug for the design of the poster.  
We are grateful to R. Eisberg, O. Knak and S. K\"onig for recording the talks and all technical help. We thank M. Mayer, K. Sachs and M. Stein for their help in  printing the proceedings.
We are grateful to the CERN and DESY directorates for financial support
of the workshops  and for their encouragement to investigate the HERA - LHC connection in detail.

\begin{flushleft} 
Hannes Jung and Albert De Roeck
\end{flushleft}

\begin{picture}(0.001,0.001)
\end{picture}

\CLDP
\tableofcontents 
\CLDP

\pagestyle{fancy}
\setcounter{page}{1}
\pagenumbering{arabic}

\newcommand{\thepeg}{T\scalebox{0.8}{HE}PEG\xspace}
\newcommand{\alpgen}{{AlpGen}}
\newcommand{\sherpa}{{SHERPA}}

\chapter[\large  WG: Parton Density Functions]{ Working Group \\ Parton Density
Functions}
{\Large\bfseries Convenors:}
\vspace{10mm}

{\Large\itshape
M. Dittmar (ETH, Z\"urich, CMS), \\
S. Forte (U. Milan), \\
A. Glazov (DESY, H1)\\
S. Moch (DESY)
}

\CLDP
\begin{papers} 
 
\coltoctitle{Introduction}
\coltocauthor{M.~Dittmar,
S.~Forte,
A.~Glazov,
S.~Moch
G.~Altarelli, 
J.~Anderson, 
R.~D.~Ball, 
G.~Beuf, 
M.~Boonekamp, 
H.~Burkhardt, 
F.~Caola, 
M.~Ciafaloni, 
D.~Colferai, 
A.~Cooper-Sarkar,
A.~De~Roeck,
L.~Del~Debbio, 
J.~Feltesse, 
F.~Gelis, 
J.~Grebenyuk, 
A.~Guffanti, 
V. Halyo, 
J.~I.~Latorre, 
V.~Lendermann,
G.~Li\,
L.~Motyka, 
T.~Petersen 
A.~Piccione, 
V.~Radescu, 
M.~Rogal,
J.~Rojo, 
C.~Royon, 
G.~P.~Salam, 
D.~\v S\'alek, 
A.~M.~Sta\'sto,
R.~S.~Thorne, 
M.~Ubiali, 
J.~A.~M.~Vermaseren,
A.~Vogt,
G.~Watt,
C.~D.~White}
\Includeart{\CAUT}{\CTIT}{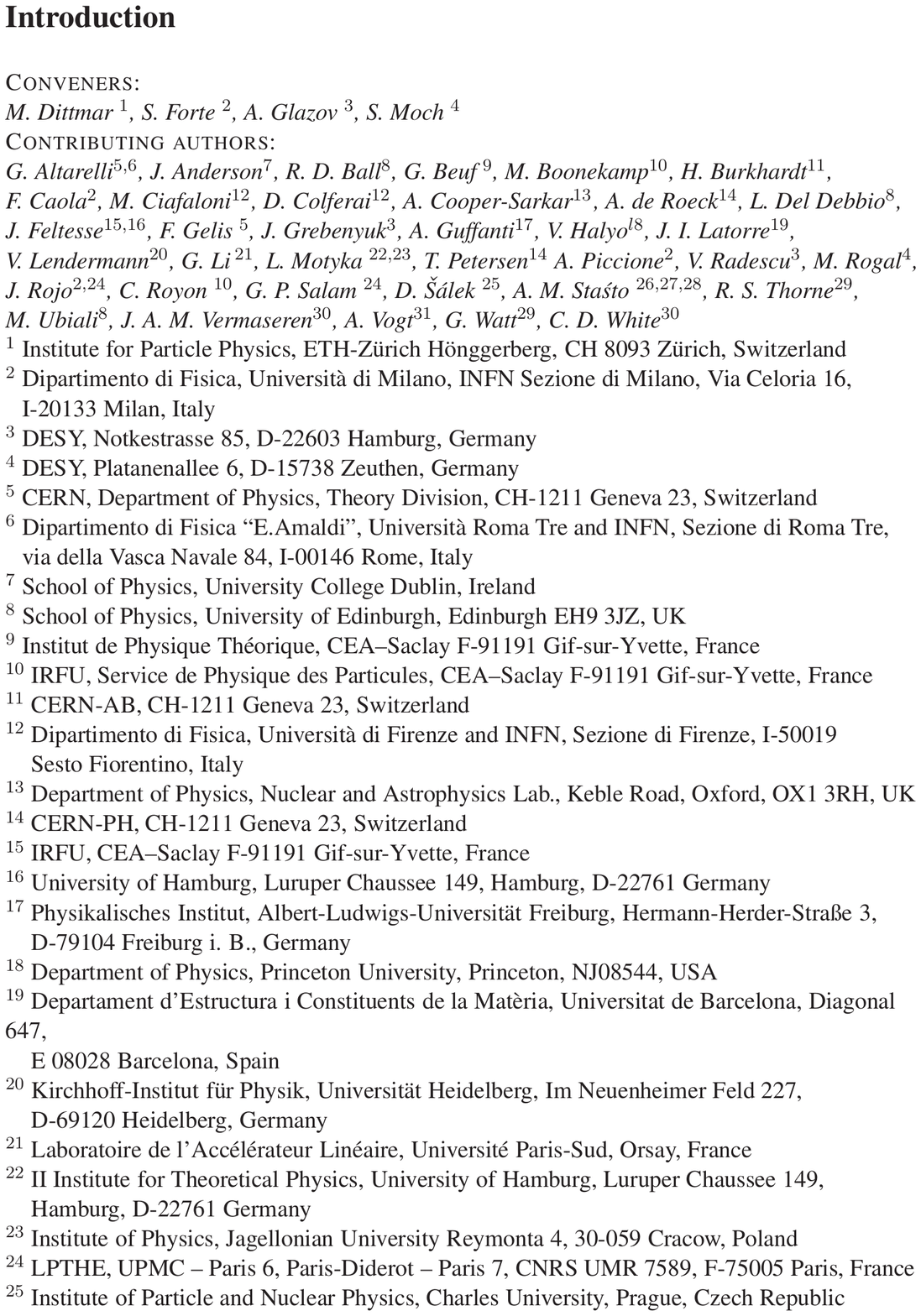}

\coltoctitle{Theoretical issues}
\coltocauthor{S.~Moch, M.~Rogal, J.~A.~M.~Vermaseren, A.~Vogt,
 G.~Altarelli, R.~D.~Ball,
  M.~Ciafaloni, D.~Colferai, G.~P.~Salam, 
A.~Sta\'sto, R.~S.~Thorne, C.~D.~White, G.~Beuf, F.~Caola, F.~Gelis, 
L.~Motyka, 
Ch.~Royon, D.~\v S\'alek, A.~M.~Sta\'sto}
\index{Moch, S.}
\index{Rogal, M.}
\index{Vermaseren, J.A.M.}
\index{Vogt, A.}
\index{Altarelli, G.}
\index{Ball, R.D.}
\index{Ciafaloni, M.}
\index{Colferai, D.} 
\index{Salam, G.P.}
\index{Sta\'sto, A.M.} 
\index{Thorne, R.S.} 
\index{White, Ch.}
\index{Beuf, G.} 
\index{Caola, F.}
\index{Gelis, F.}
\index{Motyka, L.}
\index{Royon, Ch.}
\index{S\'alek, D.}
\index{Sta\'sto, A.M.}
\Includeart{\CAUT}{\CTIT}{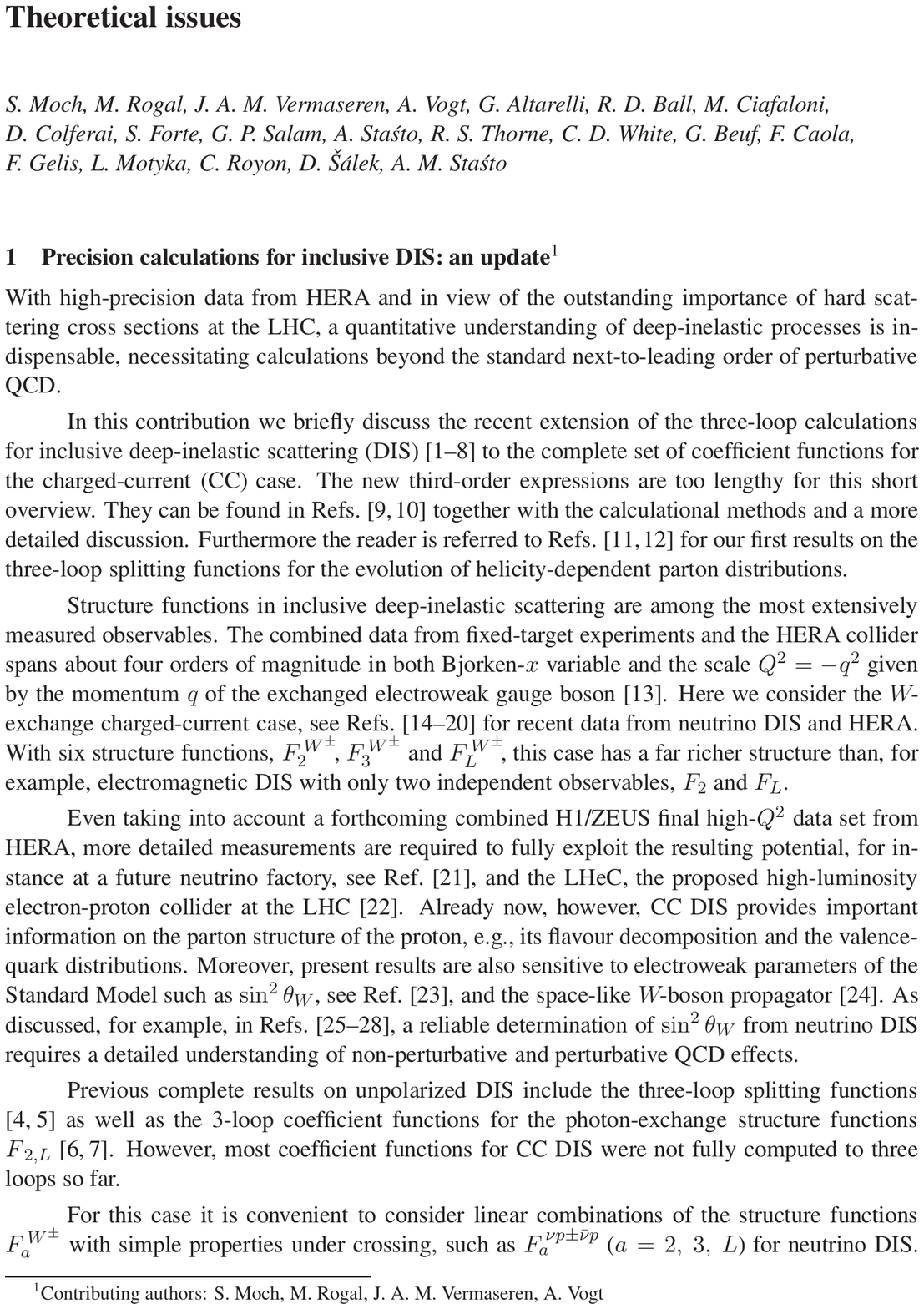}

\coltoctitle{Benchmarking of parton distributions and their uncertainties}
\coltocauthor{R.~D.~Ball, L.~Del~Debbio, J.~Feltesse, 
S.~Forte, A.~Glazov, A.~Guffanti, J.~I.~Latorre,
A.~Piccione, V.~Radescu, J.~Rojo, R.~S.~Thorne, M.~Ubiali, G.~Watt}
\index{Ball, R.D.}
\index{Del~Debbio, L.}
\index{Feltesse, J.}
\index{Forte, S.}
\index{Glazov, A.}
\index{Guffanti, A.}
\index{Latorre, J.I.}
\index{Piccione, A.}
\index{Radescu, V.}
\index{Rojo, J.}
\index{Thorne, R.S.} 
\index{Ubiali, M.}
\index{Watt, G.}
\Includeart{\CAUT}{\CTIT}{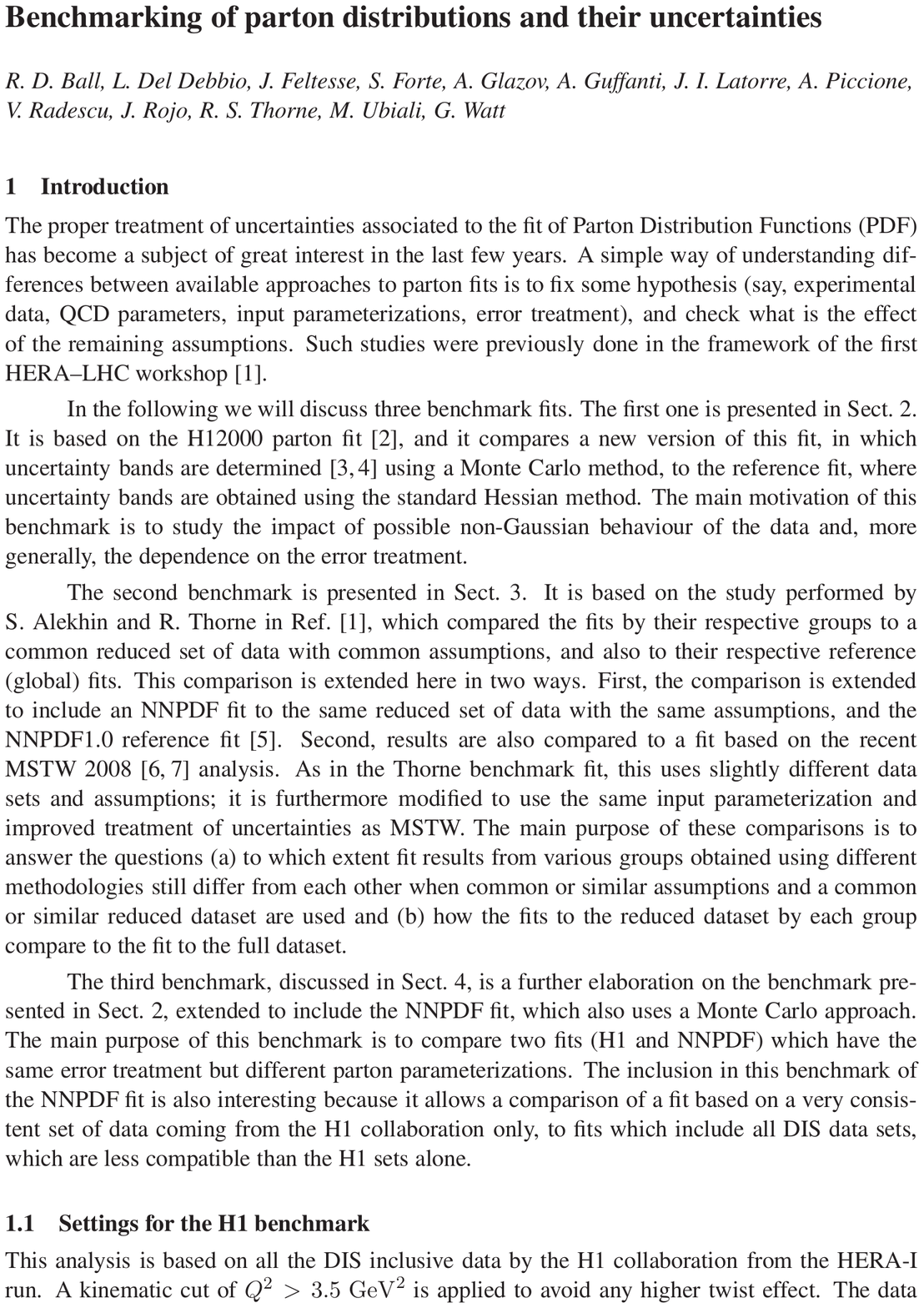}

\coltoctitle{Determination of parton distributions }
\coltocauthor{A.~Cooper-Sarkar,  A.~Glazov, G.~Li,
 J.~Grebenyuk, V.~Lendermann}
 \index{Cooper-Sarkar, A.}
 \index{Glazov, A.}
 \index{Li, G.}
 \index{Grebenyuk, J.} 
 \index{Lendermann, V.}
\Includeart{\CAUT}{\CTIT}{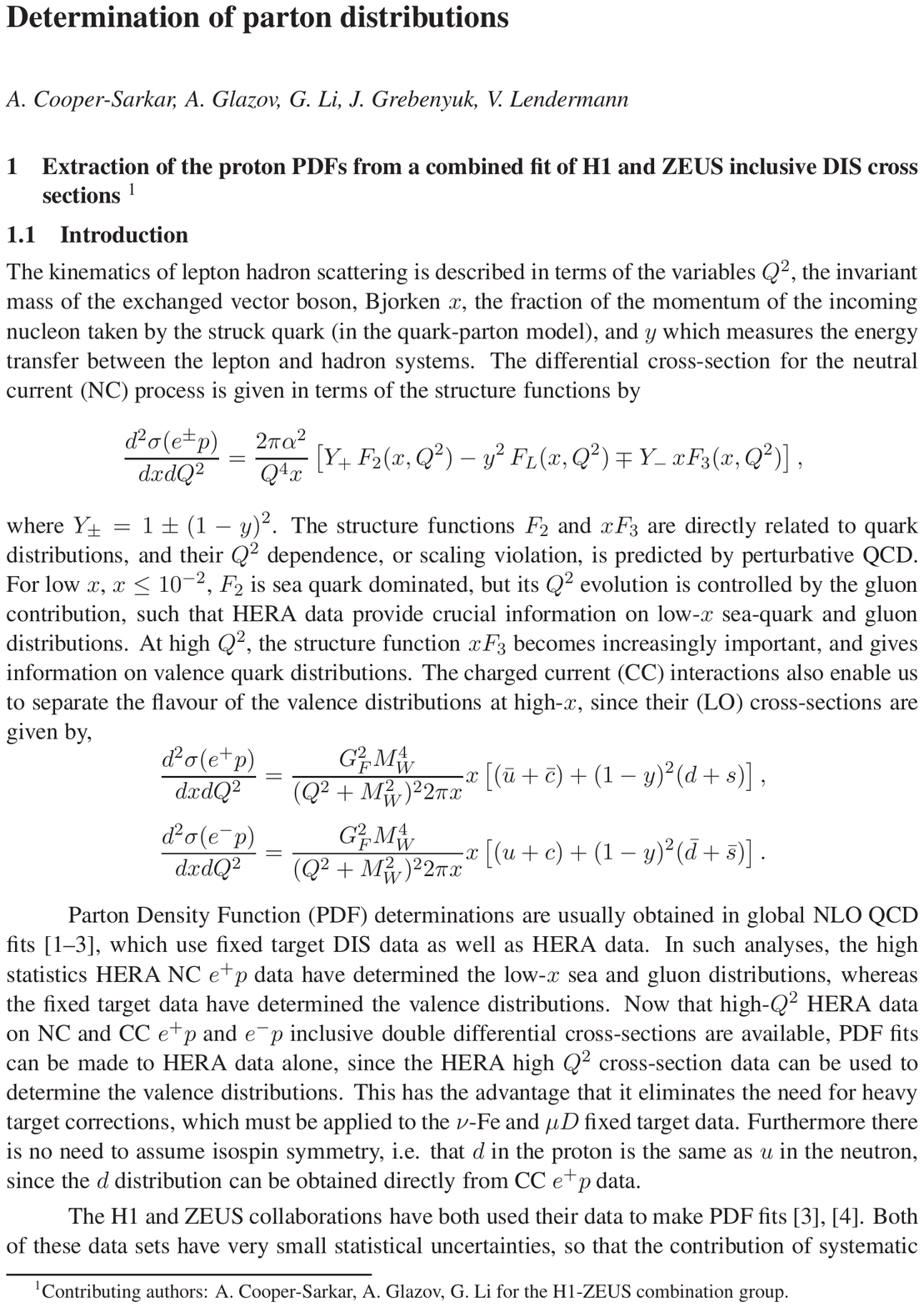}

\coltoctitle{Proton-proton luminosity, standard candles and PDFs at the LHC}
\coltocauthor{J. Anderson, M. Boonekamp,
  H. Burkhardt, M. Dittmar, V. Halyo, T. Petersen}
 \index{Anderson, J.}
 \index{Boonekamp, M.}
  \index{Burkhardt, H.}
  \index{Dittmar, M.}
  \index{Halyo, V.}
  \index{Petersen, T.}
\Includeart{\CAUT}{\CTIT}{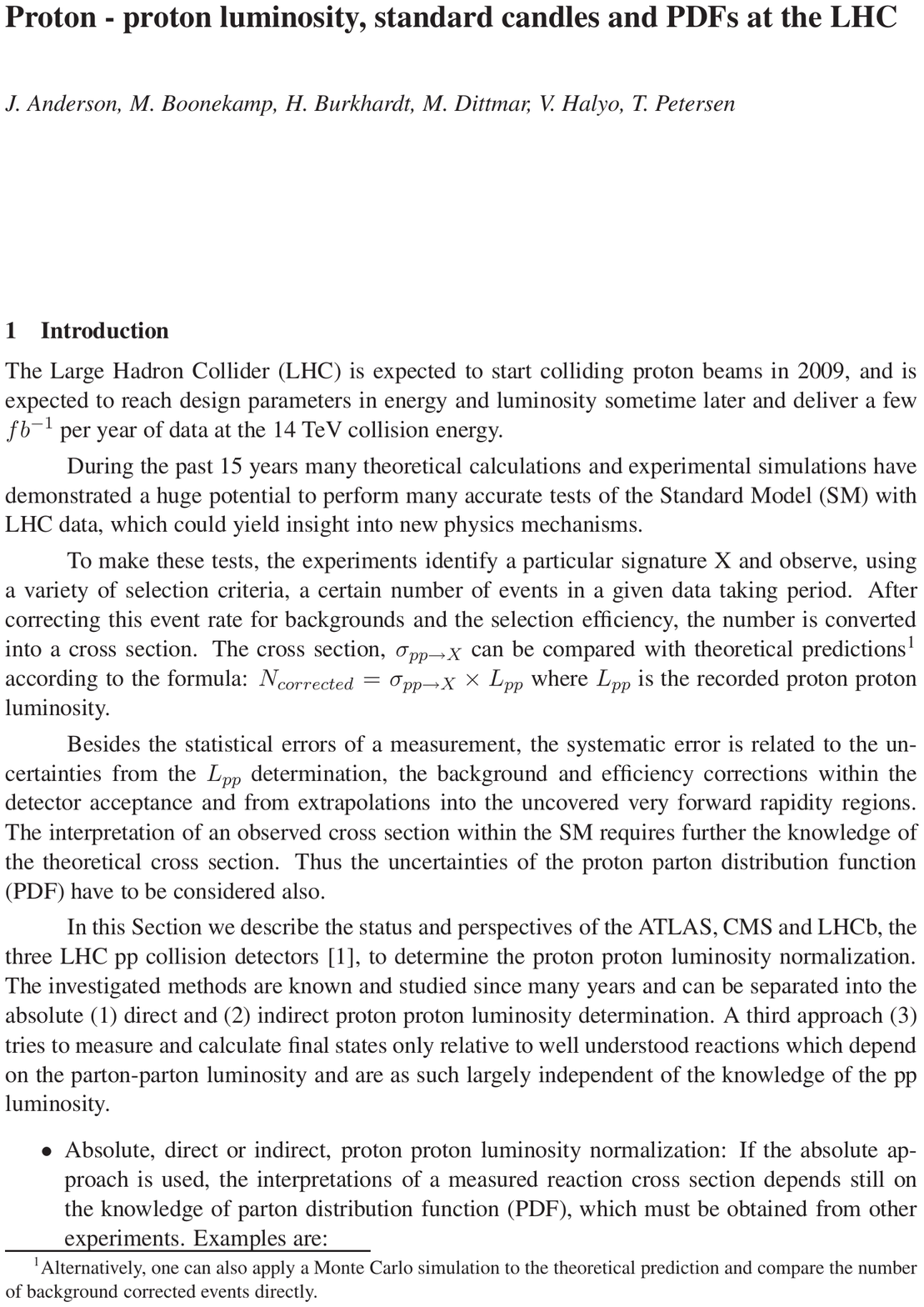}

\coltoctitle{Outlook: the PDF4LHC initiative}
\coltocauthor{A.~De~Roeck}
\index{De~Roeck, A.}
\Includeart{\CAUT}{\CTIT}{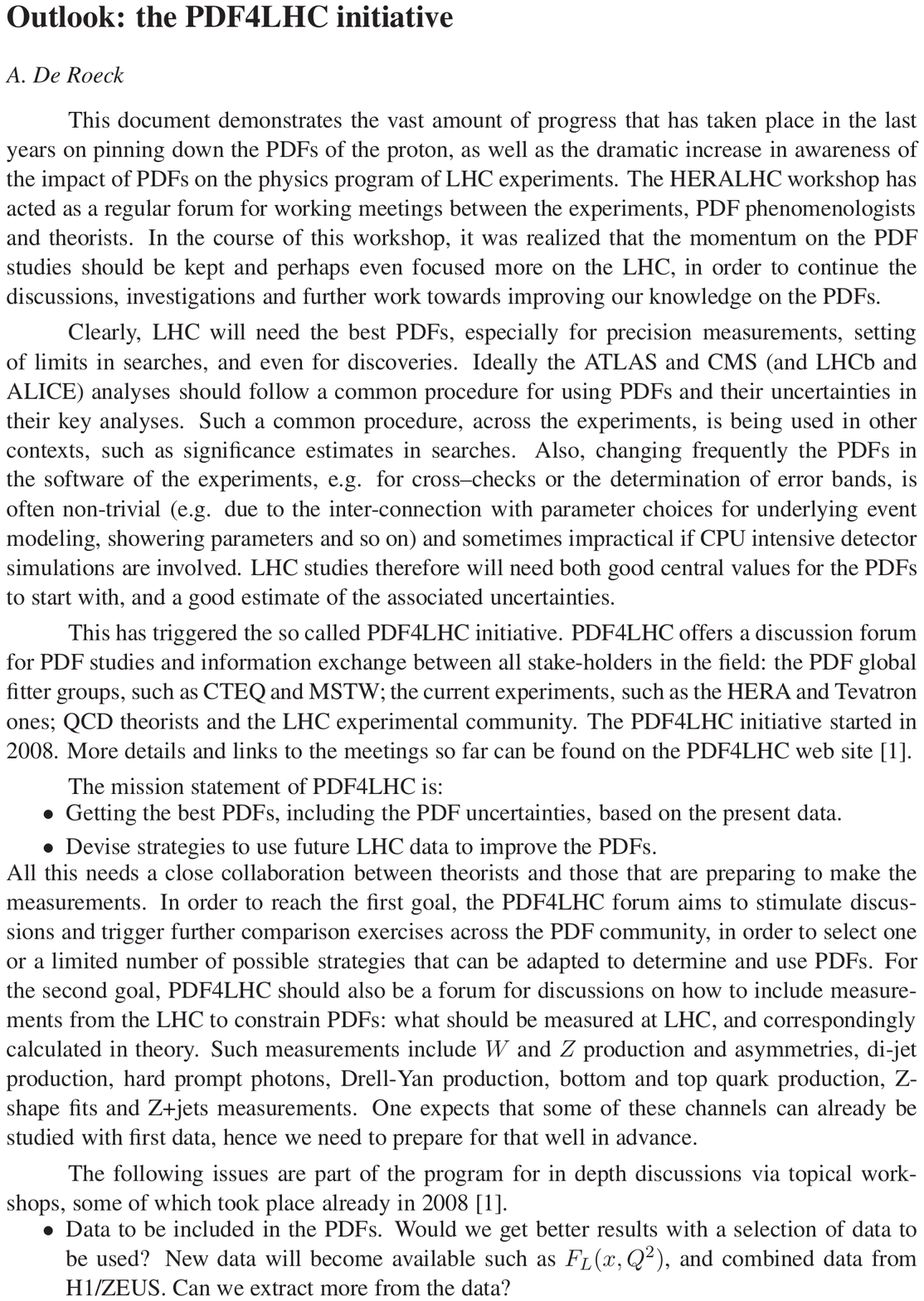}


\end{papers} 

\chapter[\large  WG: Multi-Jet Final States and Energy Flows ]{ Working Group \\
Multi-Jet Final States and Energy Flows}
{\Large\bfseries Convenors:}
\vspace{10mm}

{\Large\itshape
C. Gwenlan (UCL, ZEUS) \\
L. L\"onnblad (Lund),\\
E. Rodrigues (LHCb), \\
G. Zanderighi (CERN), \\
Contactpersons:  S. Banerjee (CMS), D. Traynor (H1)
}

\CLDP
\begin{papers} 
 
\coltoctitle{Introduction}
\coltocauthor{C.Gwenlan, .~L\"onnblad,
  E.~Rodrigues, G.~Zanderighi (Eds.),\\
  A. Bacchetta, A.~Banfi, S.~Baranov,
  J.~Bartels, A.~Bunyatyan,
  V.~Coco, G.~Corcella,
  M.~Dasgupta, M.~De\'ak,
  P.-Antoine~Delsart, I.M.~Dremin,
  F.~Hautmann, 
  S.~Joseph,
  H.~Jung, A.~Knutsson, K.
  Kutak, A. Lipatov, G.~Luisoni,
  S.~Majhi, L.~Marti,
  K.~M\"uller, T.~Namsoo,
  S.~Osman, H.~Perrey,
  G.~Rodrigo, J.~Rojo,
  Z.~R\'urikov\'a, A.~Sabio~Vera,
  C.~Sander, Th.~Sch\"orner-Sadenius,
  F.~Schwennsen, G.~Somogyi,
  G.y~Soyez, M.~Strikman,
  M.~Treccani, D.~Treleani,
  Z.~Tr\'ocs\'anyi, B.F.L.~Ward,
  S.A.~Yost, N.~Zotov
}
\index{Gwenlan, C.}
\index{L\"onnblad, L.}
  \index{Rodrigues,  E.}
  \index{Zanderighi, G.}
  \index{Bacchetta, A.}
  \index{Banfi, A.} 
  \index{Baranov, S.}
  \index{Bartels, J.}
  \index{Bunyatyan, A.}
  \index{Coco, V.}
  \index{Corcella, G.}
  \index{Dasgupta, M.} 
  \index{De\'ak, M.}
  \index{Delsart, P.A.} 
  \index{Dremin, I.M.}
  \index{Hautmann, F.} 
  \index{Joseph, S.}
  \index{Jung, H.} 
  \index{Knutsson, A.} 
  \index{Kutak, K.} 
  \index{Lipatov, A.}
  \index{Luisoni, G.}
  \index{Majhi, A.}
  \index{Marti, L.}
 \index{M\"uller, K.} 
  \index{Namsoo, T.}
  \index{Osman, S.}
  \index{Perrey, H.}
  \index{Rodrigo, G.} 
  \index{Rojo, J.}
  \index{R\'urikov\'a, Z.} 
 \index{Sabio~Vera, A.}
  \index{Sander, C.}
  \index{Sch\"orner-Sadenius, Th.}
  \index{Schwennsen, F.}
  \index{Somogyi, G.}
  \index{Soyez, G.}
 \index{Strikman, M.}
 \index{Treccani, M.}
  \index{Treleani, D.}
 \index{Tr\'ocs\'anyi, Z.}
 \index{Ward, B.F.L.}
 \index{Yost, S.A.}
\index{Zotov, N.}
\Includeart{\CAUT}{\CTIT}{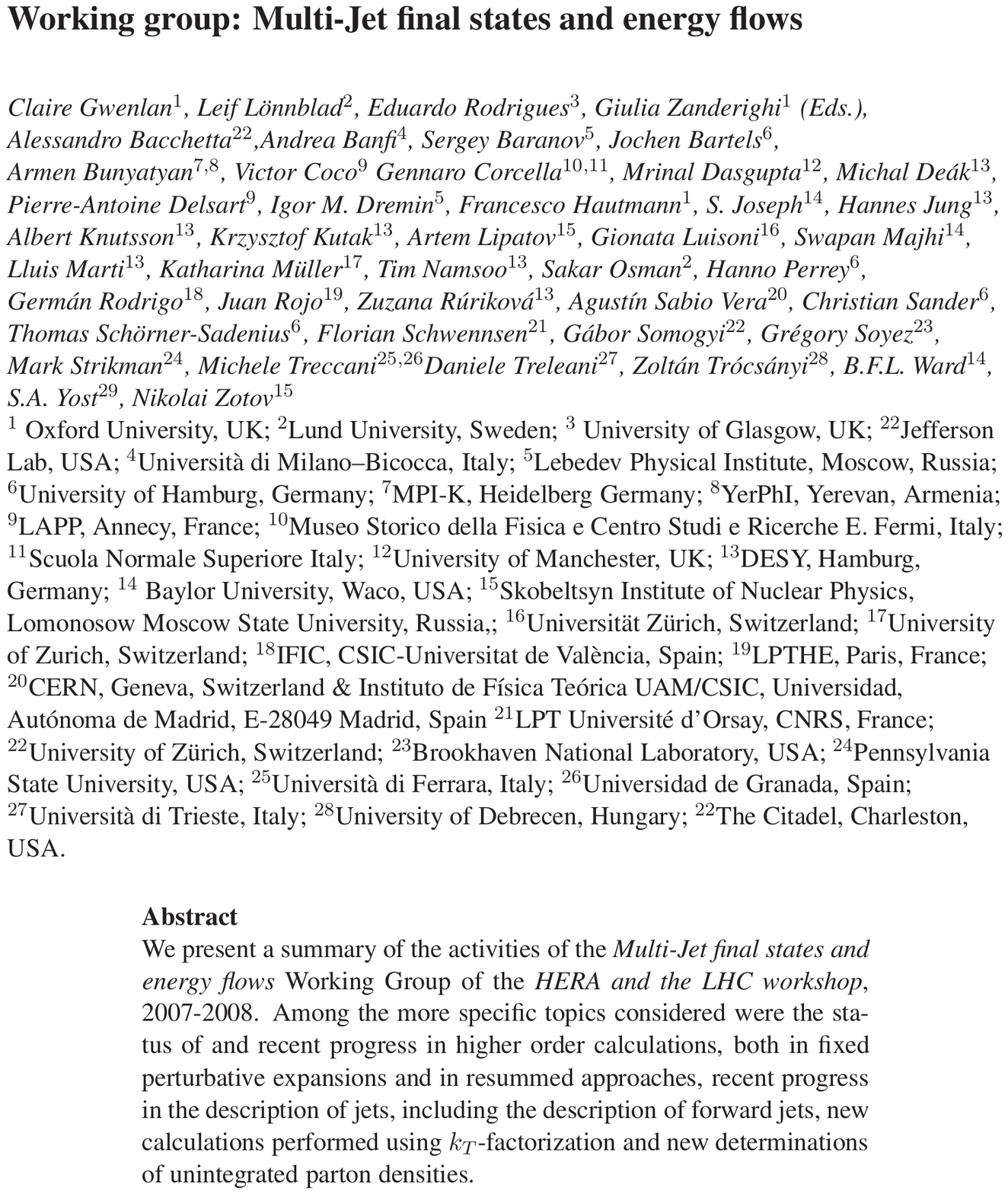}

\coltoctitle{Higher-order calculations}
\coltocauthor{G.~Zanderighi, G.~Rodrigo, M.~Treccani, G.~Somogyi}
\Includeart{\CAUT}{\CTIT}{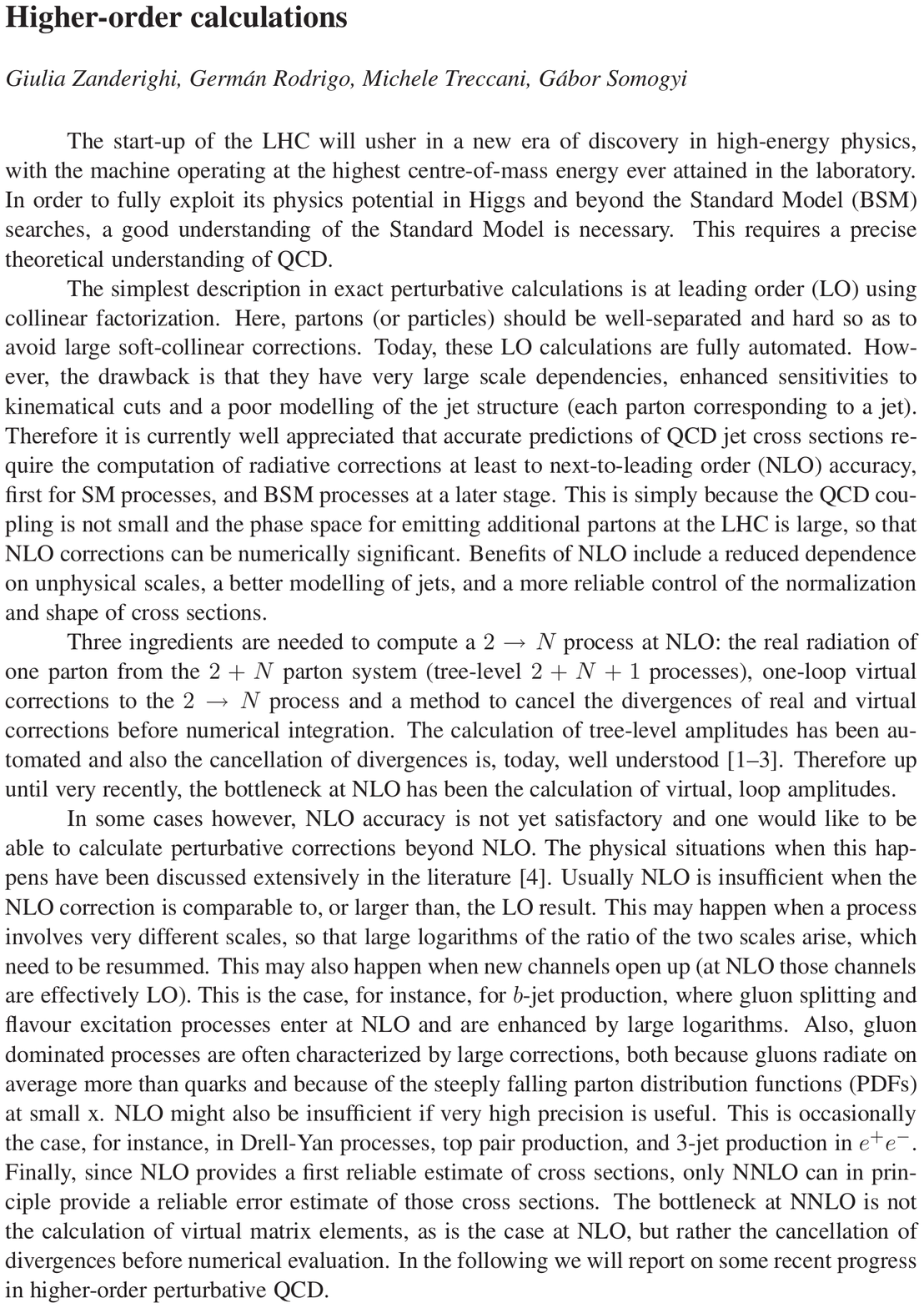}
 
\coltoctitle{Event shapes and resummation}
\coltocauthor{A.~Banfi, G.~Corcella,
  M.~Dasgupta,S.~Joseph
G.~Luisoni,
  S.~Majhi, B.F.L.~Ward,
  S.A.~Yost}
 \index{Banfi, A.}
 \index{Corcella, G.}
\index{Dasgupta, M.}
 \index{Joseph, S.}
\index{Luisoni, G.}
 \index{Majhi, S.}
\index{Ward, B.F.L.}
\index{Yost, S.A.}

\Includeart{\CAUT}{\CTIT}{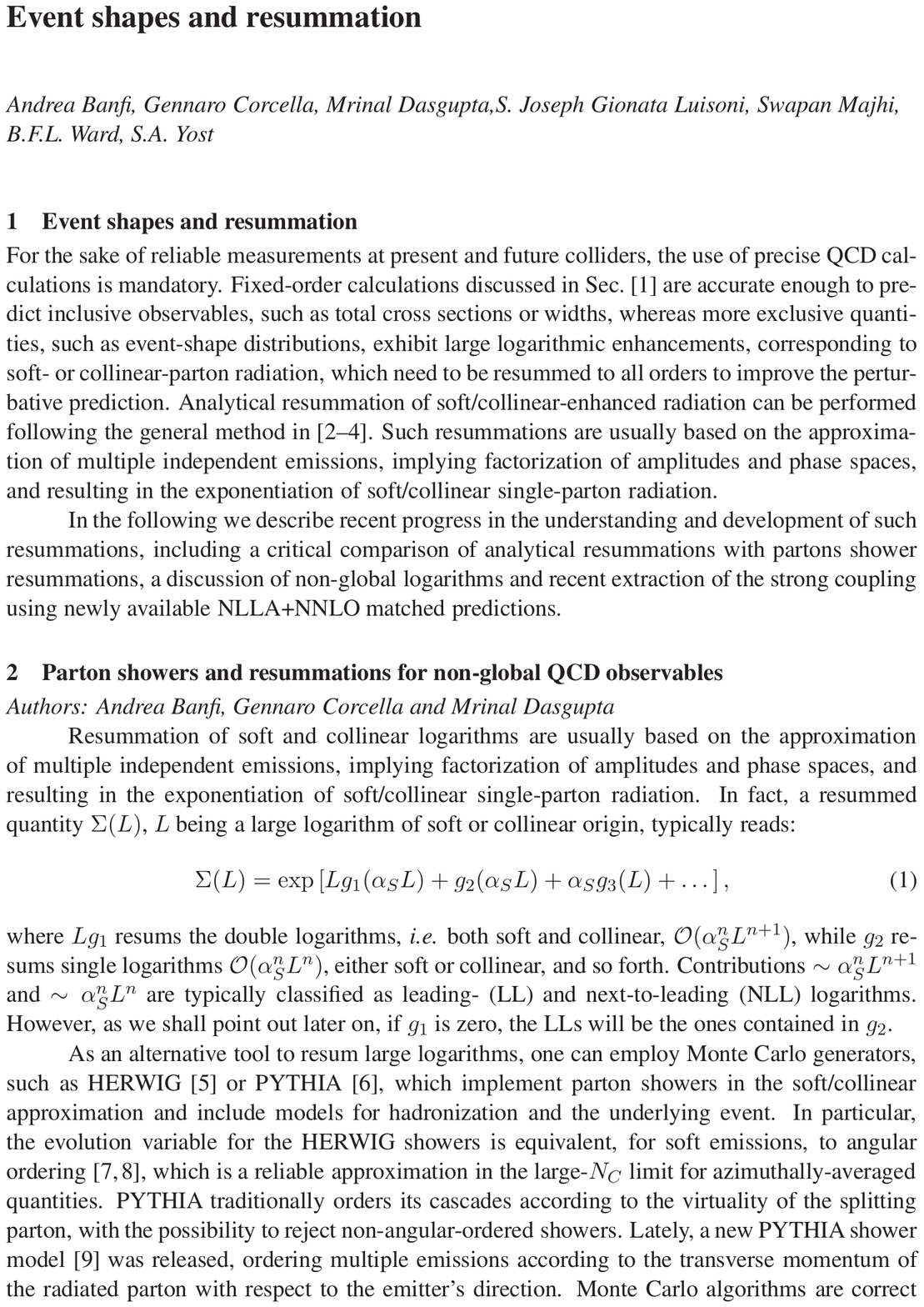}

\coltoctitle{Jets and jet algorithms}
\coltocauthor{V.~Coco, P.-Antoine~D., J.~Rojo, Ch.~Sander, G.~Soyez}
\index{Coco, V.}
\index{Delsart, P.A.} 
\index{Rojo, J.}
\index{Sander, C.}
\index{Soyez, G.}
\Includeart{\CAUT}{\CTIT}{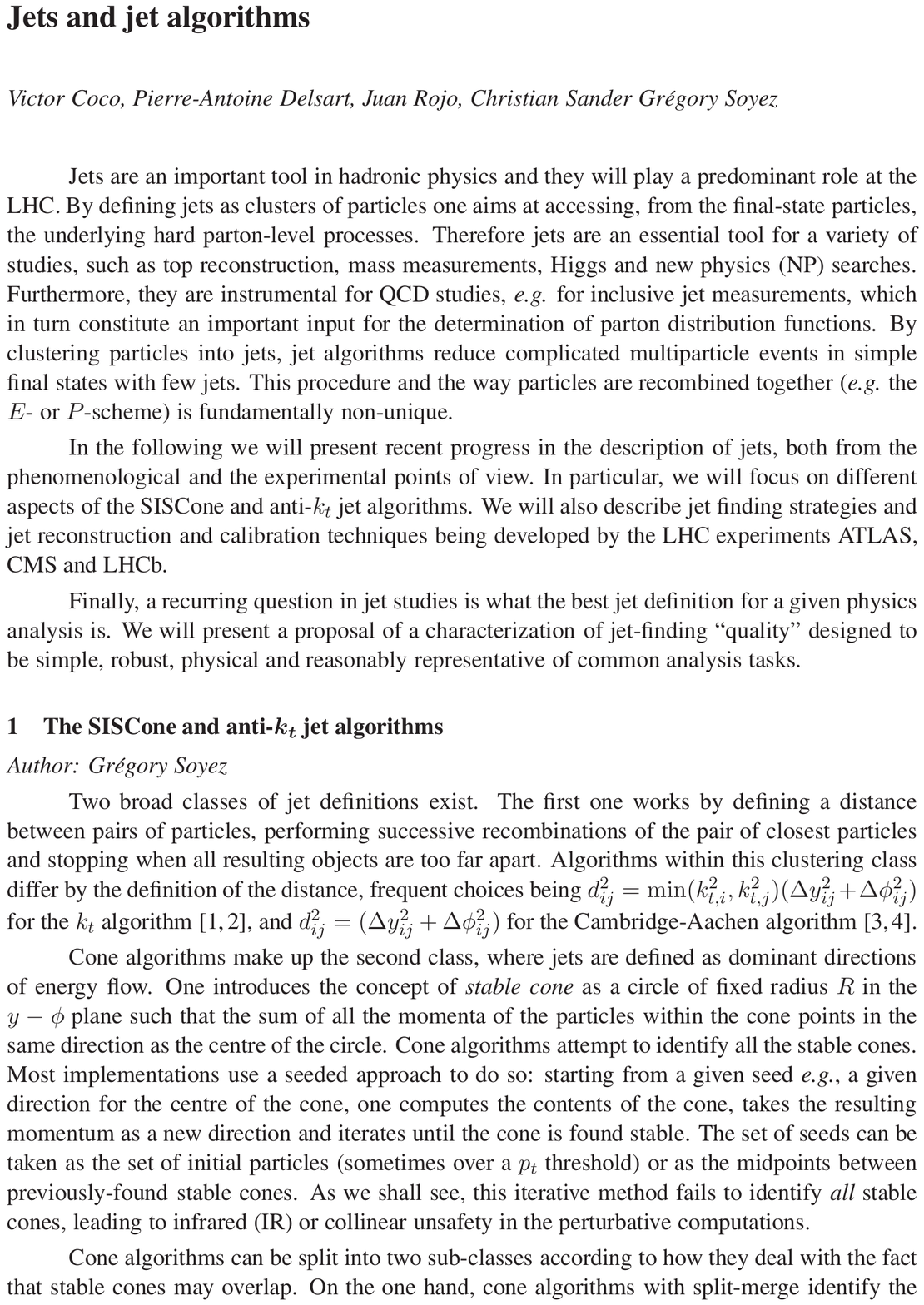}
 
\coltoctitle{\boldmath $k_\perp$-factorization and forward jets}
\coltocauthor{S.~Baranov, J.~Bartels, M.~De\'ak,
 F. Hautmann, H.~Jung, A.~Knutsson, K.~Kutak,
A.~Lipatov, Ch. Royon, A.~Sabio~Vera, F.~Schwennsen,  N.~Zotov}
\index{Baranov, S.}
\index{Bartels, J.}
\index{De\'ak, M.}
\index{Hautmann, F.} 
\index{Jung, H.}
\index{Knutsson, A.}
\index{Kutak, K.}
\index{Lipatov, A.}
\index{Royon, Ch.} 
\index{Sabio~Vera, A.} 
\index{Schwennsen, F.}
\index{Zotov, N.}

\Includeart{\CAUT}{\CTIT}{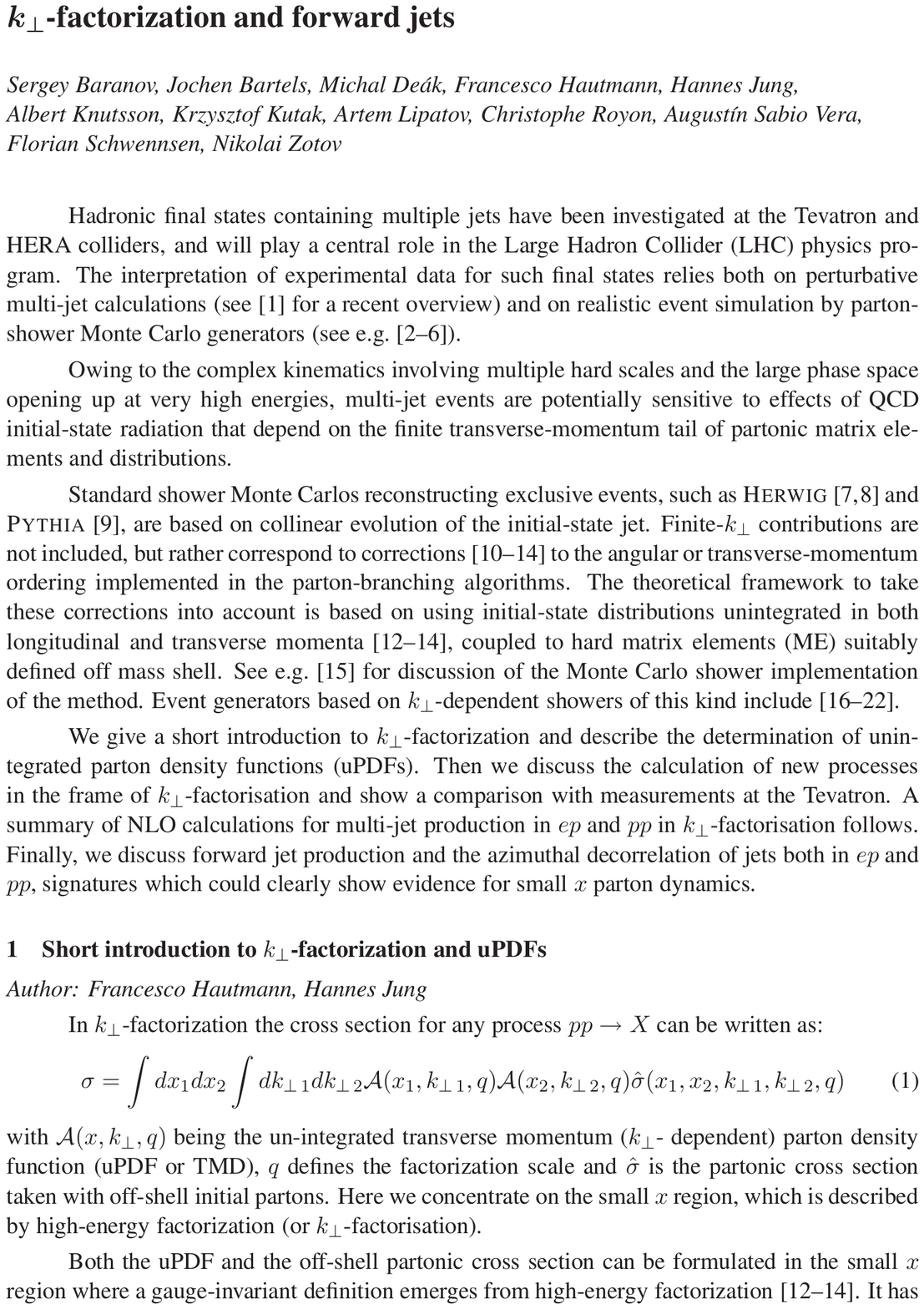}

\coltoctitle{HERA Results}
\coltocauthor{K. M\"uller, H.~Perrey,
  Th.~Sch\"orner-Sadenius}
  \index{M\"uller, K.}
   \index{Perrey, H.}
   \index{Sch\"orner-Sadenius, Th.}
\Includeart{\CAUT}{\CTIT}{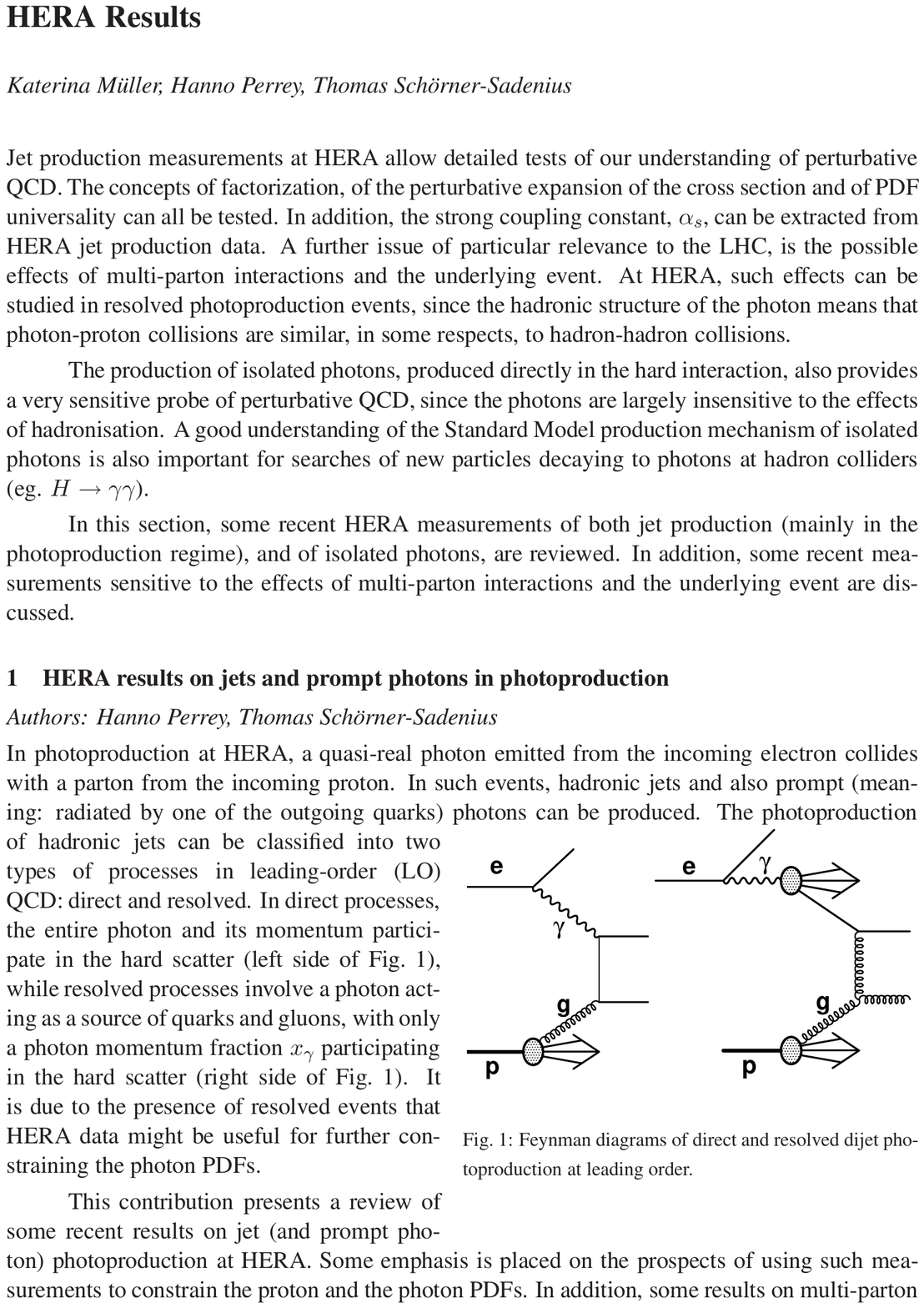}

\coltoctitle{Interactions at high gluon densities}
\coltocauthor{M.~Strikman, I.~M.~Dremin}
\index{Strikman, M.}
\index{Dremin, I.M.}
\Includeart{\CAUT}{\CTIT}{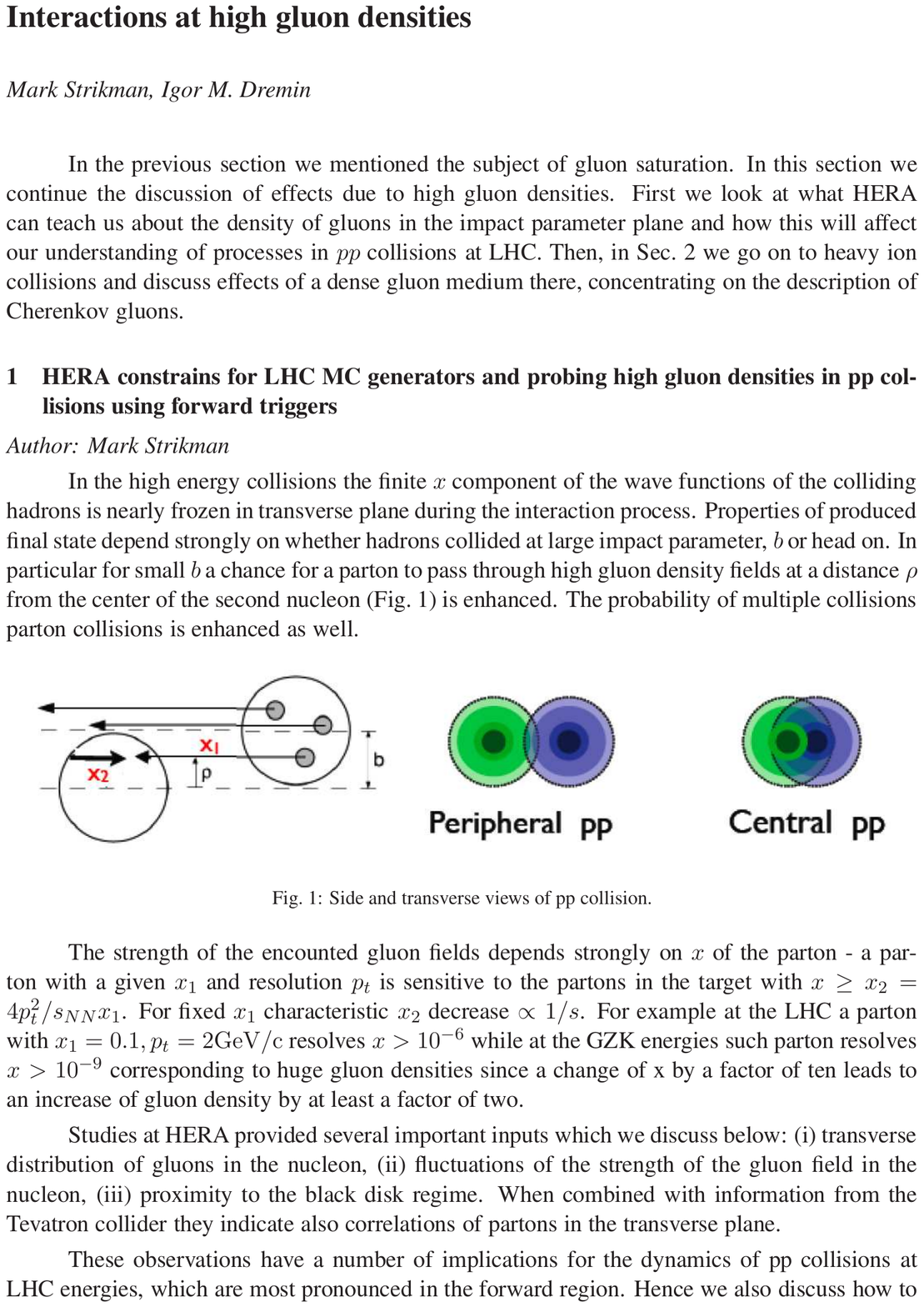}





\end{papers} 



\chapter[\large  WG: Heavy Quarks (Charm and Beauty)]{ Working Group \\ 
Heavy Quarks (Charm and Beauty)}
{\Large\bfseries Convenors:}
\vspace{10mm}

{\Large\itshape
M Cacciari (Paris VI \& VII.),\\
A. Dainese (INFN, ALICE), \\
A. Geiser (DESY, ZEUS)\\
H. Spiesberger (U. Mainz),  \\
Contactpersons:  K. Lipka (U. Hamburg, H1), Ulrich Uwer (CERN)
}

\begin{papers} 


\coltoctitle{Introduction}
\coltocauthor{M. Cacciari, A. Dainese, A. Geiser, H. Spiesberger}
\Includeart{\CAUT}{\CTIT}{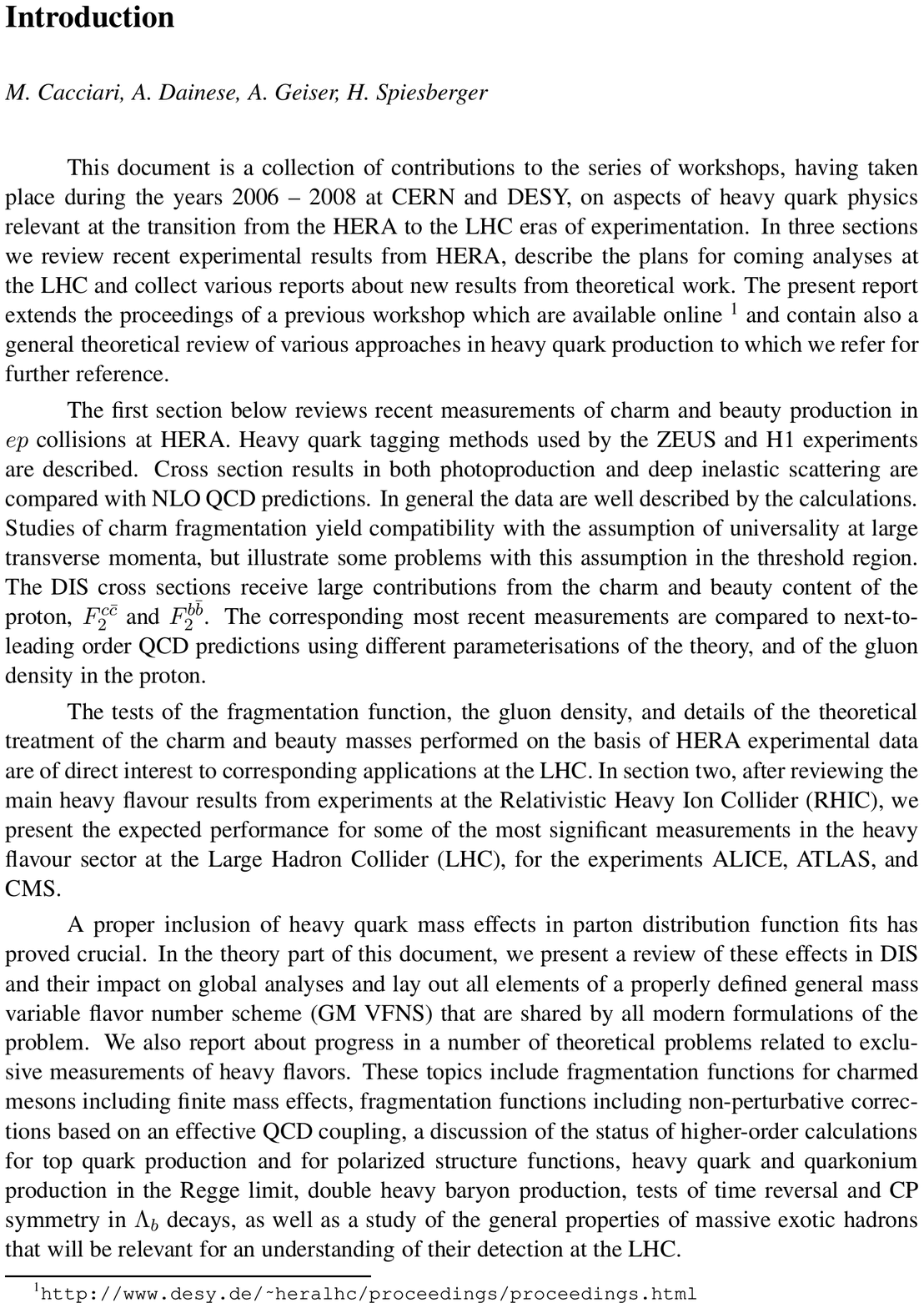}
\index{Cacciari, M.}
\index{Dainese, A.}
\index{Geiser. A.}
\index{Spiesberger, H.}

\coltoctitle{Experimental study of heavy flavour production at HERA}
\coltocauthor{S. Boutle, J. Bracinik, A. Geiser, G. Grindhammer, A.W. Jung, 
P. Roloff, Z. R\'urikov\'a, M.~Turcato, A. Yag\"ues-Molina}
\index{Boutle, S.}
\index{Bracinik, J.}
\index{Geiser, A.}
\index{Grindhammer, G.}
\index{Jung, A.W.}
\index{Roloff, P.}
\index{R\'urikov\'a, Z.} 
\index{Turcato, M.}
\index{Yag\"ues-Molina, A.}
\Includeart{\CAUT}{\CTIT}{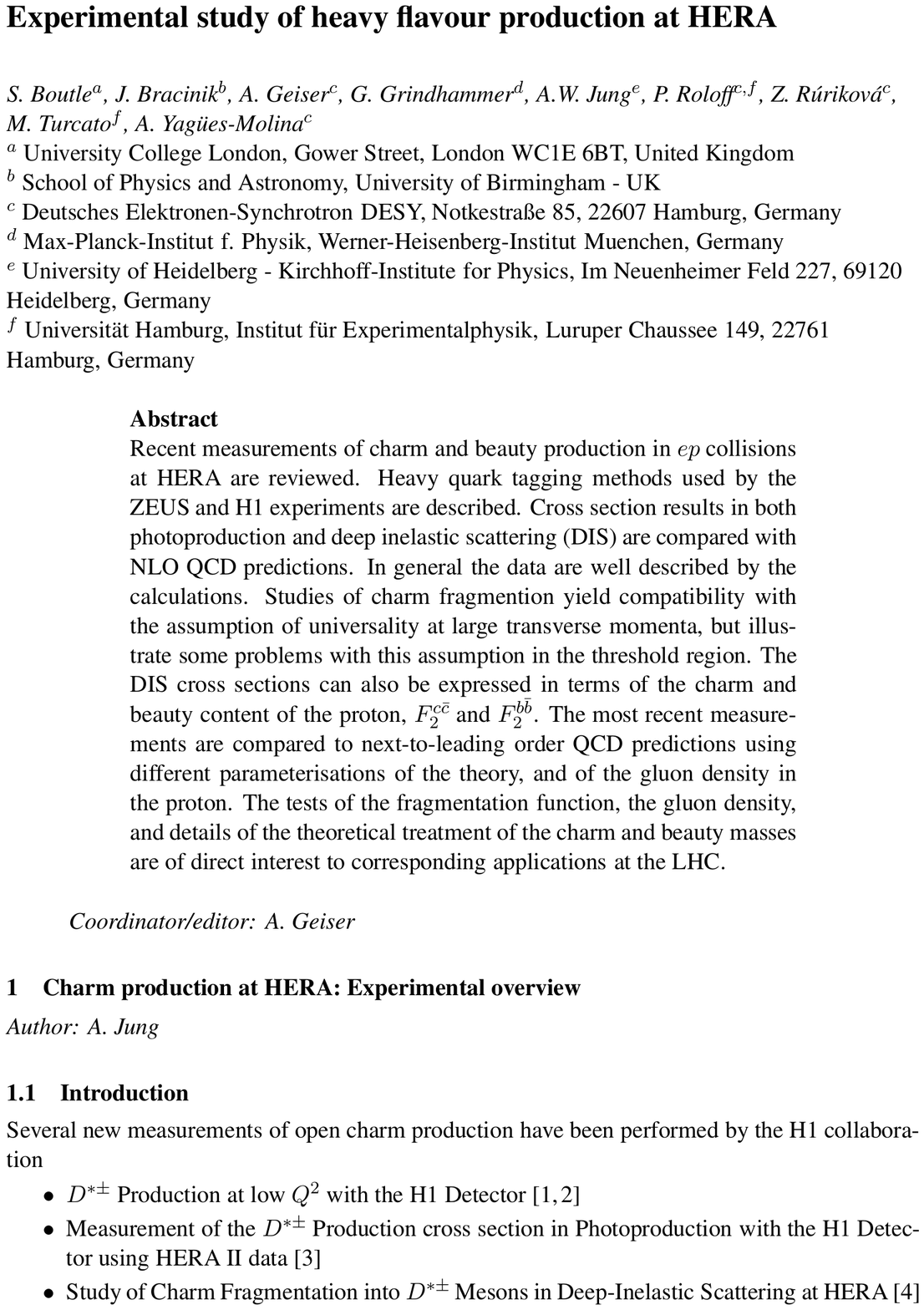}

\coltoctitle{Experimental study of heavy flavour production at RHIC and LHC}
\coltocauthor{
  M.~Biasini,
  C. Bombonati, 
  G.E. Bruno,
  E. Lytken,      
  A. Mischke,
  C. Rosemann, 
  A.~Starodumov,
  D. Stocco,
  R. Wolf,
  M. zur Nedden
        }
\index{Biasini, M.}
\index{Bombonati, C.}
\index{Bruno, G.E.}
\index{Lytken,  E.}
\index{Mischke, A.}
\index{Rosemann, C.}
\index{Starodumov, A.}
\index{Stocco, D.}
\index{Wolf, R.}
\index{zur Nedden, M.}

\Includeart{\CAUT}{\CTIT}{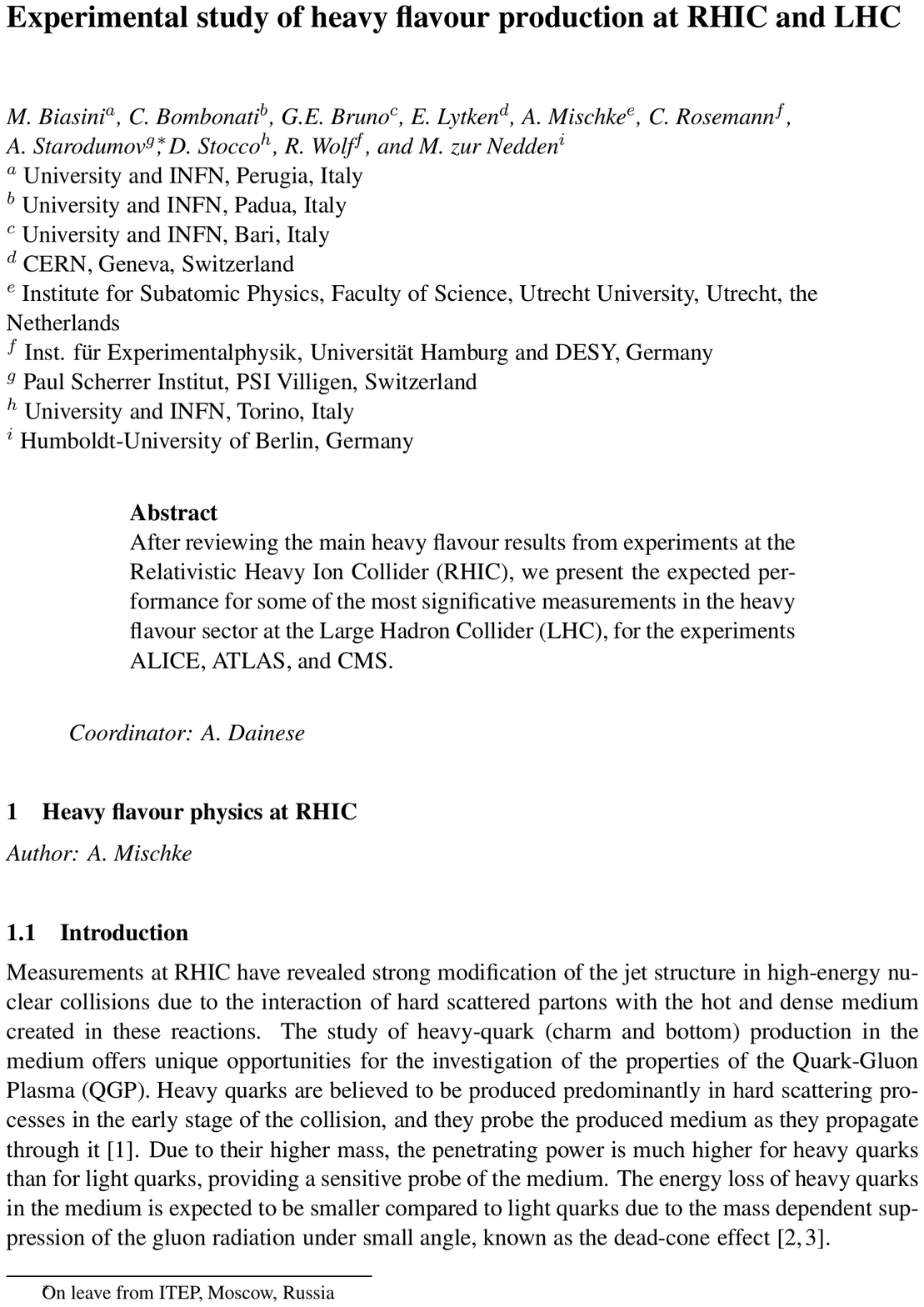}

\coltoctitle{Heavy flavour production at the LHC: Theoretical Aspects}
\coltocauthor{
  Z.J.~Ajaltouni,
  A.~Banfi,
  S.~Baranov,
  I.~Bierenbaum,
  J.~Bl\"umlein,
  G.~Corcella,
  M.~Czakon, 
  G.~Ferrera, 
  S.~Klein, 
  B.~A.~Kniehl,
  G.~Kramer, 
  A.~Likhoded, 
  D.~A.~Milstead,
  O.~I.~Piskounova,
  V.~A.~Saleev,
  I.~Schienbein,
  H.~Spiesberger,
  R.S.~Thorne,
  W.K.~Tung,
  G.~Zanderighi,
  N.~Zotov
        }
 \index{Ajaltouni, Z.J.}
\index{Banfi, A.}
\index{Baranov, S.}
\index{Bierenbaum, I.}
\index{Bl\"umlein, J.}
\index{Corcella, G.}
\index{Czakon, M.}
\index{Ferrera, G.}
\index{Klein,  S.}
\index{Kniehl, B.A.}
\index{Kramer, G.}
\index{Likhoded, A.}
\index{Milstead, D.A.}
\index{Piskounova, O.I.}
\index{Saleev, V.A.}
\index{Schienbein, I.}
\index{Spiesberger, H.}
\index{Thorne, R.S.}
\index{Tung, W.K.}
\index{Zanderighi, G.}
\index{Zotov, N.}

\Includeart{\CAUT}{\CTIT}{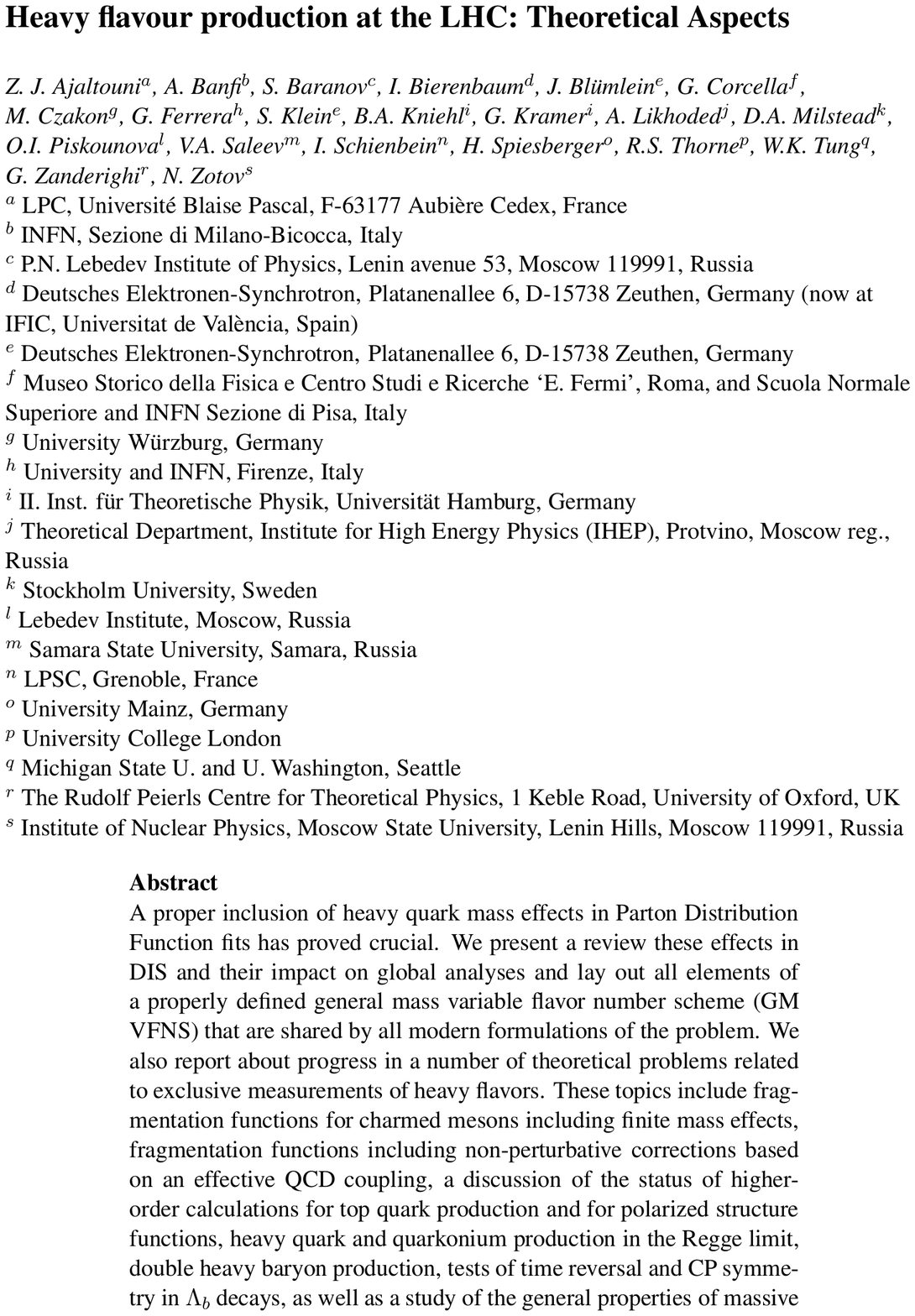}

\end{papers} 

\chapter[\large  WG: Diffraction]{ Working Group \\ Diffraction}
{\Large\bfseries Convenors:}
\vspace{10mm}

{\Large\itshape
M. Arneodo (U. Piemonte Orientale, Novara, INFN, CMS, ZEUS), \\
M. Diehl (DESY), 
P. Newman (U. Birmingham, H1)
V. A. Khoze (U. Durham) \\
\vspace{5mm}\\
Contactpersons: A. Bruni (INFN, ZEUS), B. Cox (ATLAS), R. Orava (U. Helsinki)
}\\
\vspace{10mm}

{\Large\bfseries Working Group Members:}
\vspace{5mm}

V.~Andreev,
M.~Arneodo,
J.~Bartels,
A.~Bonato,
K.~Borras,
A.~Bruni,
A.~Bunyatyan,
P.~Bussey,
F.A.~Ceccopieri,
S.~Cerci,
T.~Coughlin,
B.~Cox,
M.~Diehl,
S.~Erhan,
C.~Ewerz,
K.~Golec-Biernat,
K.~Goulianos,
M.~Grothe,
K.~Hiller,
J.~Hollar,
X.~Janssen,
M.~Kapishin,
J.~Kaspar,
V.A.~Khoze,
M.~Klasen,
G.~Kramer,
V.~Kundrat,
J.~\L{}ukasik,
A.~v.~Manteuffel,
P.~Marage,
U.~Maor,
I.~Melzer-Pellman,
A.D.~Martin,
L.~Motyka,
M.~Mozer,
P.~Newman,
H.~Niewiadomski,
C.~Nockles,
J.~Nystrand,
R.~Orava,
K.~{\"O}sterberg,
A.~Panagiotou,
A.~Pilkington,
J.L.~Pinfold,
W.~Plano,
X.~Rouby,
C.~Royon,
M.~Ruspa,
P.~Ryan,
M.G.~Ryskin,
R.~Schicker,
F.-P.~Schilling,
W.B.~Schmidke,
G.~Shaw,
W.~S\l{}omi\'n{}ski,
H.~Stenzel,
M.~Strikman,
M.~Ta\v{s}evski,
K.~Terashi,
T.~Teubner,
L.~Trentadue,
A.~Valk\'a{}rov\'a,
P.~Van~Mechelen,
A.~Vilela Pereira,
G.~Watt,
S.~Watts,
C.~Weiss,
R.~Wolf

\CLDP
\begin{papers} 

 \coltoctitle{Working Group on Diffraction: Executive Summary}
 \coltocauthor{M. Arneodo, M. Diehl, V.A.~Khoze, P. Newman}
 \index{Arneodo, M.}
  \index{Diehl, M.}
 \index{Khoze, V.A.} 
  \index{Newman, P.}
 \Includeart{\CAUT}{\CTIT}{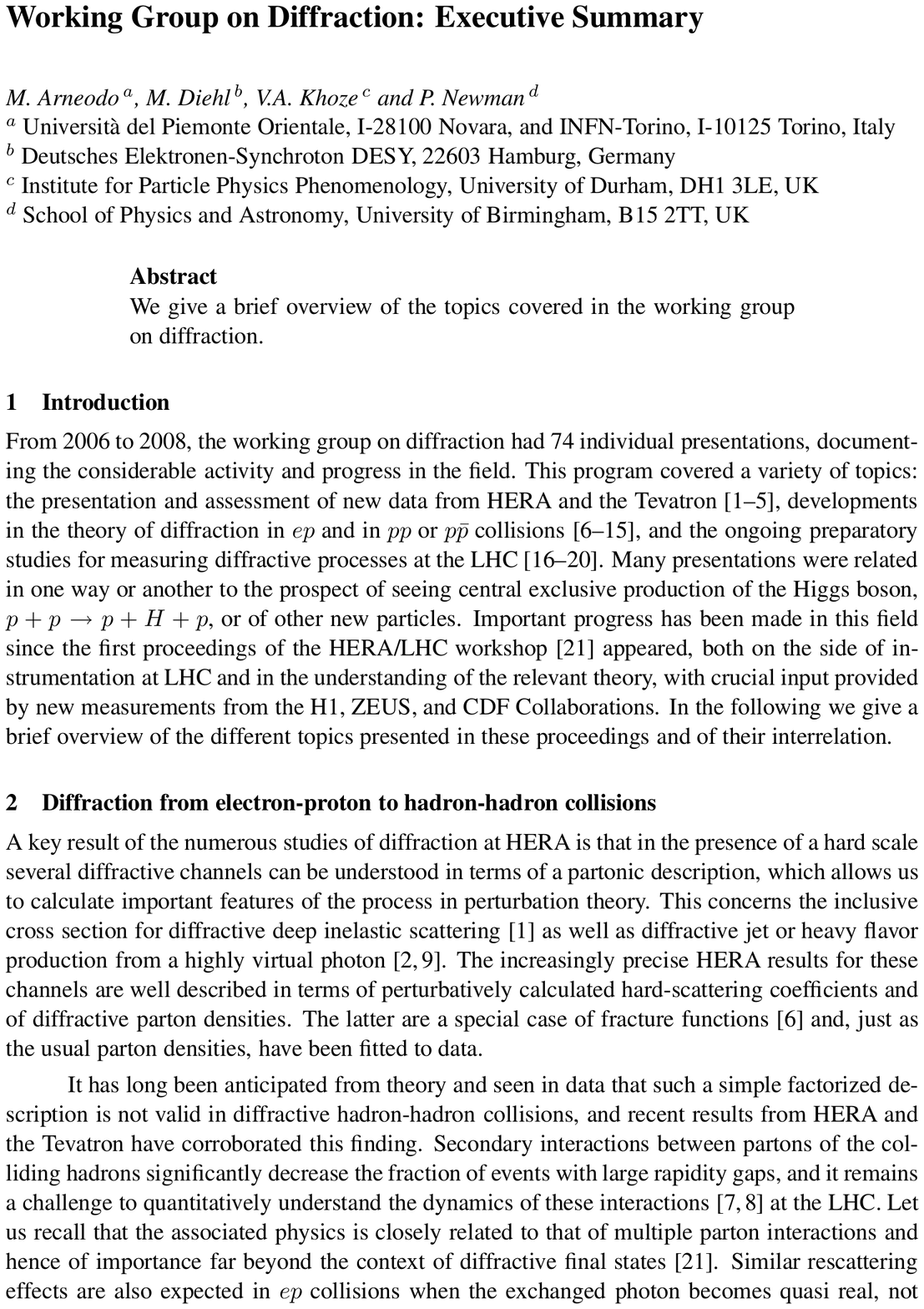}

\coltoctitle{Towards a Combined HERA Diffractive Deep Inelastic Scattering Measurement}
\coltocauthor{P. Newman, M. Ruspa}
 \index{Newman, P.}
  \index{Ruspa, M.}
 \Includeart{\CAUT}{\CTIT}{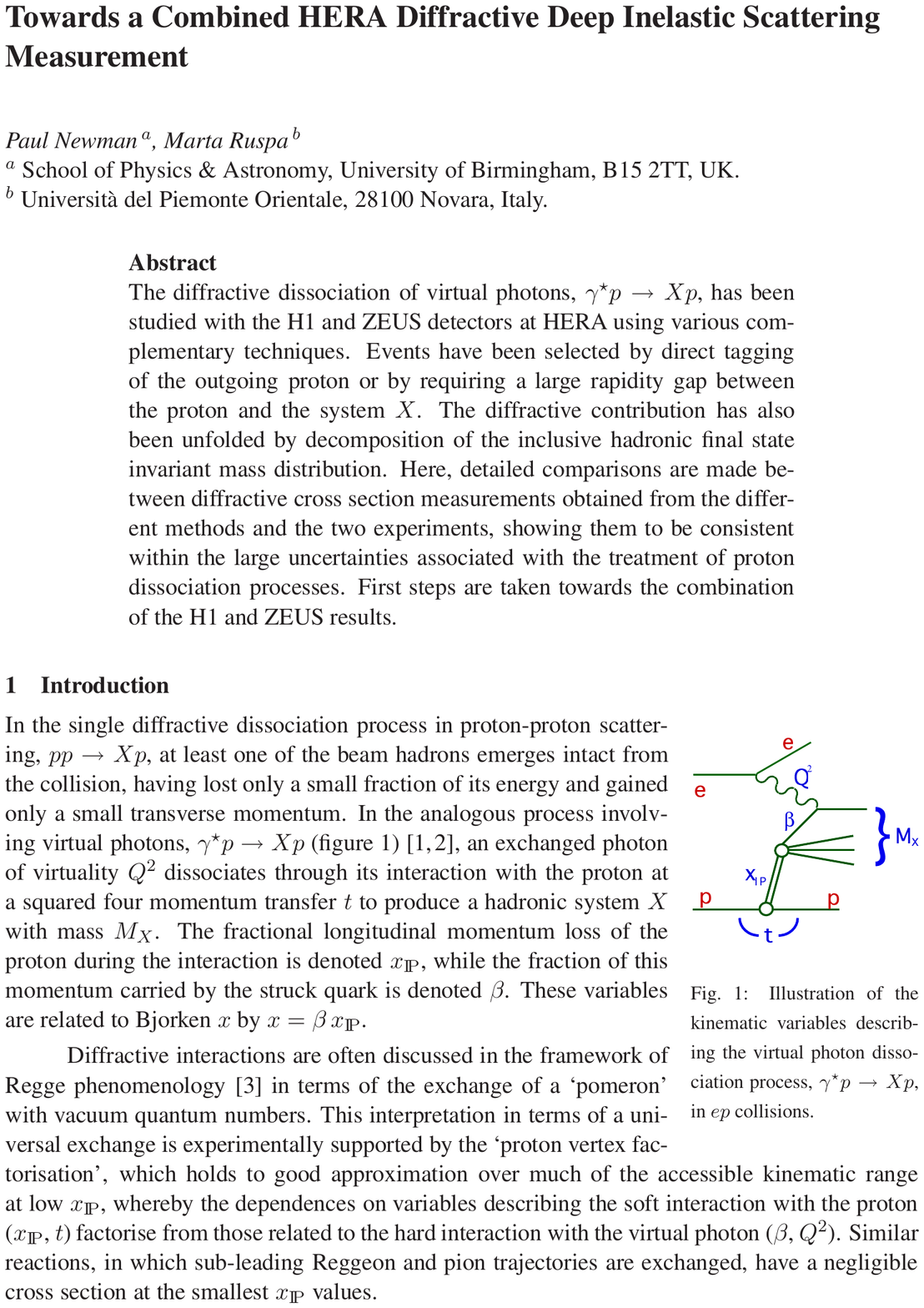}

 \coltoctitle{Diffractive Final States and Factorisation at HERA}
 \coltocauthor{W. S\l omi\'nski,  A. Valk\'arov\'a}
  \index{S\l omi\'nski, W.}
   \index{Valk\'arov\'a, A.}
 \Includeart{\CAUT}{\CTIT}{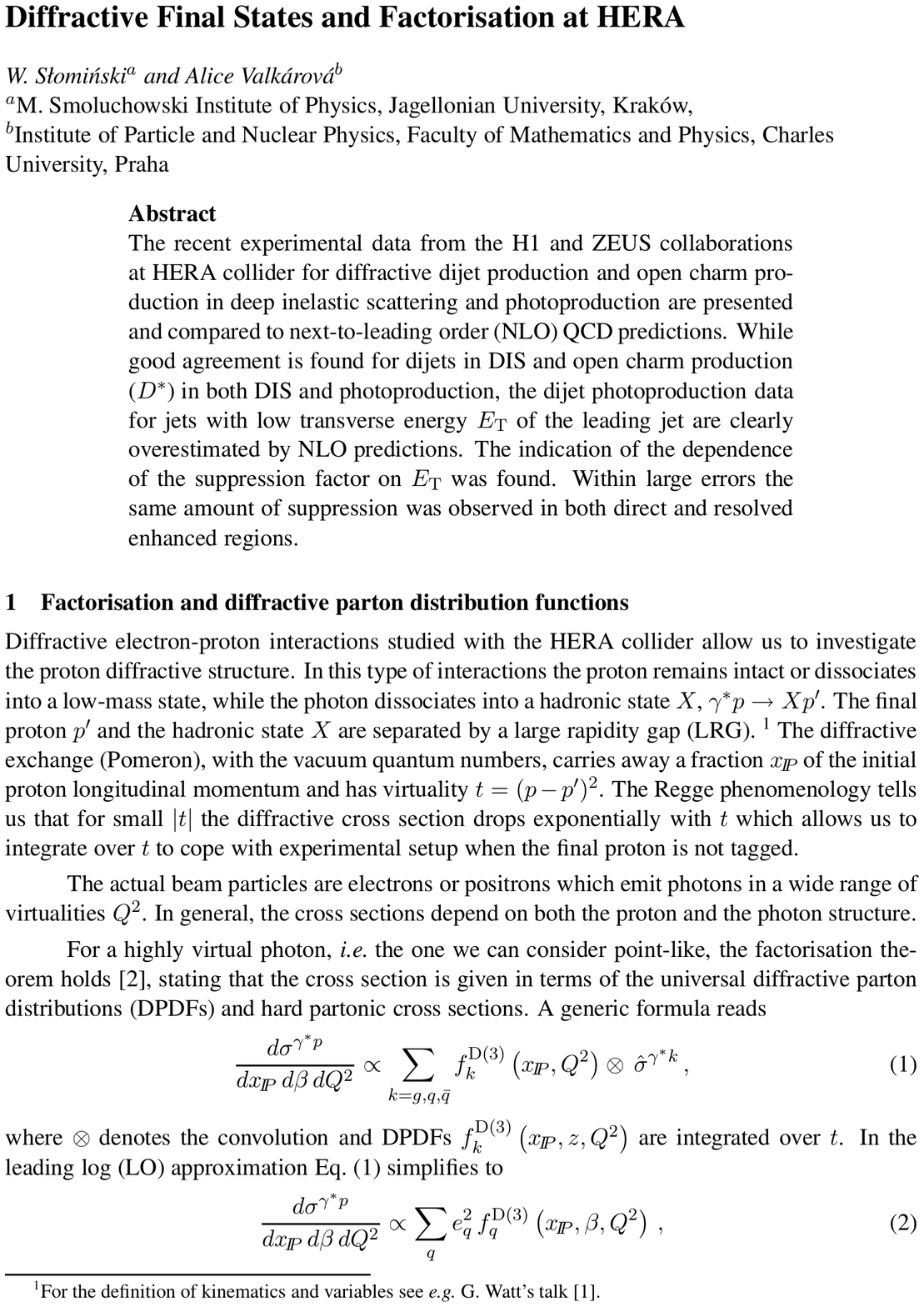}

 \coltoctitle{Leading Baryon Production at HERA}
 \coltocauthor{W.B. Schmidke, A. Bunyatyan}
 \index{Schmidke, W.B.}
 \index{Bunyatyan, A.}
 \Includeart{\CAUT}{\CTIT}{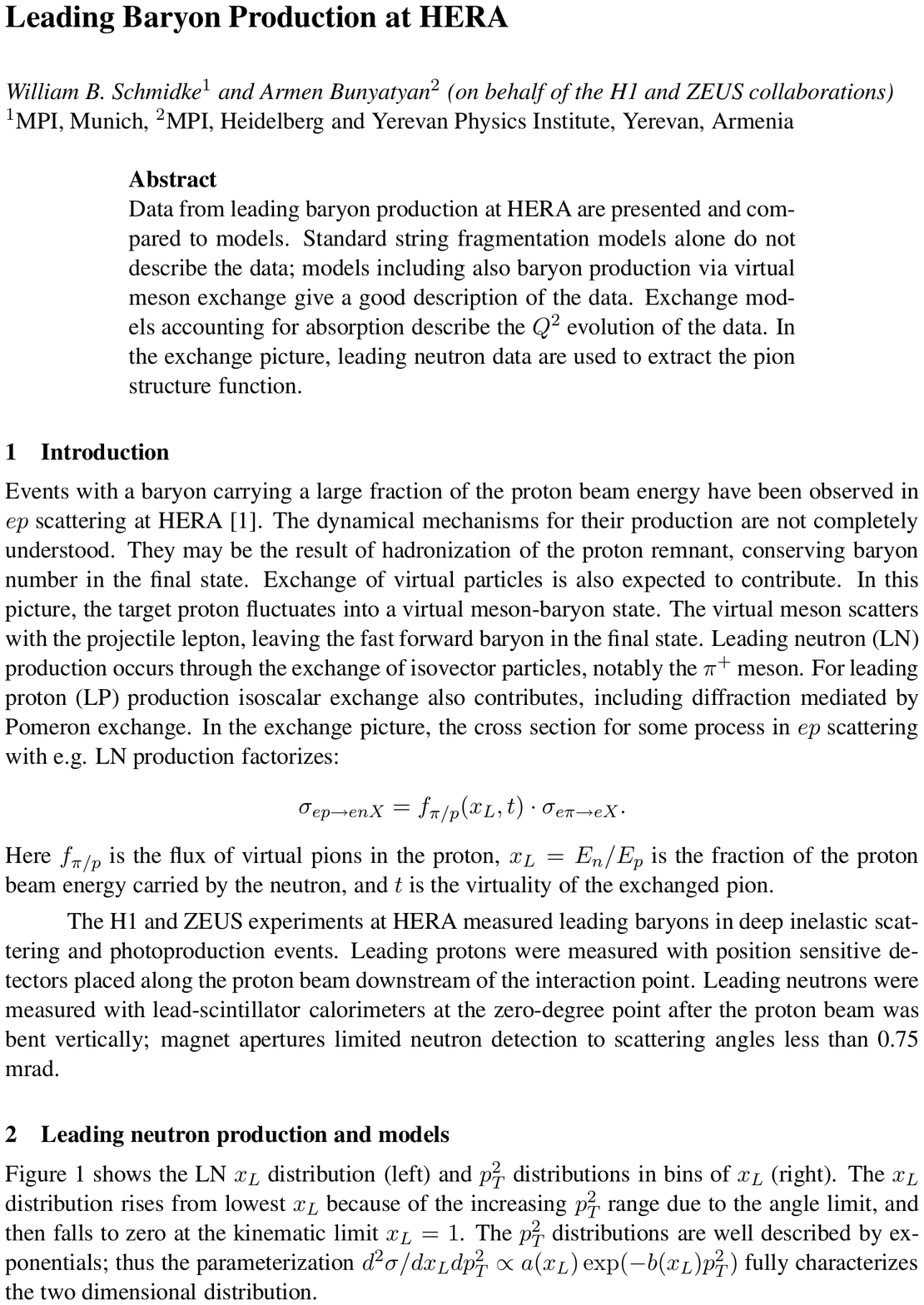}

 \coltoctitle{Exclusive Vector Meson Production and Deeply Virtual Compton Scattering at HERA}
 \coltocauthor{A. Bruni, X. Janssen, P. Marage}
 \index{Bruni, A.}
 \index{Janssen, X.} 
  \index{Marage, P.}
 \Includeart{\CAUT}{\CTIT}{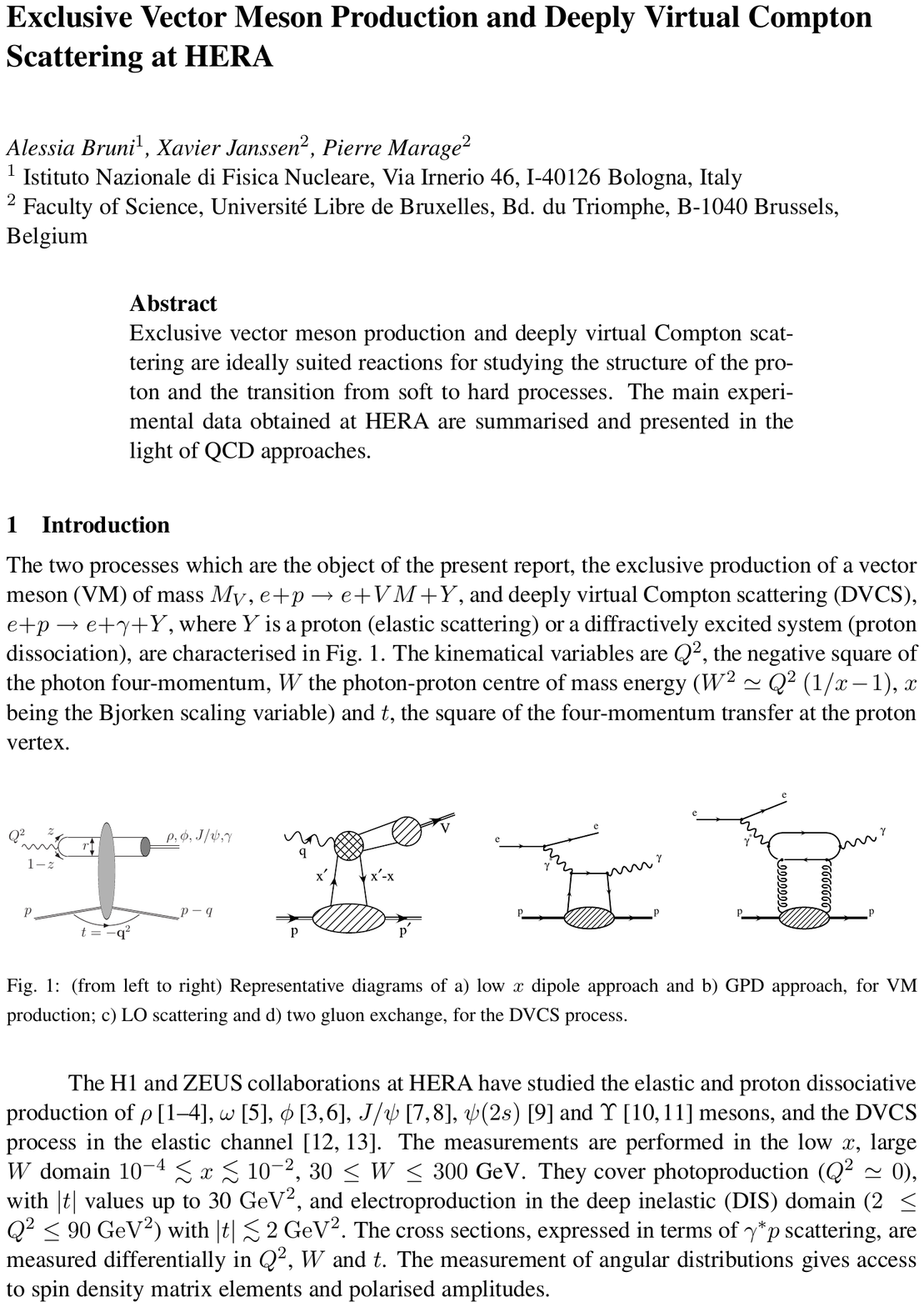}

 \coltoctitle{Exclusive Central Production and Diffractive W/Z Results from CDF~II}
 \coltocauthor{K. Goulianos, J. L. Pinfold}
 \index{Goulianos, K.}
 \index{Pinfold, J.L.}
 \Includeart{\CAUT}{\CTIT}{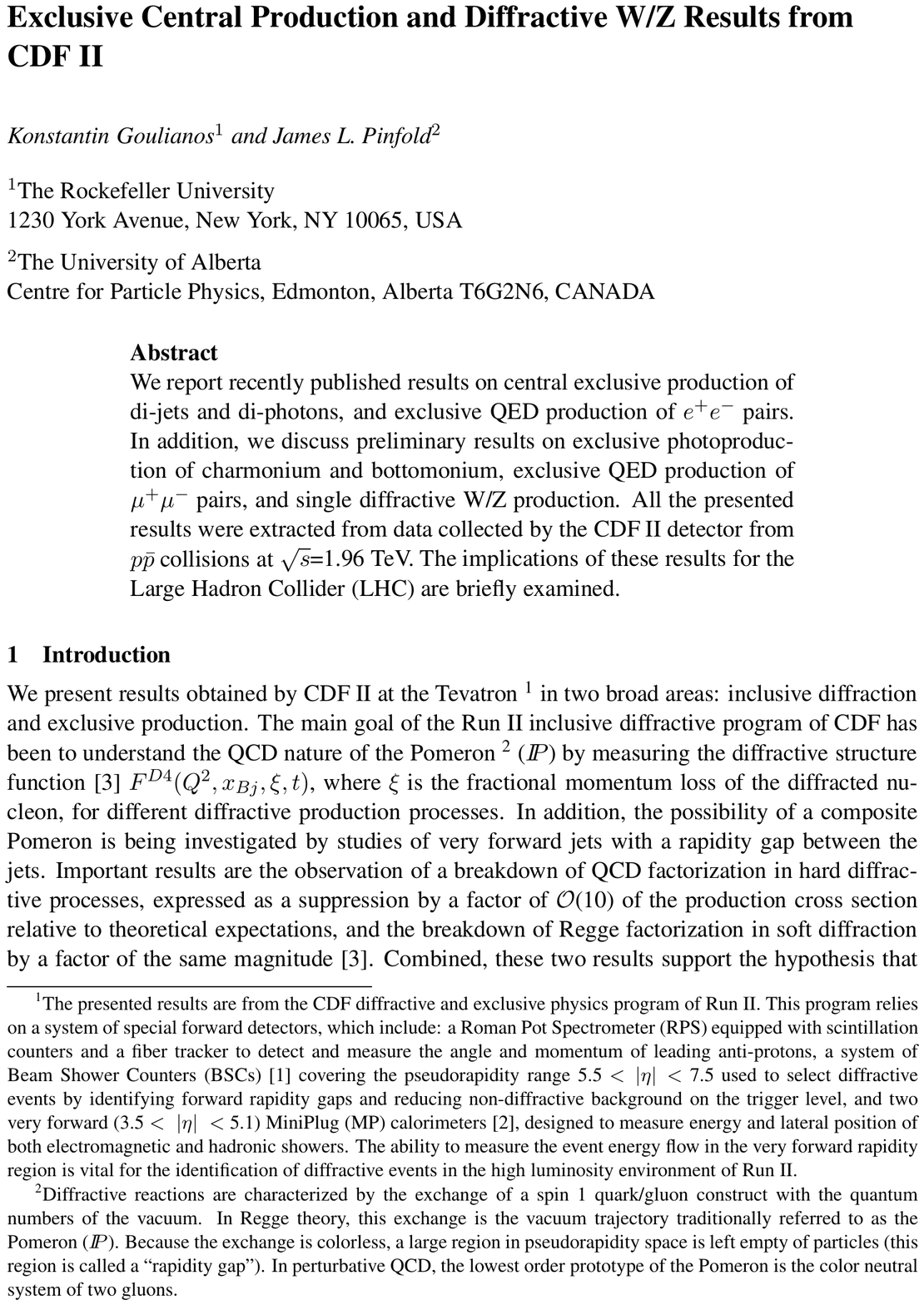}

 \coltoctitle{Survival probability in diffractive dijet photoproduction}
 \coltocauthor{M. Klasen, G. Kramer}
 \index{Klasen, M.}
 \index{Kramer, G.}
 \Includeart{\CAUT}{\CTIT}{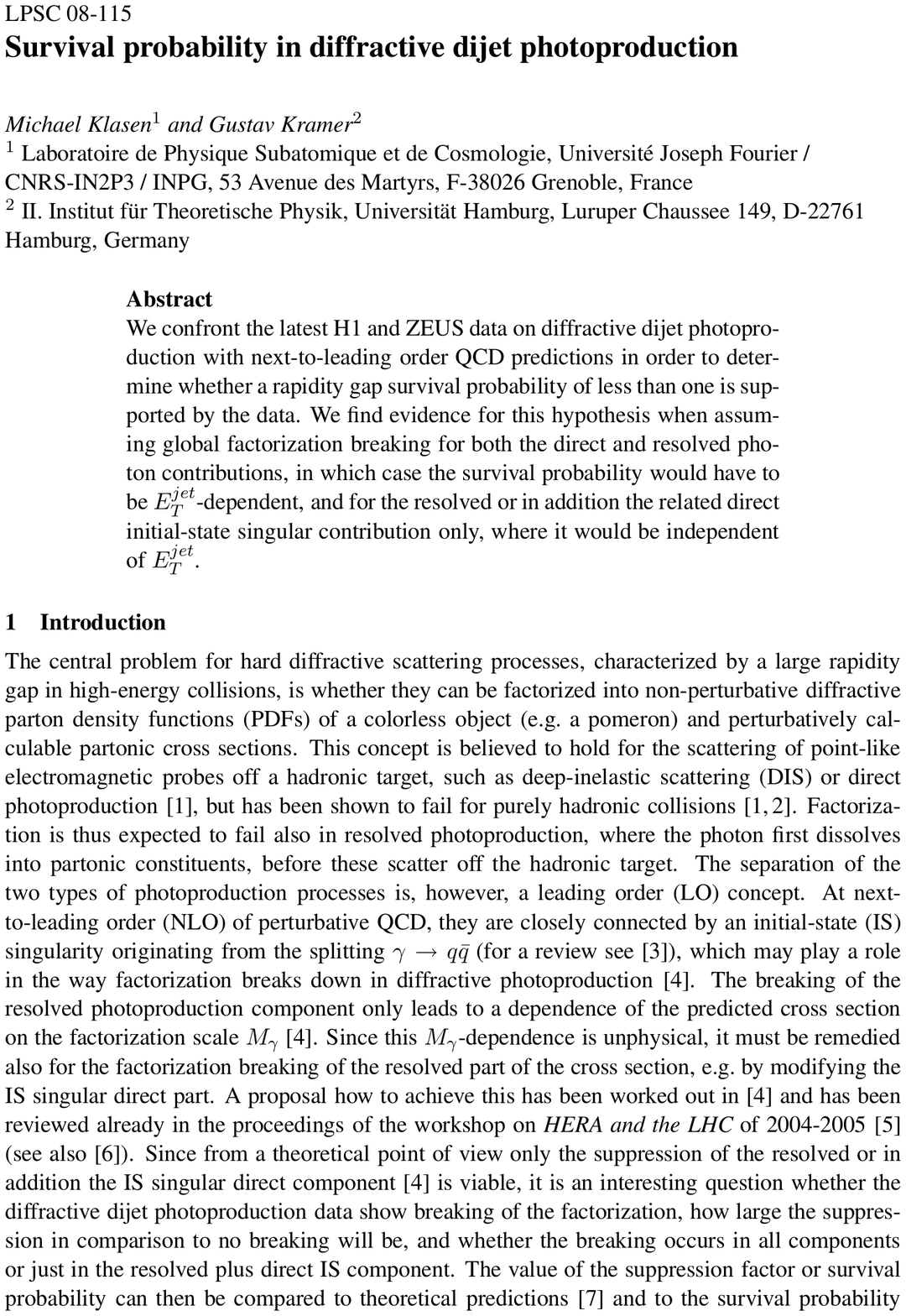}

 \coltoctitle{Fracture Functions at HERA and LHC}
 \coltocauthor{F. A. Ceccopieri, L. Trentadue}
 \index{Ceccopieri, F.A.}
 \index{Trentadue, L.}
 \Includeart{\CAUT}{\CTIT}{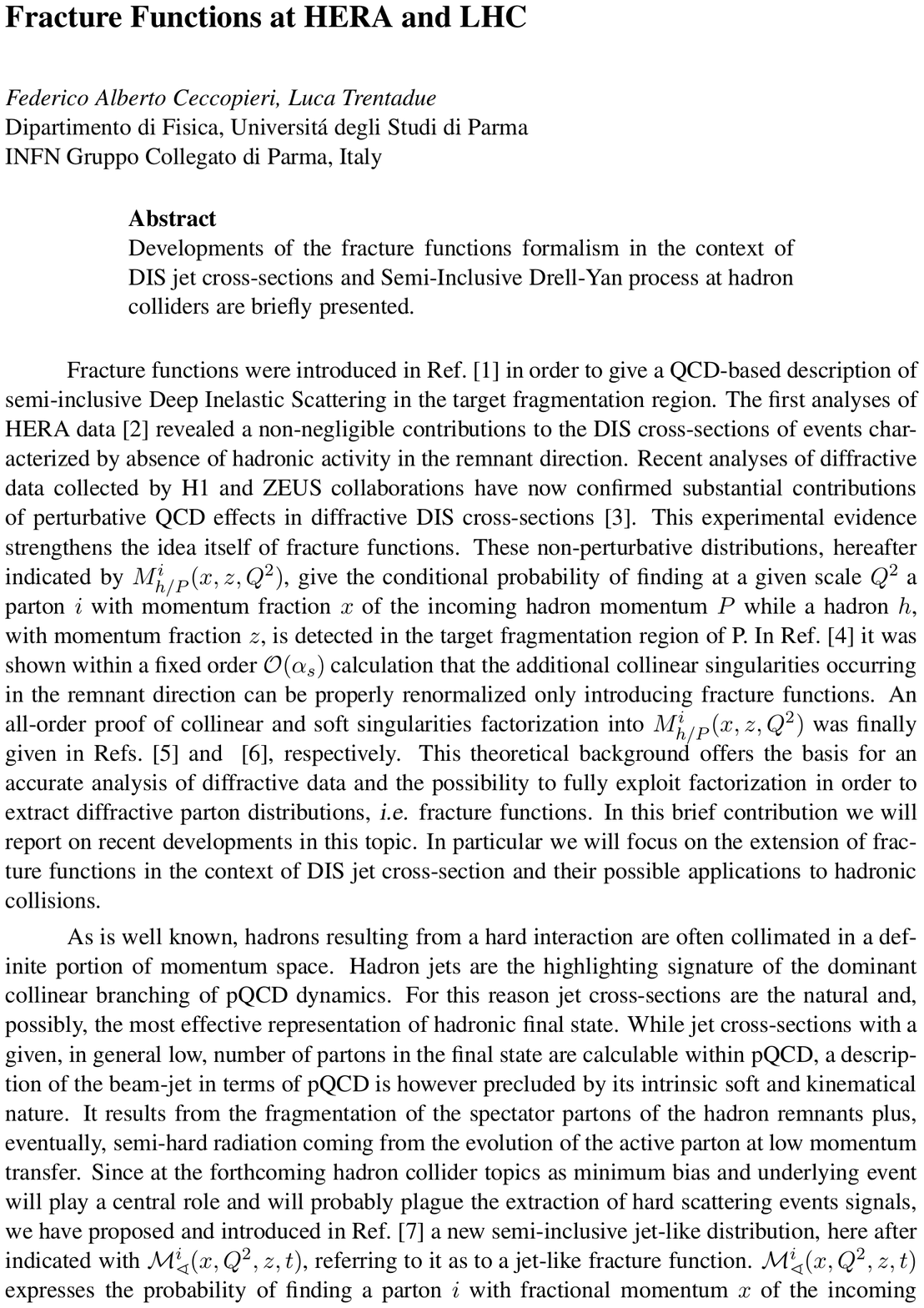}

 \coltoctitle{Generalised parton distributions and exclusive vector meson production}
 \coltocauthor{C. Nockles, T. Teubner}
 \index{Nockles, C.} 
 \index{Teubner, T.}
 \Includeart{\CAUT}{\CTIT}{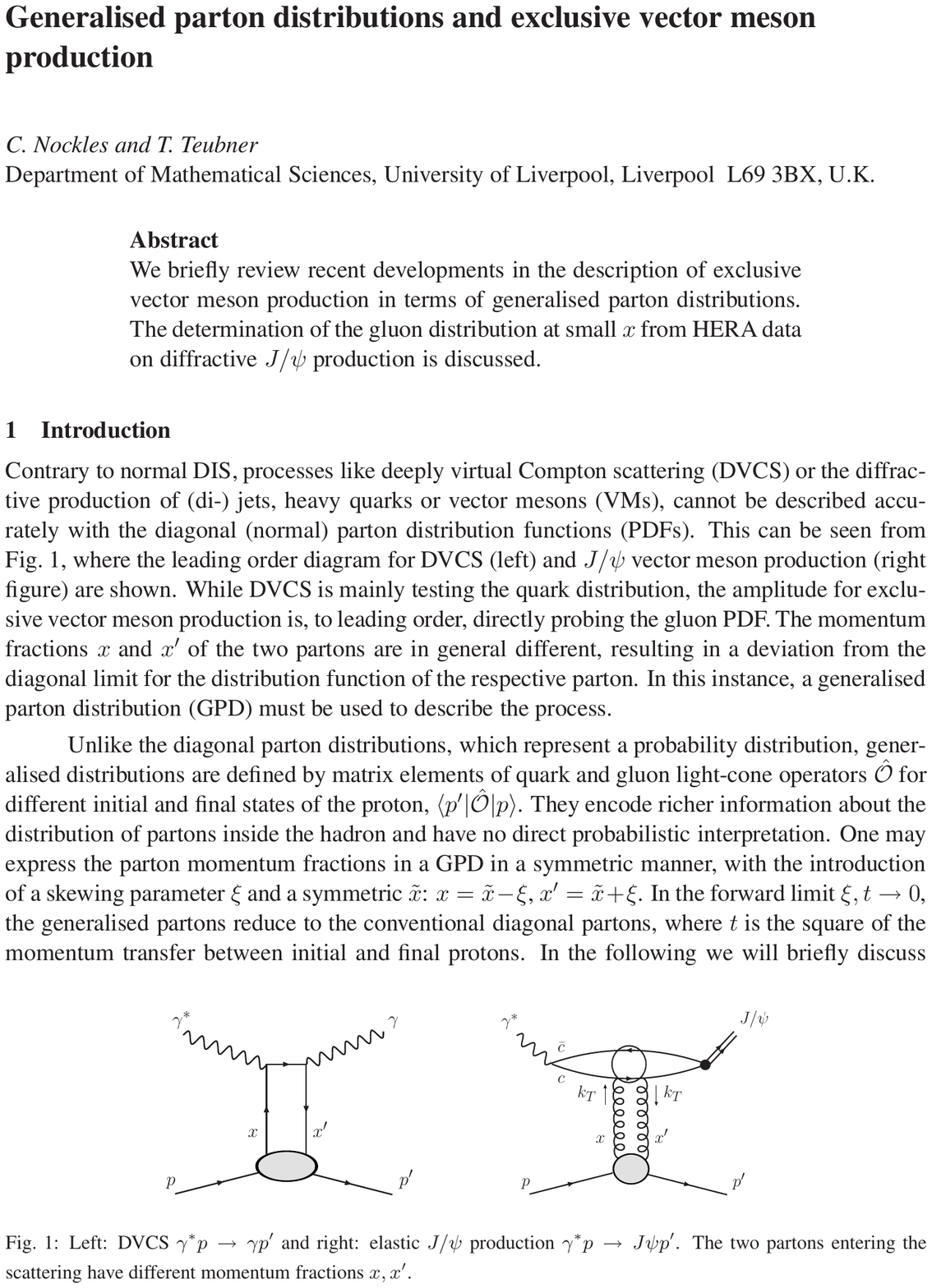}

 \coltoctitle{Dipole models and parton saturation in $ep$ scattering}
 \coltocauthor{L.~Motyka, K.~Golec-Biernat, G. Watt}
 \index{Motyka, L.} 
 \index{Golec-Biernat, K.} 
 \index{Watt, G.}
 \Includeart{\CAUT}{\CTIT}{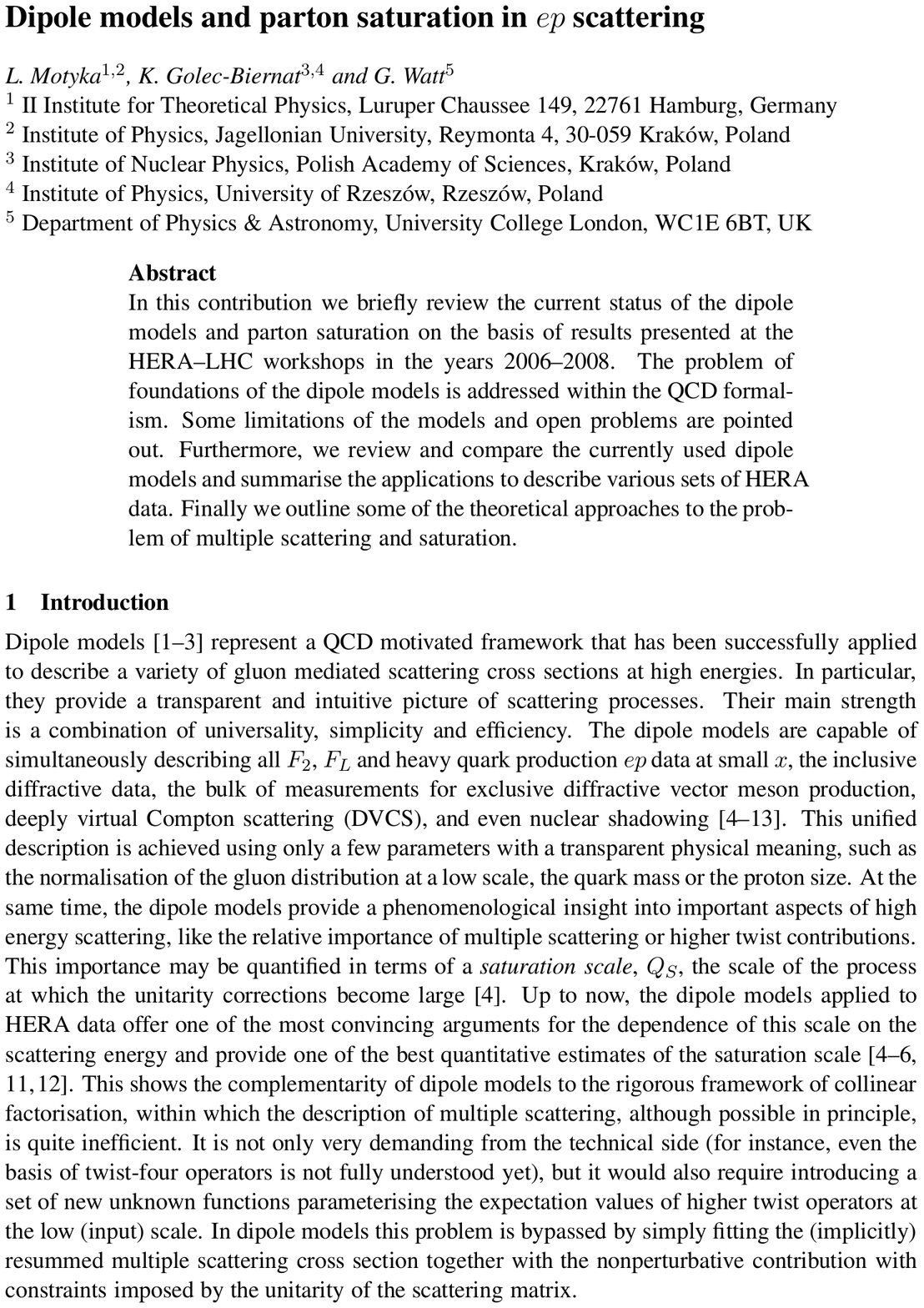}

 \coltoctitle{Checking formalism for central exclusive production in the first LHC runs}
 \coltocauthor{A.D.~Martin, V.A.~Khoze, M.G.~Ryskin}
 \index{Martin, A.D.}
 \index{Khoze, V.A.}
 \index{Ryskin, M.G.}
 \Includeart{\CAUT}{\CTIT}{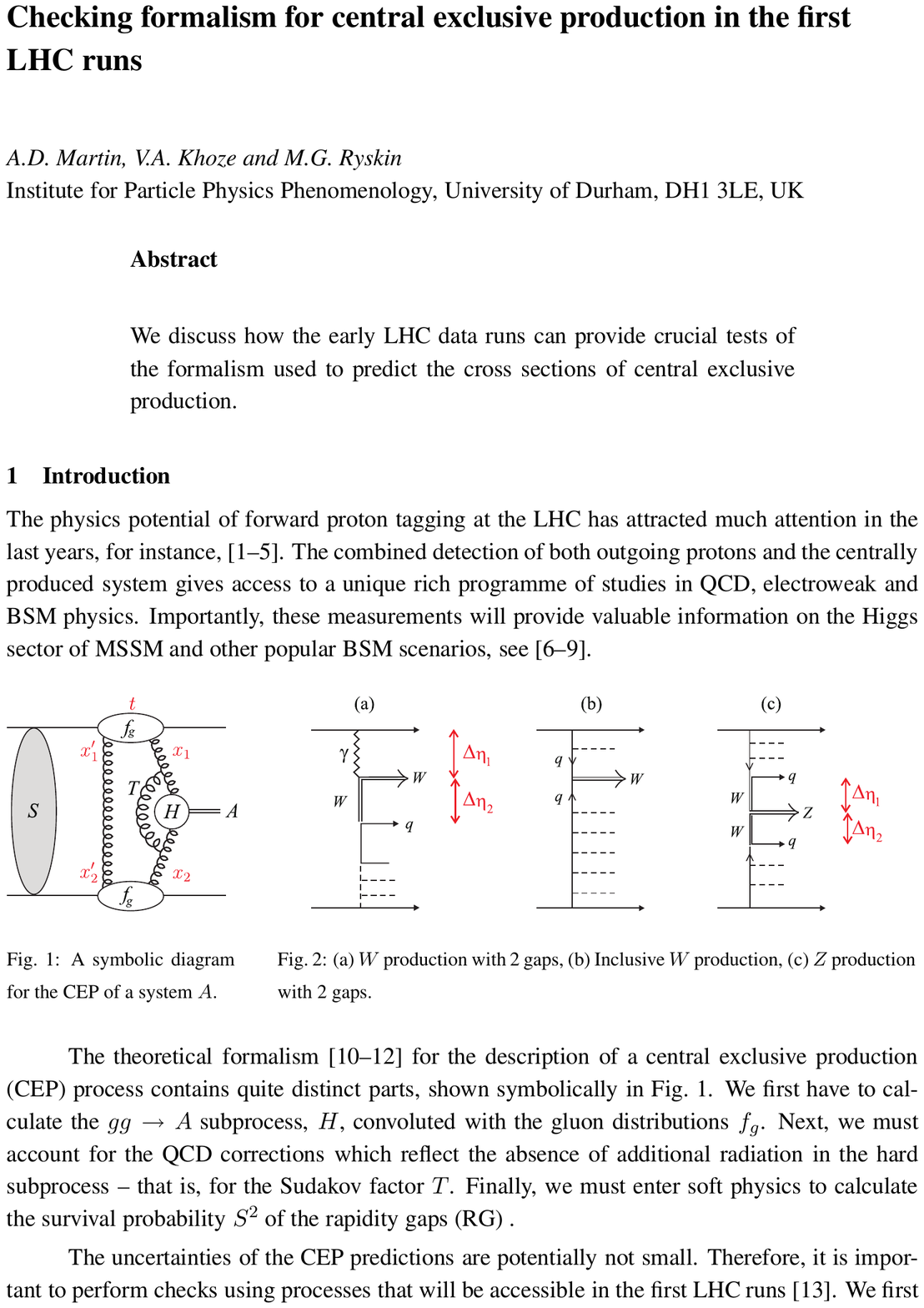}

 \coltoctitle{Rapidity gap survival probability and total cross sections}
 \coltocauthor{A.D.~Martin, V.A.~Khoze, M.G.~Ryskin}
\index{Martin, A.D.}
 \index{Khoze, V.A.}
 \index{Ryskin, M.G.}
  \Includeart{\CAUT}{\CTIT}{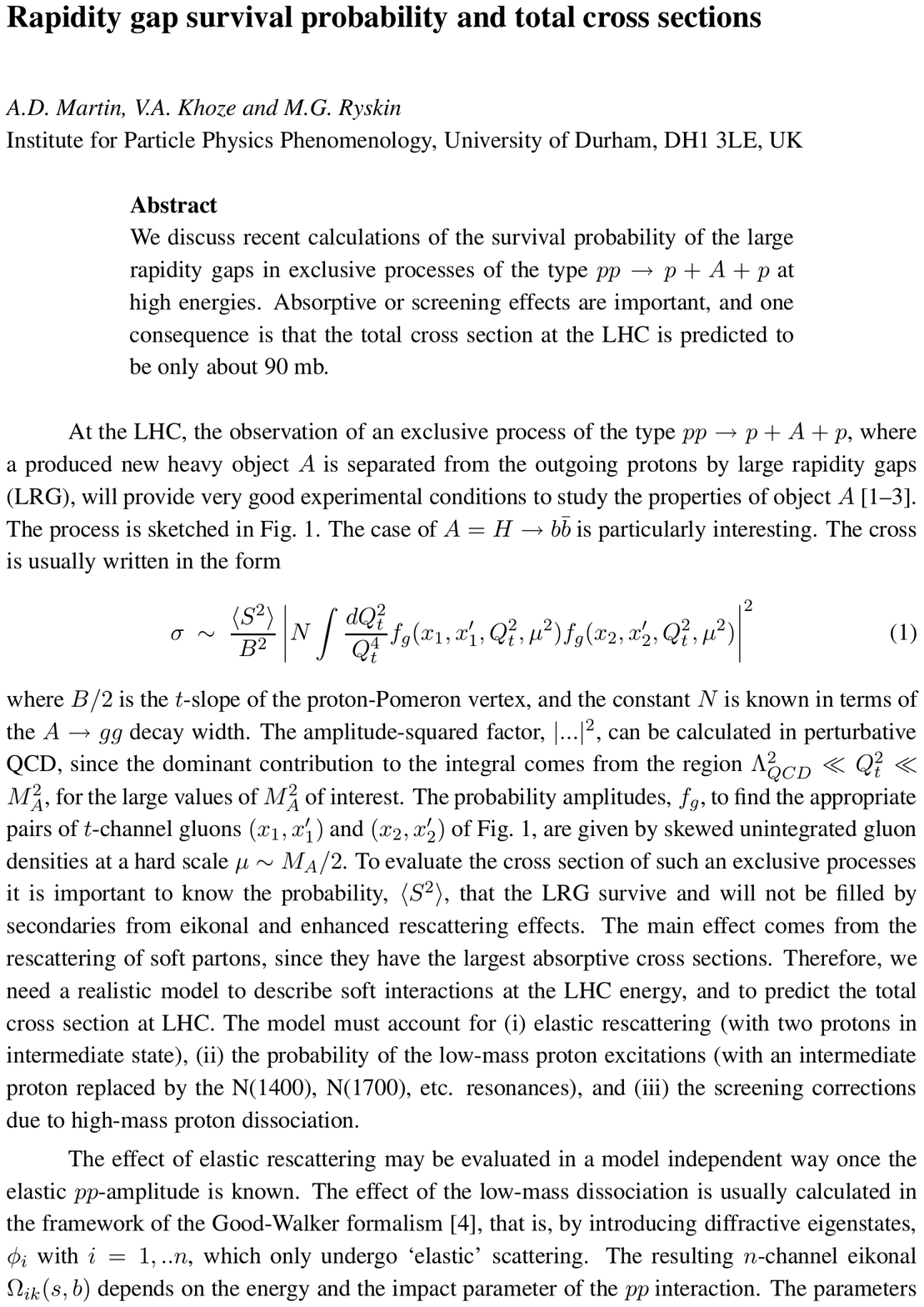}

 \coltoctitle{Rapidity gap survival in central exclusive diffraction: 
Dynamical mechanisms and uncertainties}
 \coltocauthor{M. Strikman, Ch.  Weiss}
 \index{Strikman, M.}
 \index{Weiss, Ch.}
 \Includeart{\CAUT}{\CTIT}{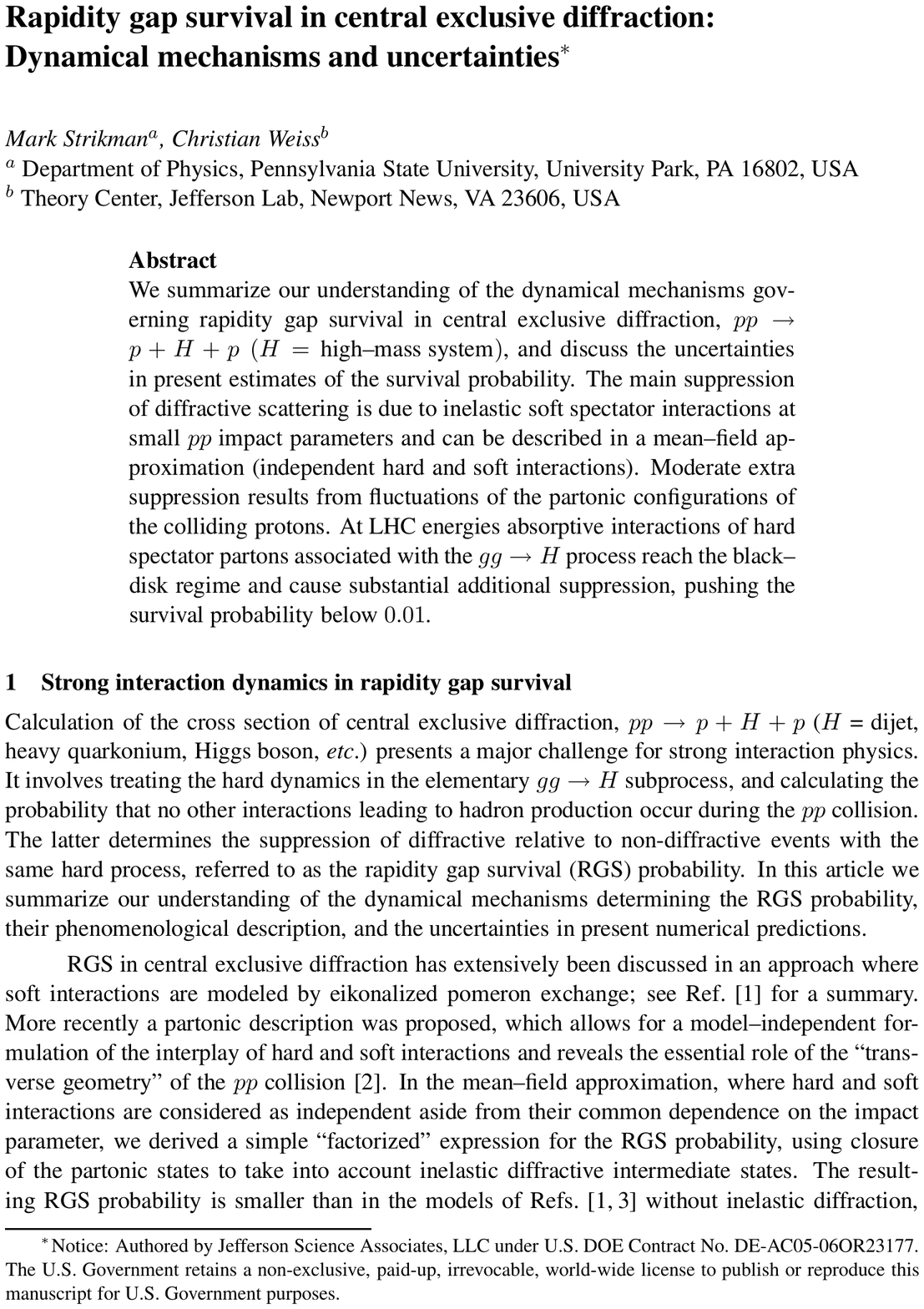}

 \coltoctitle{Two-photon and photon-hadron interactions at the LHC}
 \coltocauthor{J. Nystrand}
 \index{Nystrand, J.}
 \Includeart{\CAUT}{\CTIT}{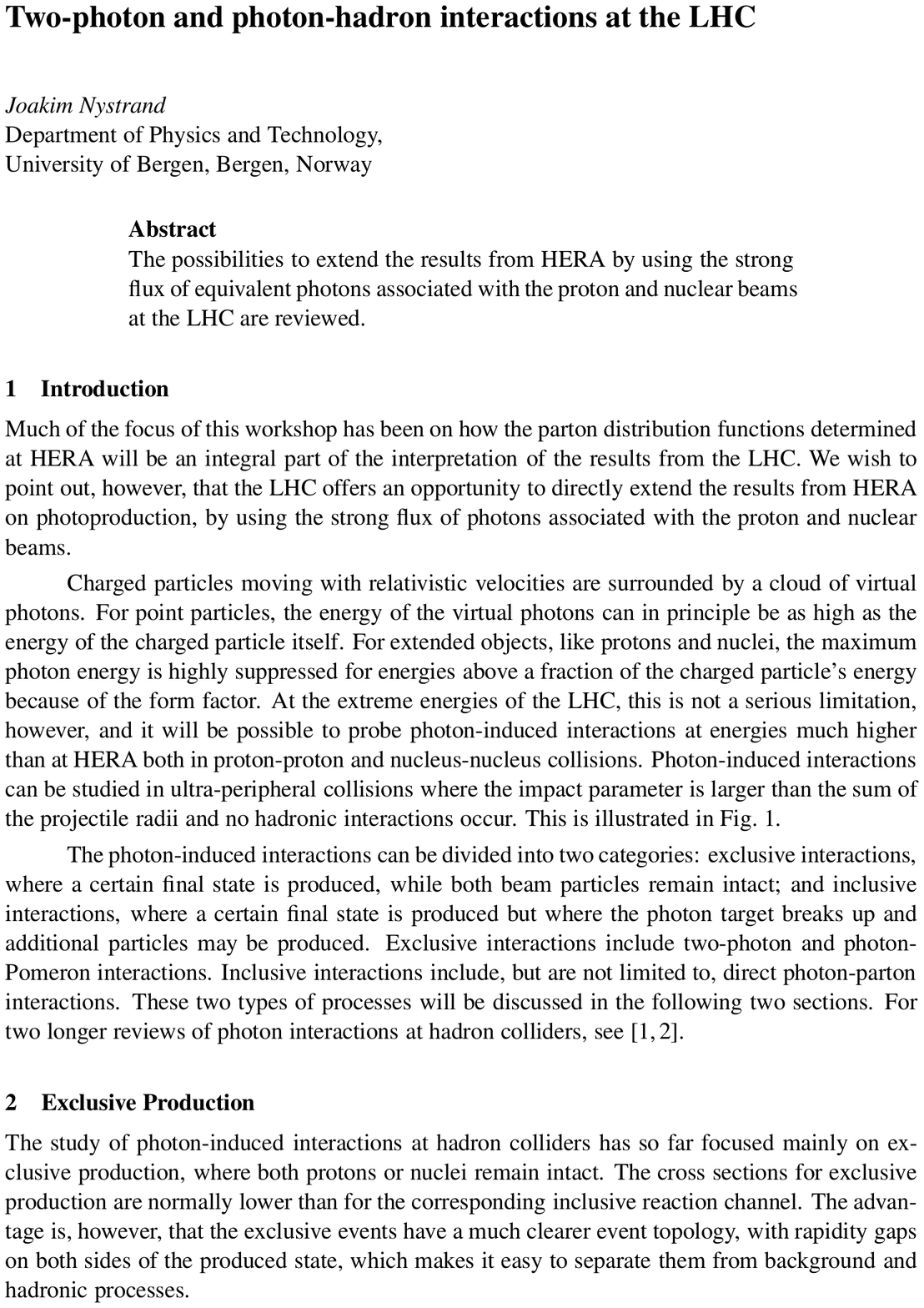}
 
 \coltoctitle{Searching for the Odderon at HERA and the LHC}
 \coltocauthor{C. Ewerz}
 \index{Ewerz, C.}
 \Includeart{\CAUT}{\CTIT}{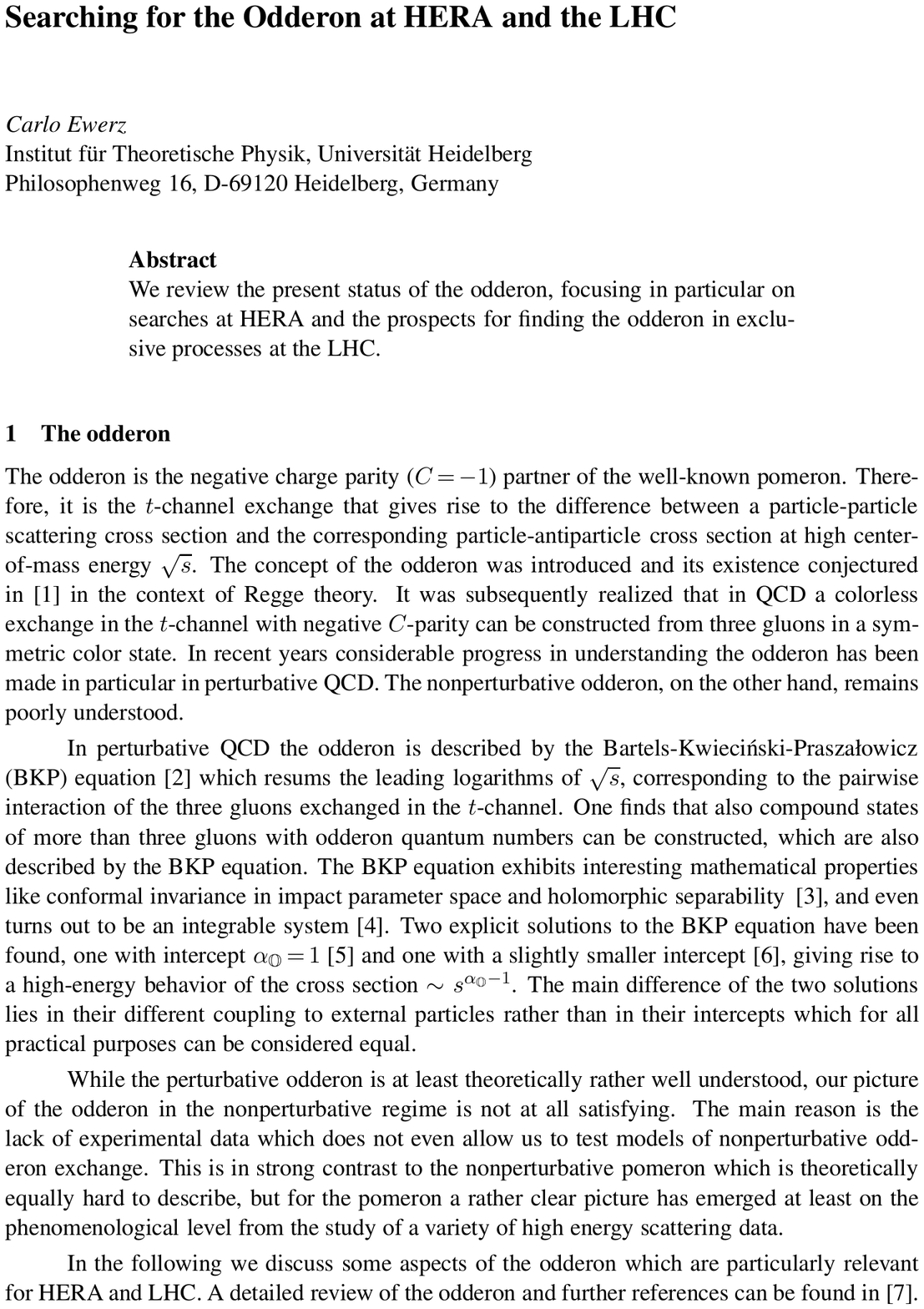}

 \coltoctitle{Forward physics with CMS}
 \coltocauthor{S. Erhan, S. Cerci, M. Grothe, J. Hollar, A. Vilela Pereira}
 \index{Erhan, S.}
 \index{Cerci, S.}
 \index{Grothe, M.}
 \index{Hollar, J.}
 \index{Pereira, A.V.}
 \Includeart{\CAUT}{\CTIT}{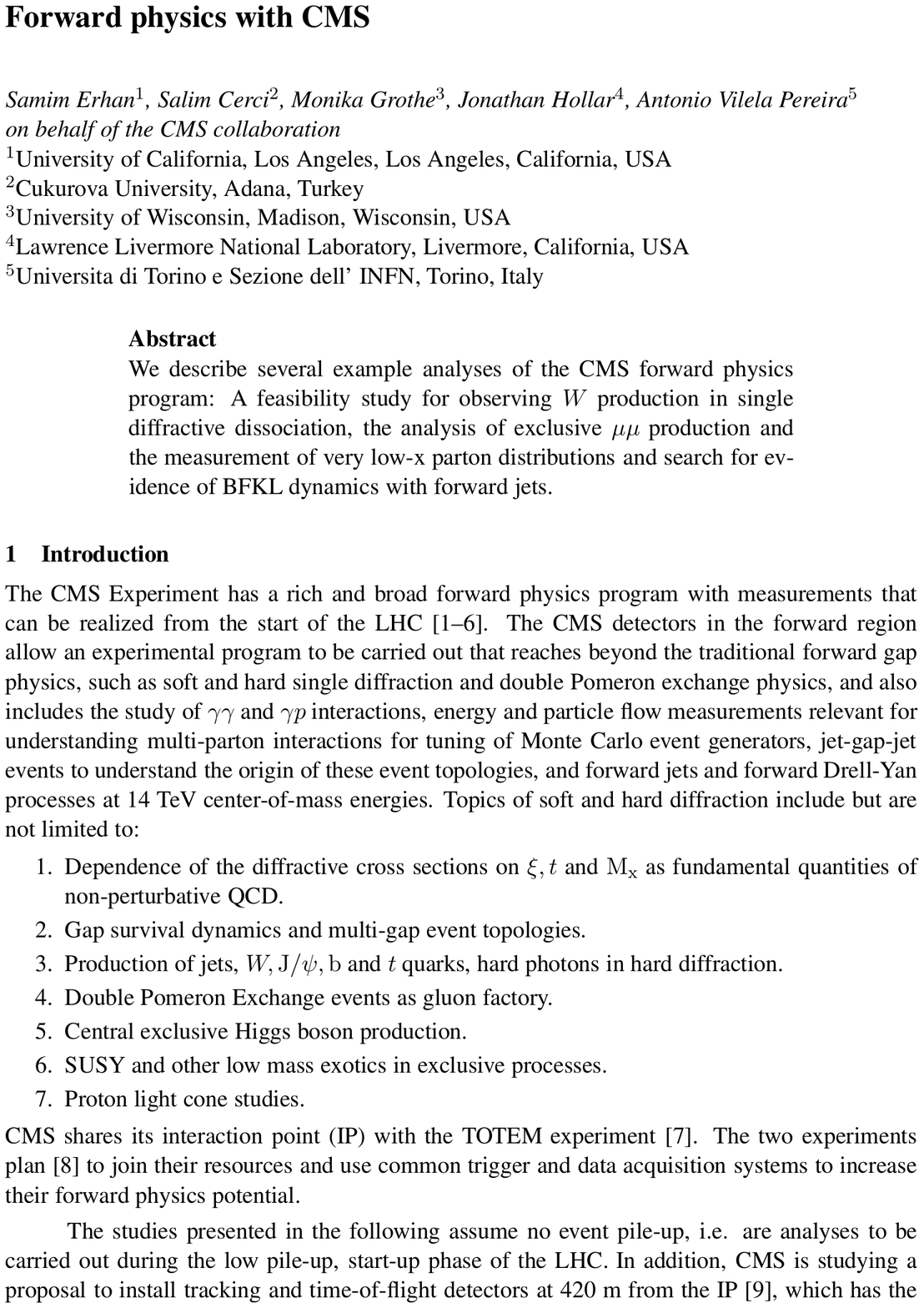}

 \coltoctitle{Diffraction at TOTEM}
 \coltocauthor{G.~Antchev, P.~Aspell V.~Avati, M.G.~Bagliesi,
V.~Berardi, M.~Berretti, U.~Bottigli,
M.~Bozzo,
E.~Br\"{u}cken, A. Buzzo, F.~Cafagna, M.~Calicchio,
M.G.~Catanesi,
P.L.~Catastini, R.~Cecchi,
M.A.~Ciocci,
M.~Deile, E.~Dimovasili,
K.~Eggert, V.~Eremin, F.~Ferro,
F.~Garcia, S.~Giani, V.~Greco, J.~Heino, T.~Hild\'en,
J.~Ka\v{s}par, J.~ Kopal, V.~Kundr\'{a}t, K.~Kurvinen, S.~Lami, G.~Latino,
R.~Lauhakangas, E.~Lippmaa,
M.~Lokaj\'{\i}\v{c}ek, M.~Lo~Vetere, F.~Lucas~Rodriguez, M.~Macr\'{\i}, G.~Magazz\`{u},
M.~Meucci S.~Minutoli,
H.~Niewiadomski, E.~Noschis,
G.~Notarnicola,
E.~Oliveri, F.~Oljemark, R.~Orava, M.~Oriunno, K.~\"{O}sterberg,
P.~Palazzi,  E.~Pedreschi,
J.~Pet\"{a}j\"{a}j\"{a}rvi, M.~Quinto,
E.~Radermacher, E.~Radicioni,
F.~Ravotti, G.~Rella, E.~Robutti,
L.~Ropelewski, G.~Ruggiero, A.~Rummel,
H.~Saarikko, G.~Sanguinetti, A.~Santroni,
A.~Scribano, G.~Sette, W.~Snoeys, F.~Spinella,
P.~Squillacioti, A.~Ster, C.~Taylor, A.~Trummal,
N.~Turini, J.~Whitmore, J.~Wu}
 \index{Antchev, G. }
  \index{Aspell, P.}
 \index{Avati, V. }
\index{Bagliesi, M.G}
\index{Berardi, V.}
 \index{Berretti, M.}
 \index{Bottigli, U.}
 \index{Bozzo, M.}
 \index{Br\"{u}cken, E.}
 \index{Buzzo, A.}
 \index{Cafagna, F.}
 \index{Calicchio, M.}
 \index{Catanesi, M.G.}
 \index{Catastini, P.L.}
 \index{Cecchi, R.}
 \index{Ciocci, M.A.}
 \index{Deile, M.}
 \index{Dimovasili, E.}
 \index{Eggert, K.}
 \index{Eremin, V.}
 \index{Ferro, F.}
 \index{Garcia, F.}
 \index{Giani, S.} 
 \index{Greco, V.} 
 \index{Heino, J.}
 \index{Hild\'en, T.}
 \index{Ka\v{s}par, J.} 
 \index{Kopal,  J.} 
 \index{Kundr\'{a}t,  V.}
 \index{Kurvinen, K.}
 \index{Lami, S.}
 \index{Latino, G.}
 \index{Lauhakangas, R.}
 \index{Lippmaa, E.}
 \index{Lokaj\'{\i}\v{c}ek, M.}
 \index{Lo~Vetere, M.}
 \index{Lucas~Rodriguez, F.} \index{Macr\'{\i}, M.}
 \index{Magazz\`{u}, G.}
 \index{Meucci, M.}
 \index{Minutoli, S.}
 \index{Niewiadomski, H.} 
 \index{Noschis, E.}
 \index{Notarnicola, G.}
 \index{Oliveri, E.}
 \index{Oljemark, F.}
 \index{Orava, R.}
 \index{Oriunno, M.}
 \index{\"Osterberg, K.}
 \index{Palazzi,  P.}
 \index{Pedreschi, E.}
 \index{Pet\"{a}j\"{a}j\"{a}rvi, J.} 
 \index{Quinto, M.}
 \index{Radermacher, E.} 
 \index{Radicioni, E.}
 \index{Ravotti, F.} 
 \index{Rella,  G.}
 \index{Robutti, E.}
 \index{Ropelewski, L.}
 \index{Ruggiero, G.}
 \index{Rummel,A.}
 \index{Saarikko, H.}
 \index{Sanguinetti, G.}
 \index{Santroni, A.}
 \index{Scribano, A.}
 \index{Sette, G.}
 \index{Snoeys, W.} 
 \index{Spinella, F.}
 \index{Squillacioti, P.}
 \index{Ster, A.} 
 \index{Taylor, C.}
 \index{Trummal, A.}
 \index{Turini, N.}
 \index{Whitmore, J.} 
 \index{Wu, J.} 
  \Includeart{\CAUT}{\CTIT}{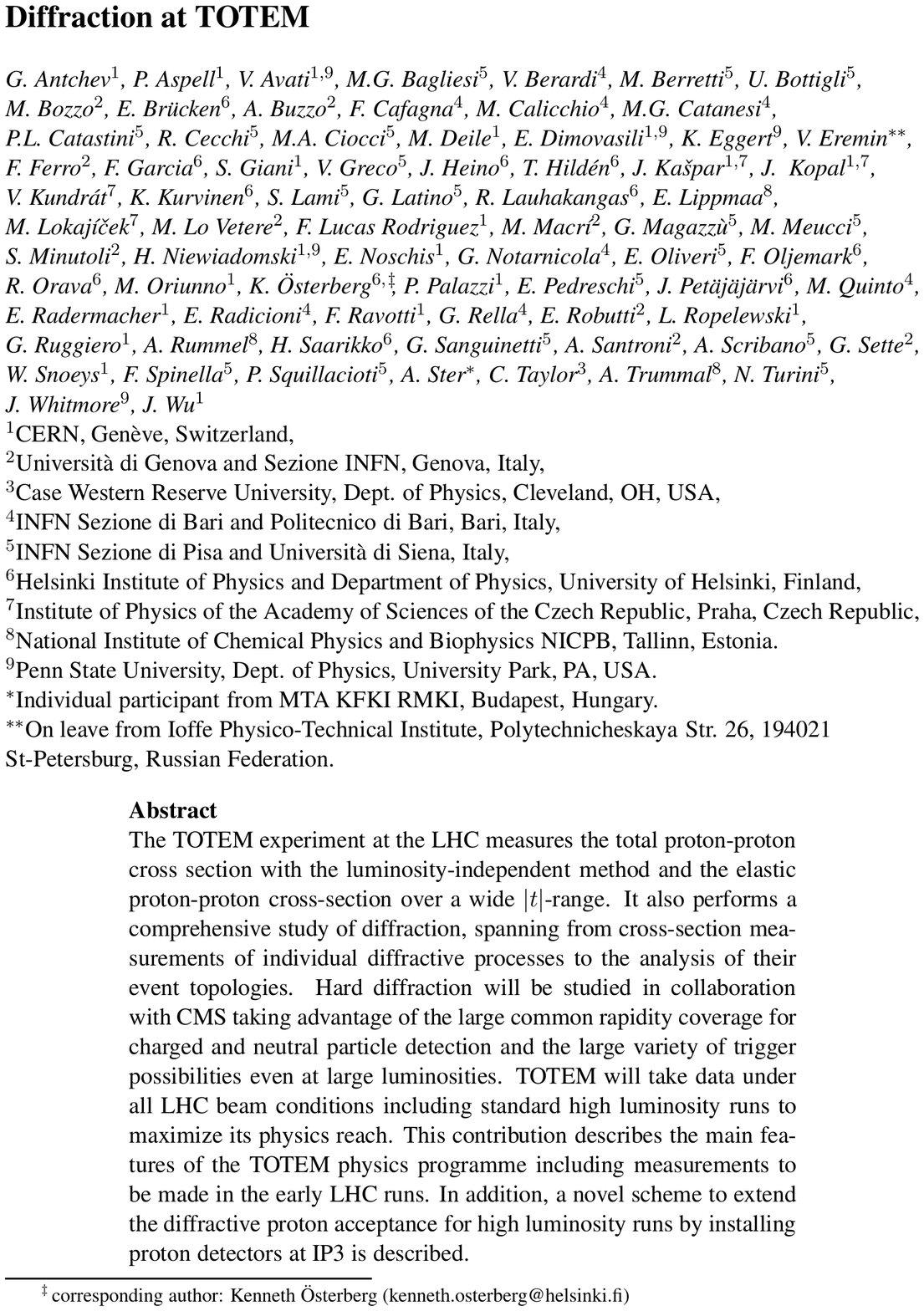}

\coltoctitle{The ALFA Detector and Physics Program}
\coltocauthor{K.~Hiller, H.~Stenzel}
 \index{Hiller, K.}
  \index{Stenzel, H.}
\Includeart{\CAUT}{\CTIT}{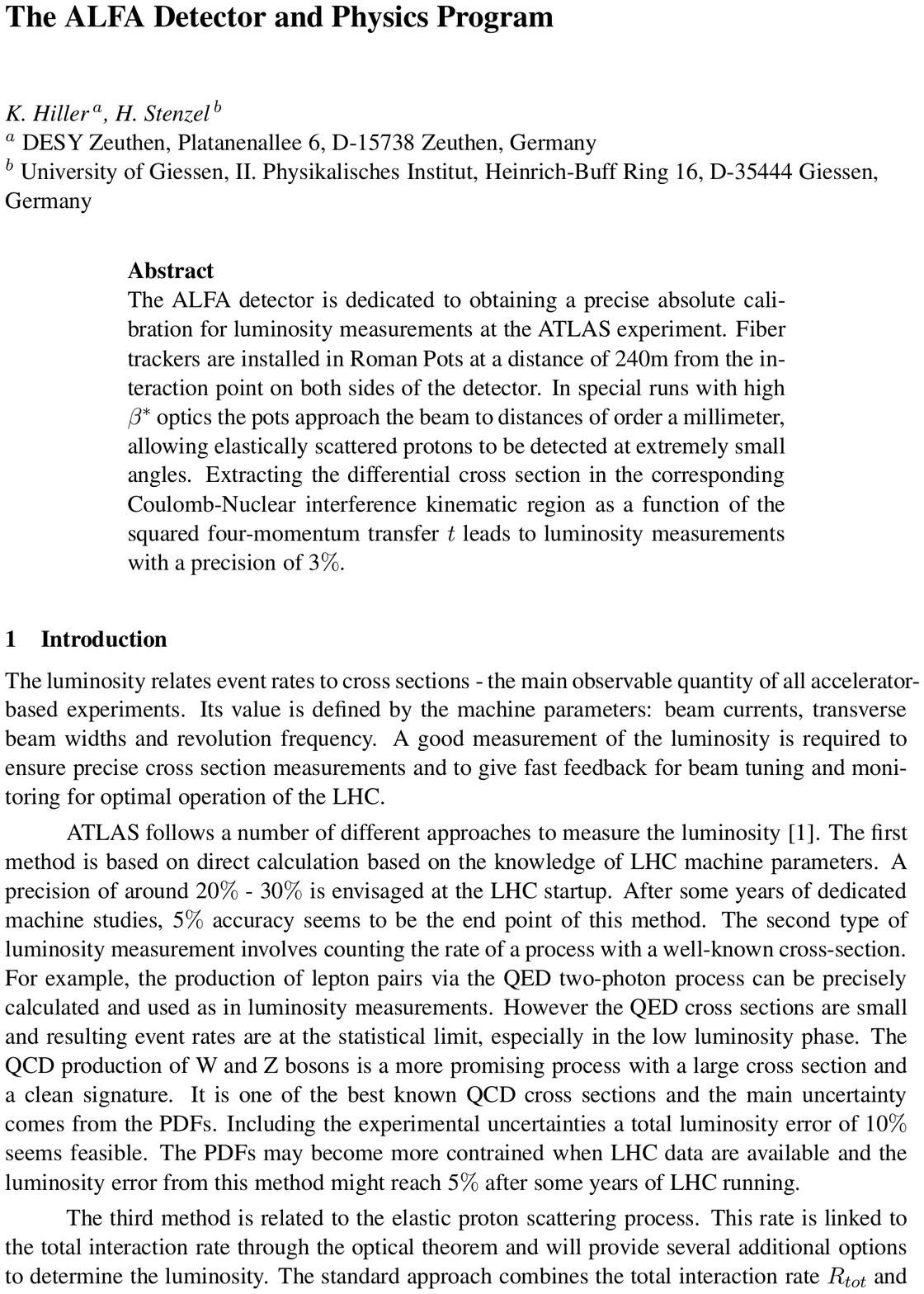}

 \coltoctitle{Diffractive Physics in ALICE}
 \coltocauthor{R. Schicker}
  \index{Schicker, R.}
 \Includeart{\CAUT}{\CTIT}{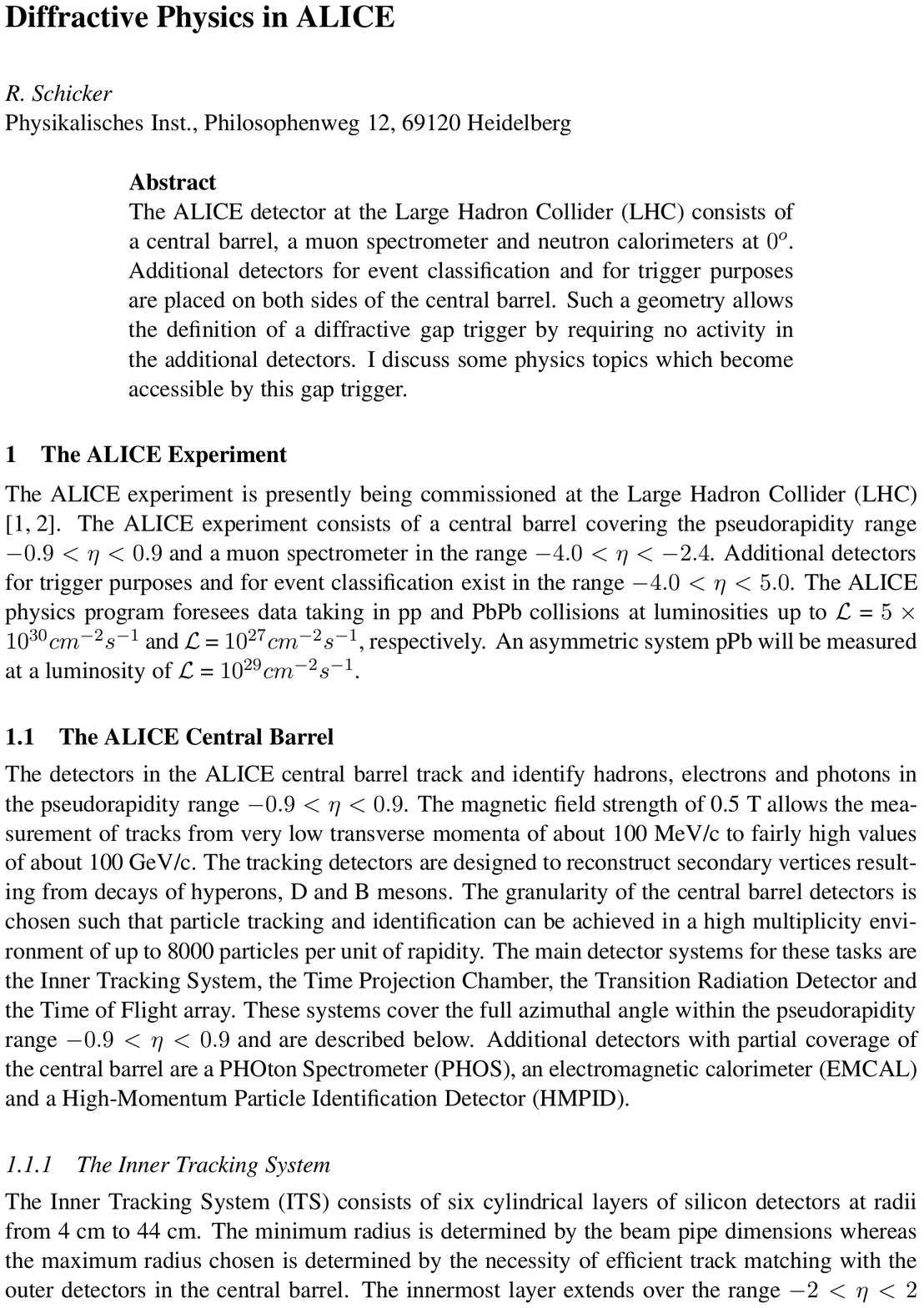}

 \coltoctitle{Physics with forward FP420/FP220 tagging systems}
 \coltocauthor{P. Bussey,  P. Van Mechelen}
  \index{Bussey, P.}
   \index{Van Mechelen, P.}
\Includeart{\CAUT}{\CTIT}{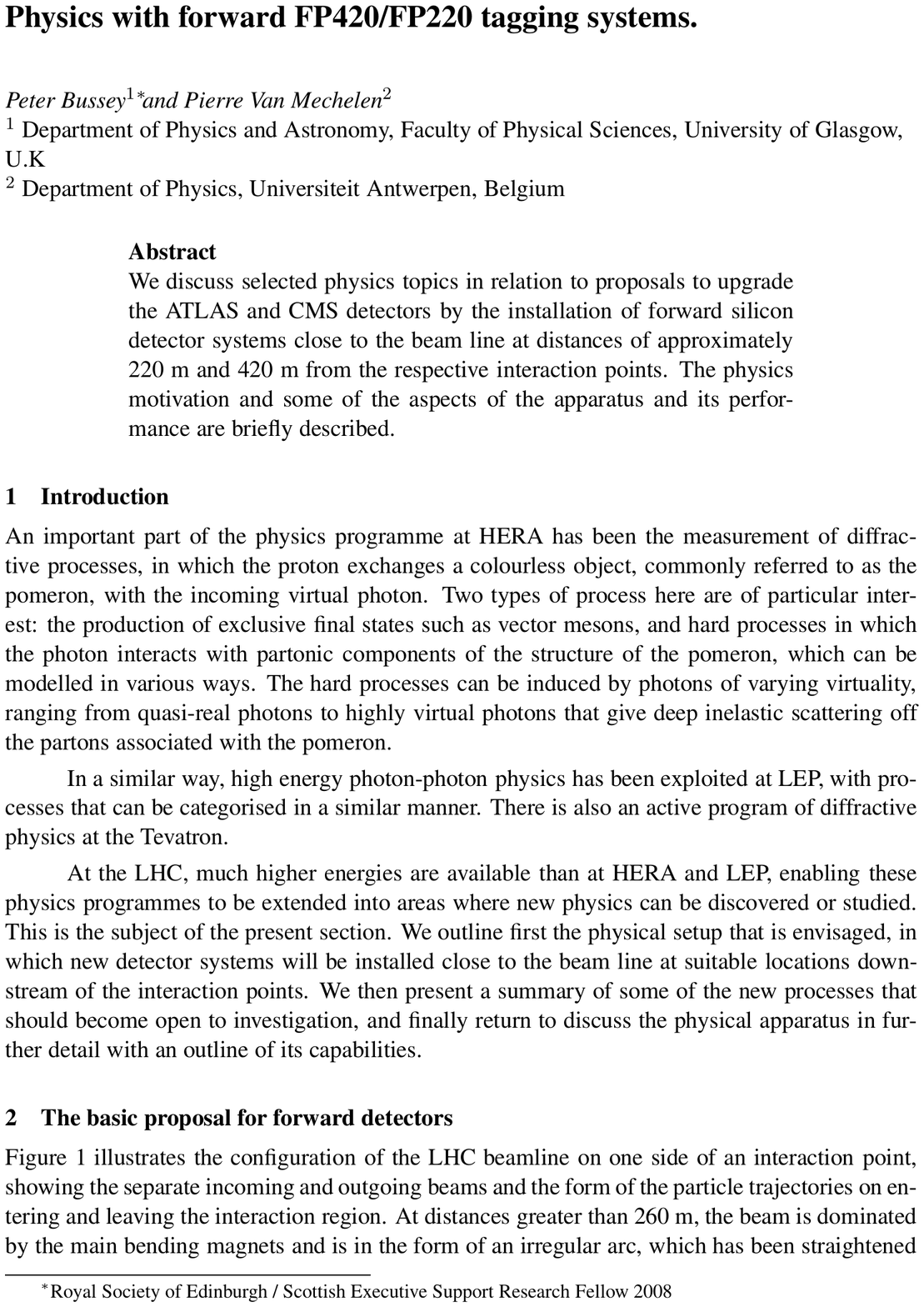}

\end{papers} 

\chapter[\large  WG: Cosmic Rays, HERA and the LHC]{ Working Group \\ Cosmic Rays, HERA and the LHC}
{\Large\bfseries Convenors:}
\vspace{10mm}

{\Large\itshape
 C. Diaconu (DESY and CPPM, Marseille), \\
 Ch. Kiesling (MPI Munich)\\
T. Pierog (FZ Karlsruhe)
}

\CLDP
\begin{papers} 
 
\coltoctitle{Introduction}
\coltocauthor{A.~Bunyatyan,
A.~Cooper-Sarkar,
C.~Diaconu,
R.~Engel,
C.~Kiesling,
K.~Kutak,
S.~Ostapchenko,
T.~Pierog,
T.C.~Rogers,
M.I.~Strikman, 
T.~Sako
}
 \index{Bunyatyan, A.}
 \index{Cooper-Sarkar, A.}
 \index{Diaconu, C.}
 \index{Engel, R.}
 \index{Kiesling, Ch.}
 \index{Kutak, K.}
 \index{Ostapchenko, S.}
 \index{Pierog, T.}
 \index{Rogers, T.C.}
 \index{Strikman, M.} 
 \index{Sako, T.}
\Includeart{\CAUT}{\CTIT}{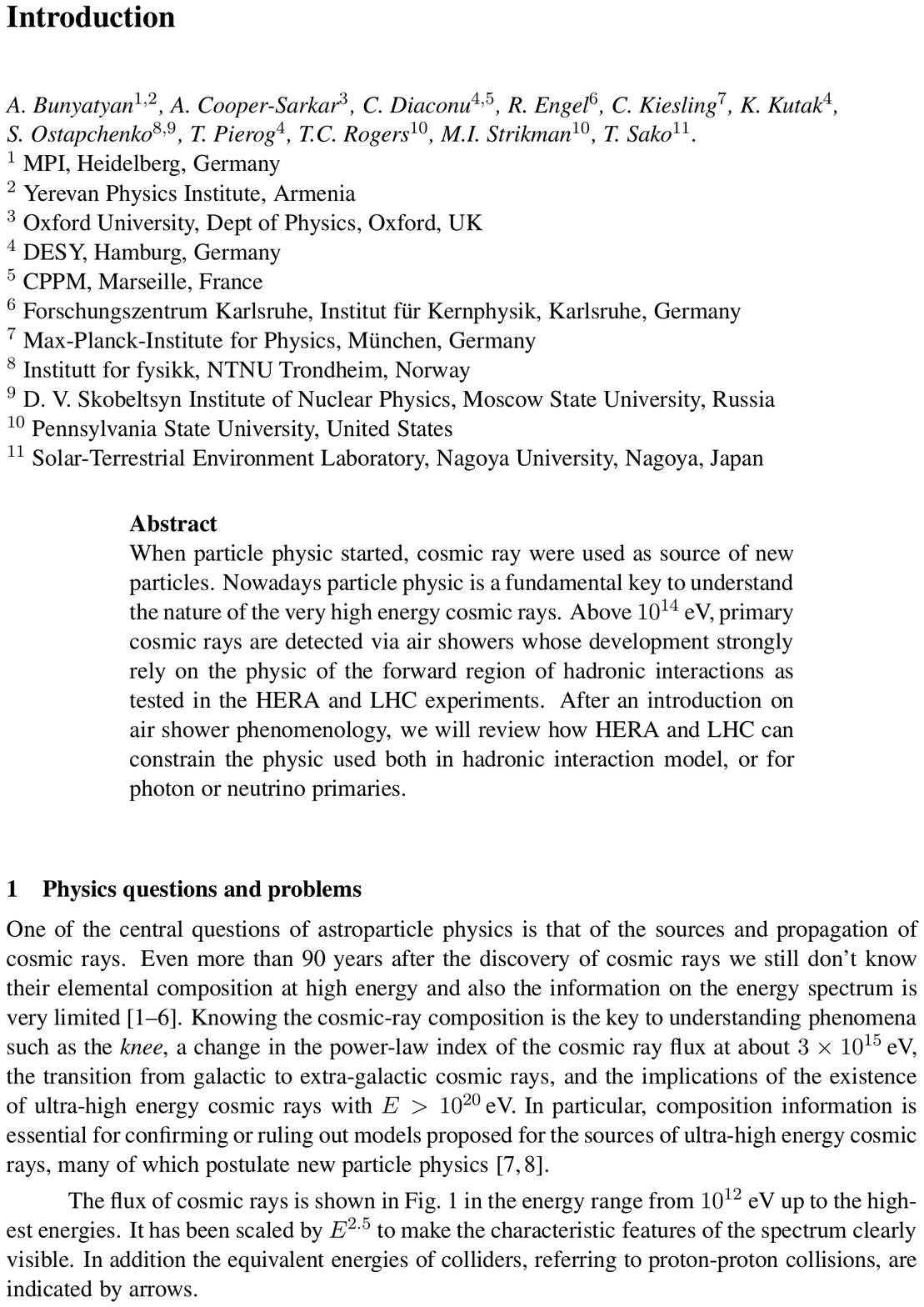}

\coltoctitle{Experimental results}
 \coltocauthor{}
\Includeart{\CAUT}{\CTIT}{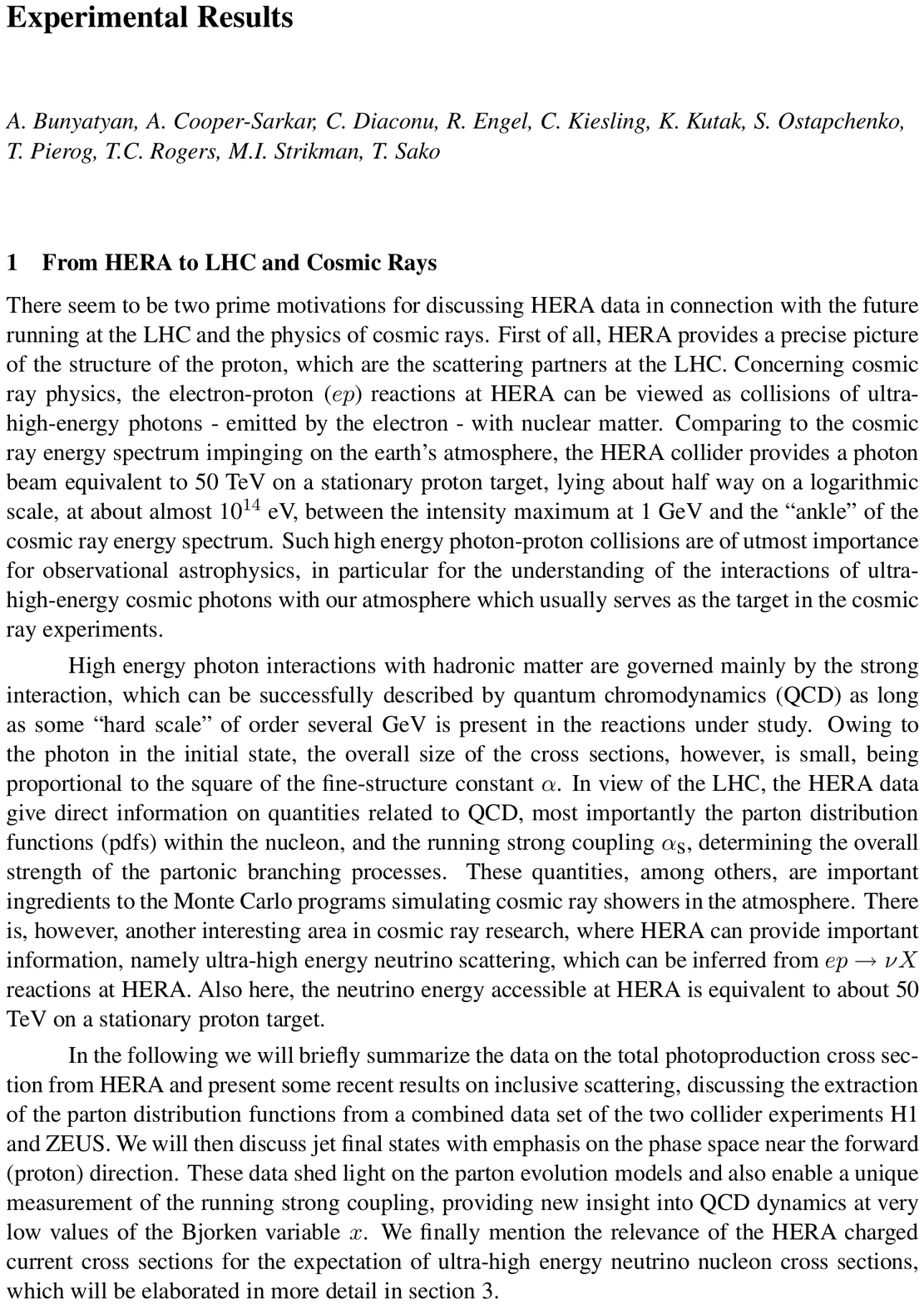}
 
\coltoctitle{Model predictions for  HERA, LHC and cosmic rays}
\Includeart{\CAUT}{\CTIT}{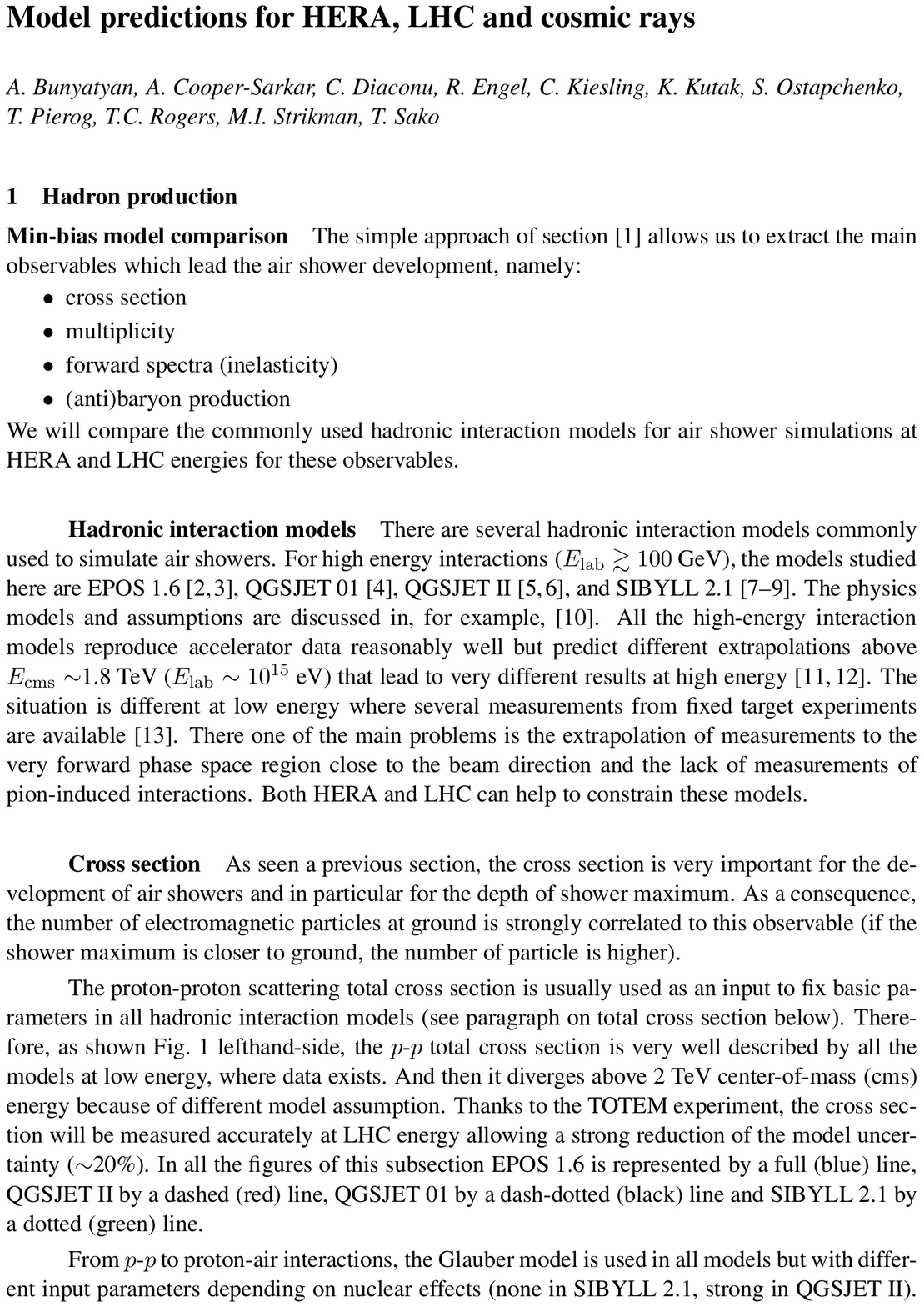}

\coltoctitle{Summary}
\Includeart{\CAUT}{\CTIT}{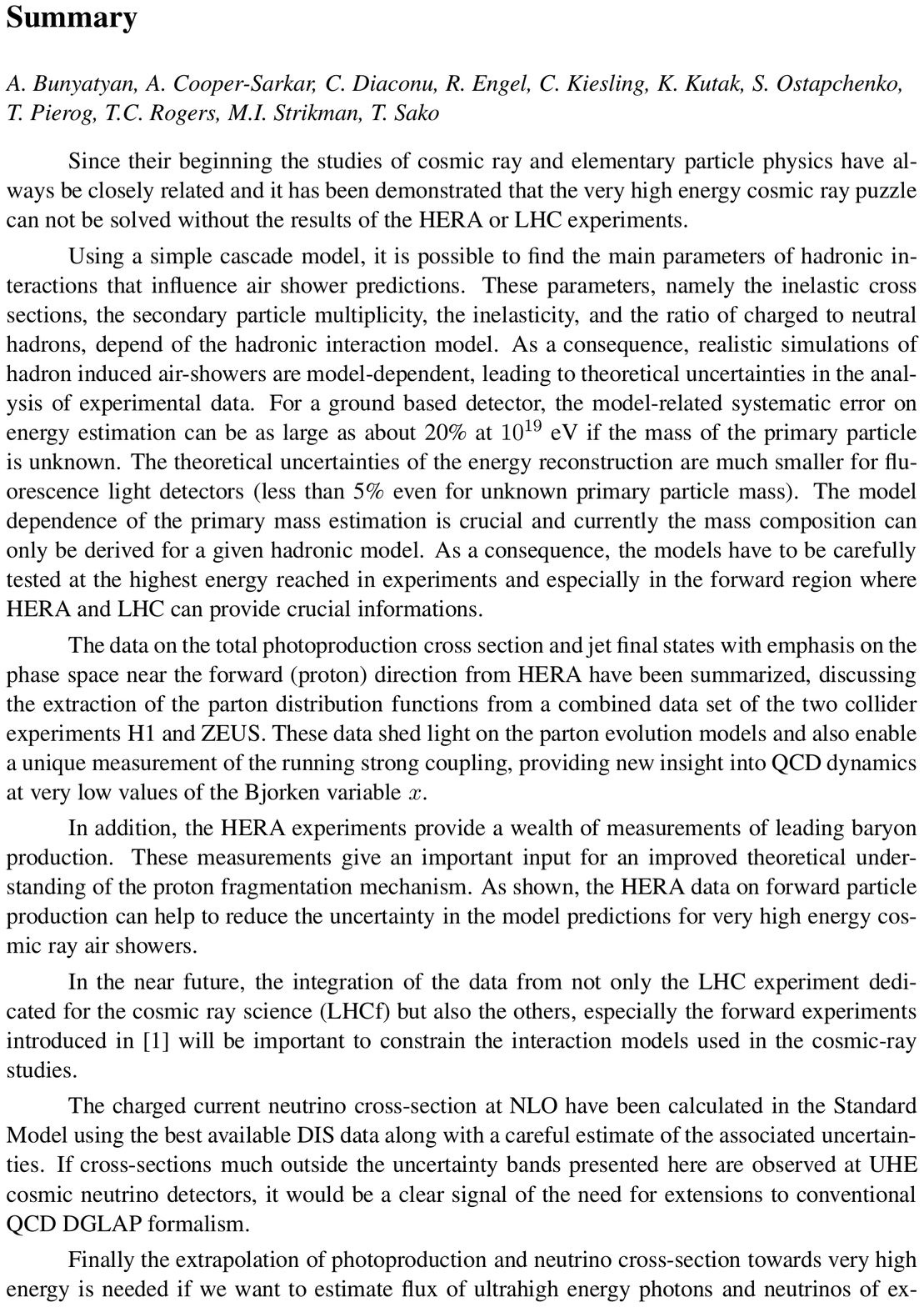}
 


\end{papers} 

\newcommand{\genname}[1]{\texttt{\textbf{#1}}}
\chapter[\large  WG: Monte Carlo and Tools]{ Working Group \\ Monte Carlo and Tools}
{\Large\bfseries Convenors:}
\vspace{10mm}

{\Large\itshape
P. Bartalini (Taiwan, CMS),\\
S.Chekanov (Argonne, ZEUS), \\
F. Krauss (IPP Durham), \\
S. Gieseke (U. Karlsruhe),
}

\CLDP
\begin{papers} 

 \coltoctitle{Introduction to Monte Carlo models and Tools working group (WG5)}
 \coltocauthor{P.~Bartalini,  S.~Chekanov,  S.~Gieseke,  F.~Krauss}
  \index{Bartalini, P.}
 \index{Chekanov, S.}
  \index{Gieseke, S.} 
 \index{Krauss, F.}
 \Includeart{\CAUT}{\CTIT}{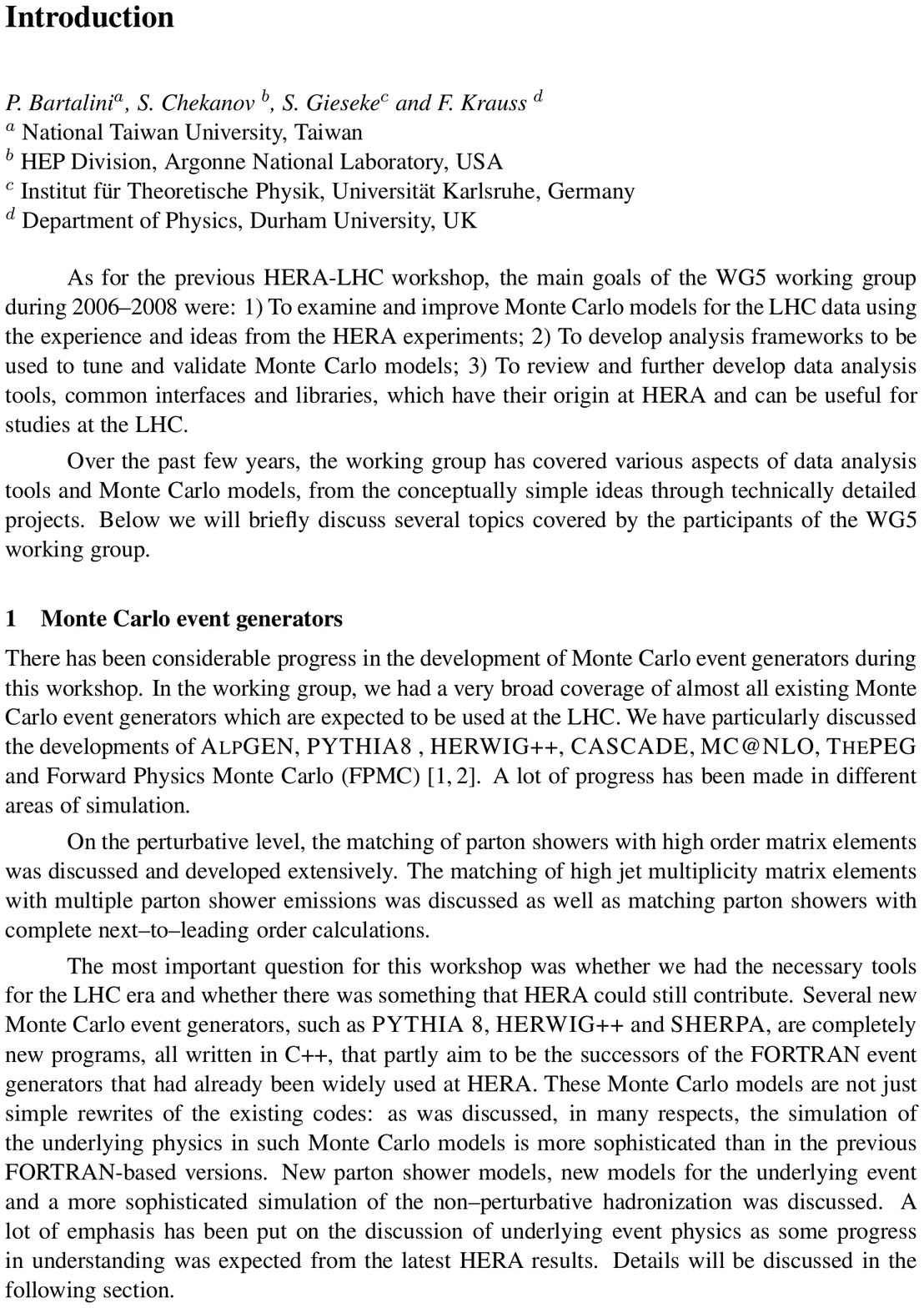}
 
 \coltoctitle{A multi-channel Poissonian model for multi-parton scatterings}
 \coltocauthor{D. Treleani}
 \index{Treleani, D.}
 \Includeart{\CAUT}{\CTIT}{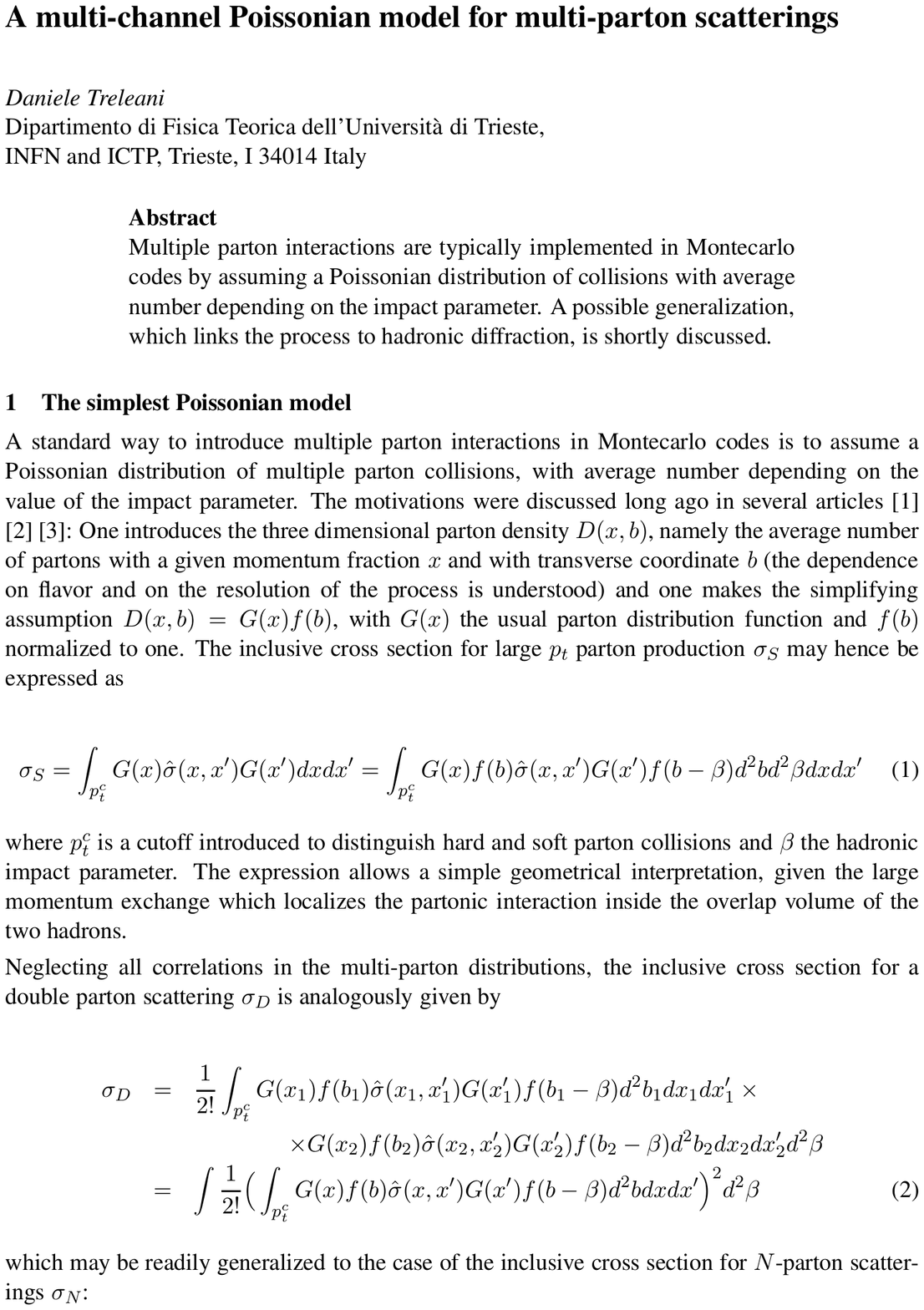}

 \coltoctitle{Underlying events in {\textrm{Herwig++}} }
 \coltocauthor{M.~B\"ahr, S.~Gieseke, 
  M.~H.~Seymour}
  \index{B\"ahr, M.}
  \index{Gieseke, S.}
   \index{Seymour, M.H.}
 \Includeart{\CAUT}{\CTIT}{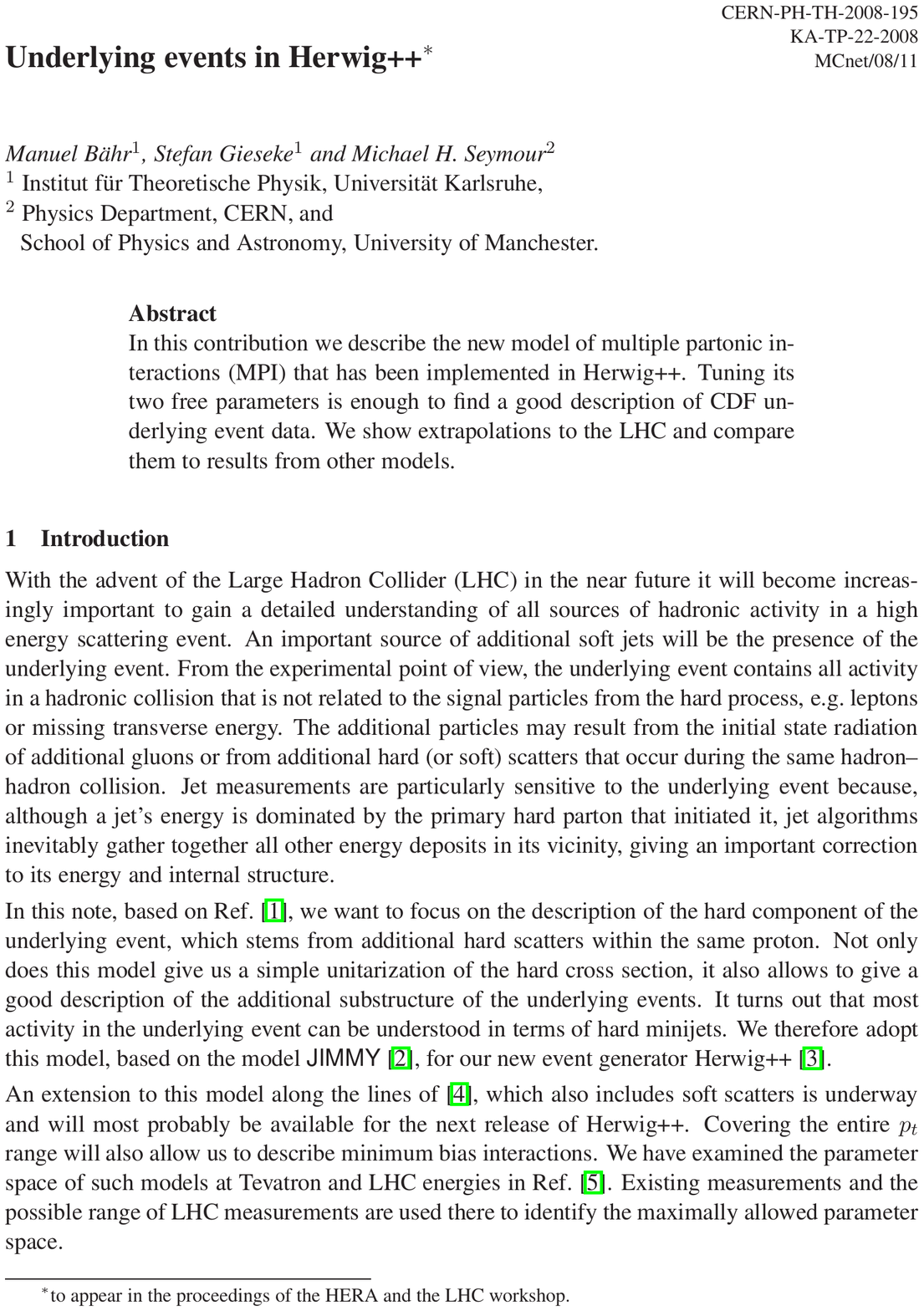}

 \coltoctitle{Multiparton Interactions at HERA }
 \coltocauthor{H. Jung, Ll. Marti, T. Namsoo, S. Osman}
  \index{Jung, H.}
  \index{Marti, L.}
   \index{Namsoo, T.}
  \index{Osman, S.}
 \Includeart{\CAUT}{\CTIT}{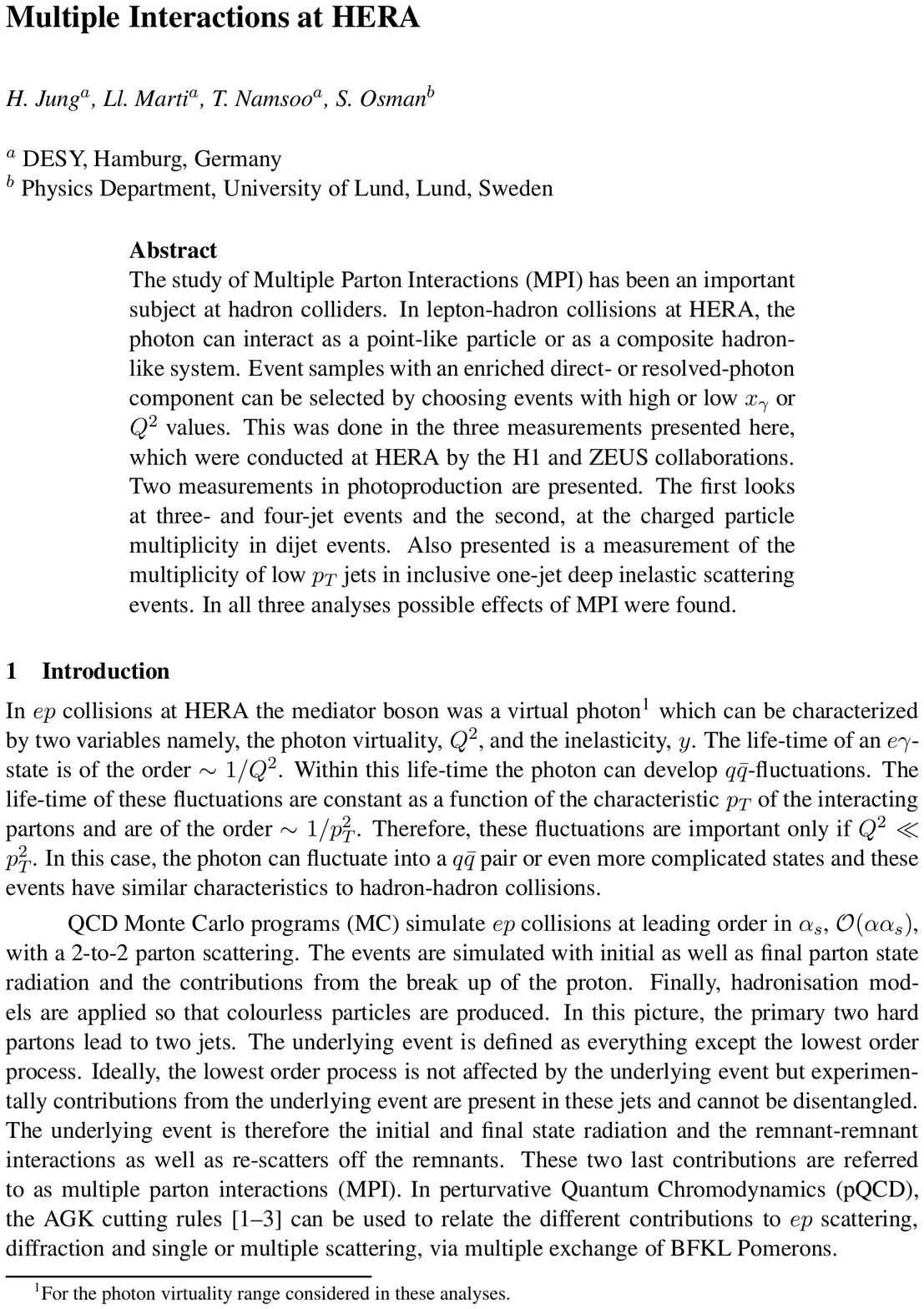}

 \coltoctitle {Modeling the underlying event: generating predictions for the LHC}
 \coltocauthor{A. Moares}
   \index{Moares, A.}
 \Includeart{\CAUT}{\CTIT}{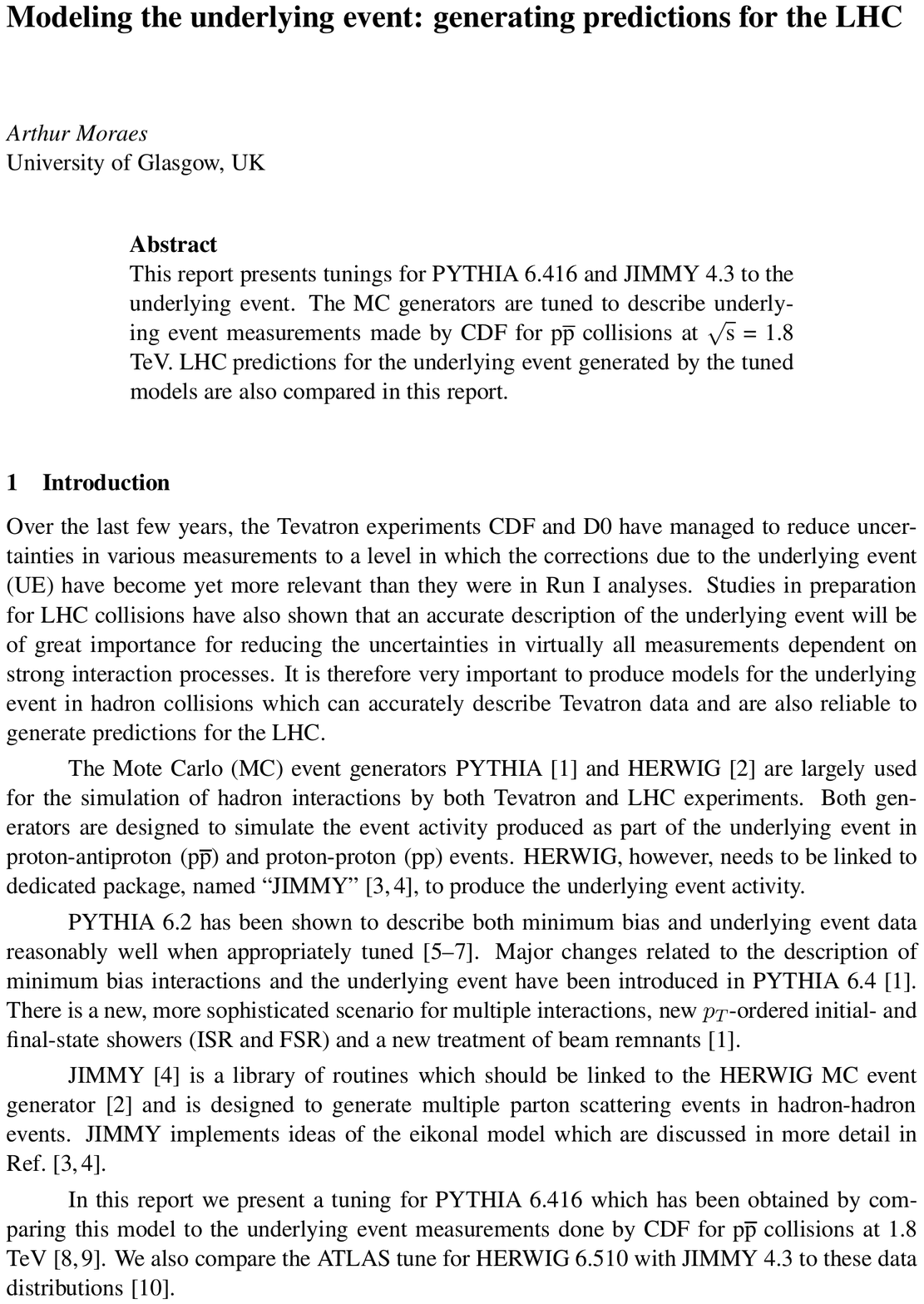}

 \coltoctitle{Measurement of the Underlying Event in Jet Topologies using Charged Particle and Momentum Densities}
 \coltocauthor{F. Ambroglini, P. Bartalini, F. Bechtel, L. Fan\`o, R. Field}
 \index{Ambroglini, F.}
 \index{Bartalini, P.}
  \index{Bechtel, F.}
 \index{Fan\`o, L.}
 \index{Field, R.}
 \Includeart{\CAUT}{\CTIT}{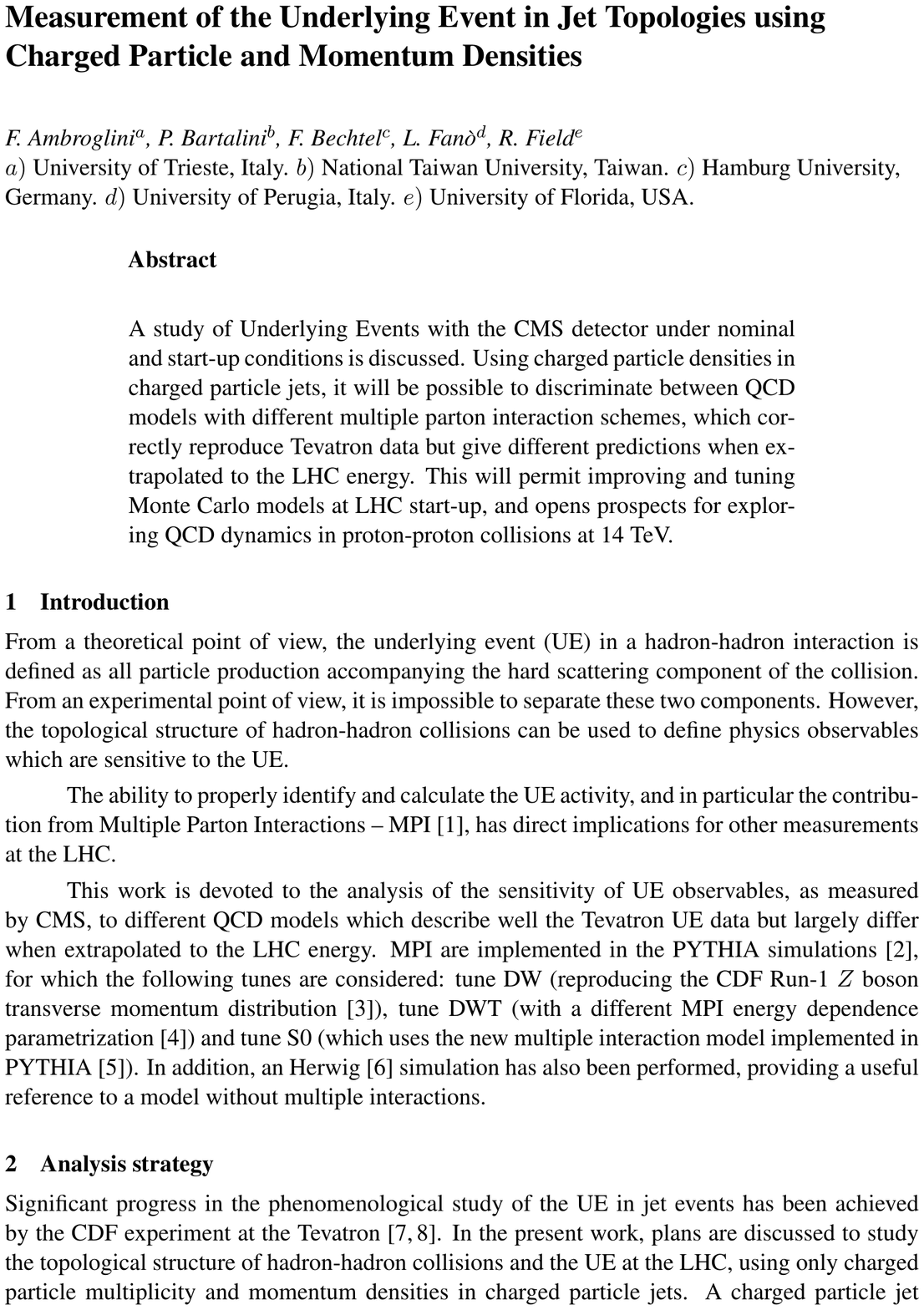}

 \coltoctitle{Double-Parton-Scattering in Photon-Three-Jet Final States at the LHC}
 \coltocauthor{F. Bechtel}
  \index{Bechtel, F.}
 \Includeart{\CAUT}{\CTIT}{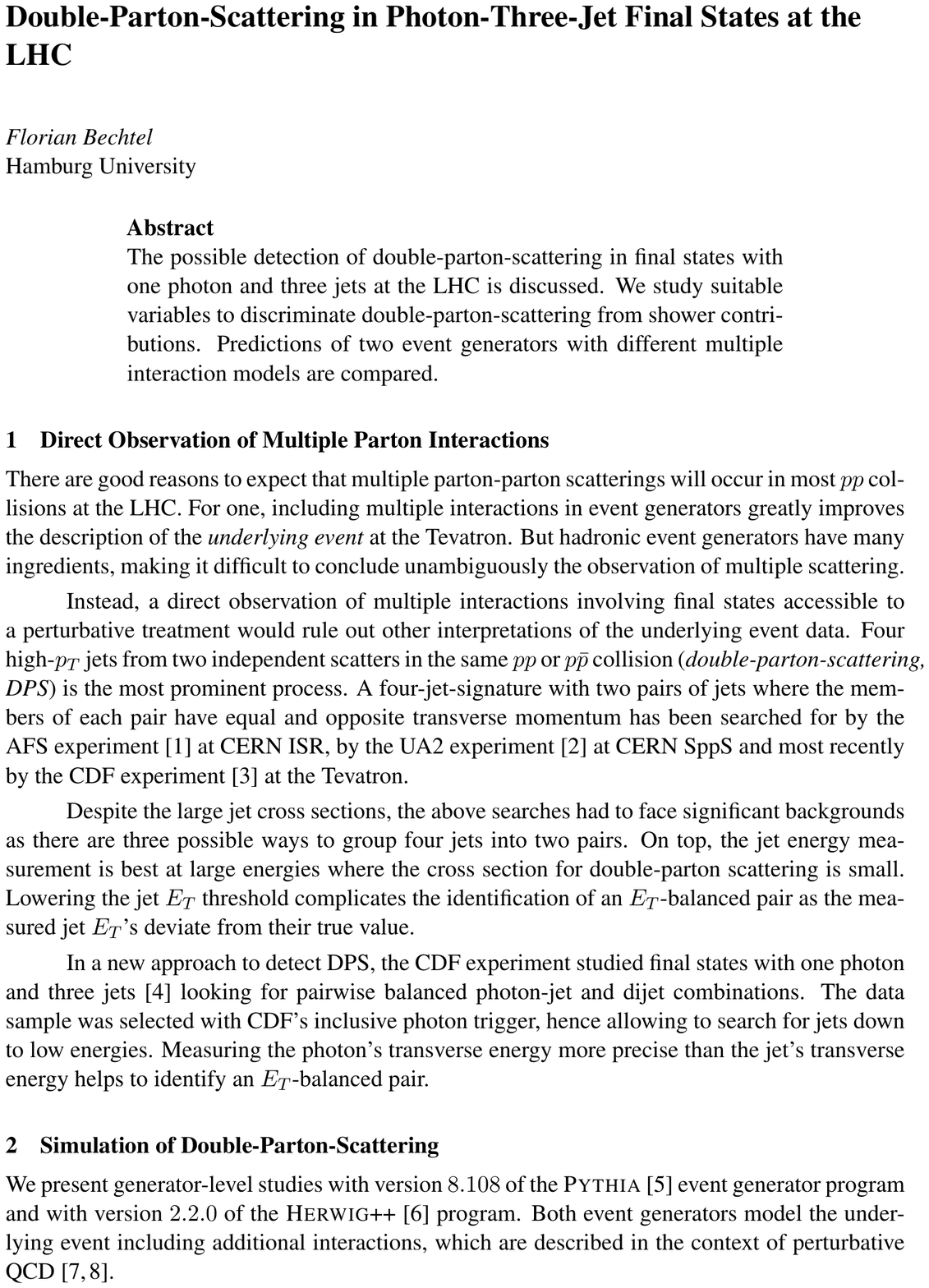}

 \coltoctitle{Underlying Event Studies with CASTOR in the CMS Experiment }
 \coltocauthor{Z. R\'urikov\'a, A. Bunyatyan}
  \index{R\'urikov\'a, Z.}
 \index{Bunyatyan, A.}
 \Includeart{\CAUT}{\CTIT}{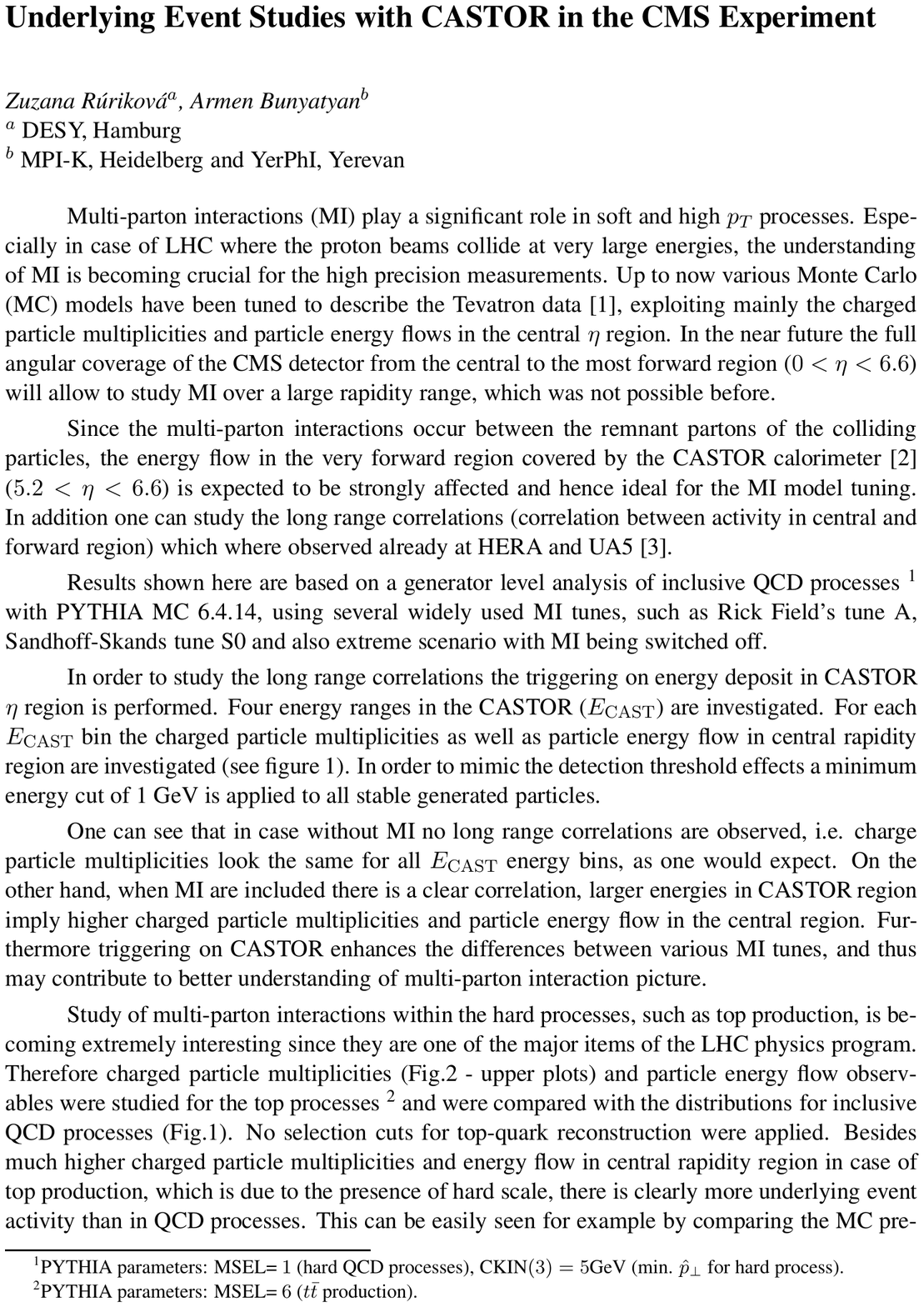}

 \coltoctitle{Direct photon production at HERA, the Tevatron and the LHC}
 \coltocauthor{R.~E.~Blair, S.~Chekanov, G.~Heinrich, A.Lipatov , N.Zotov}
 \index{Blair, R.E.}
 \index{Chekanov, S.}
\index{Heinrich, G.}
\index{Lipatov, A.} 
\index{Zotov, N.}
 \Includeart{\CAUT}{\CTIT}{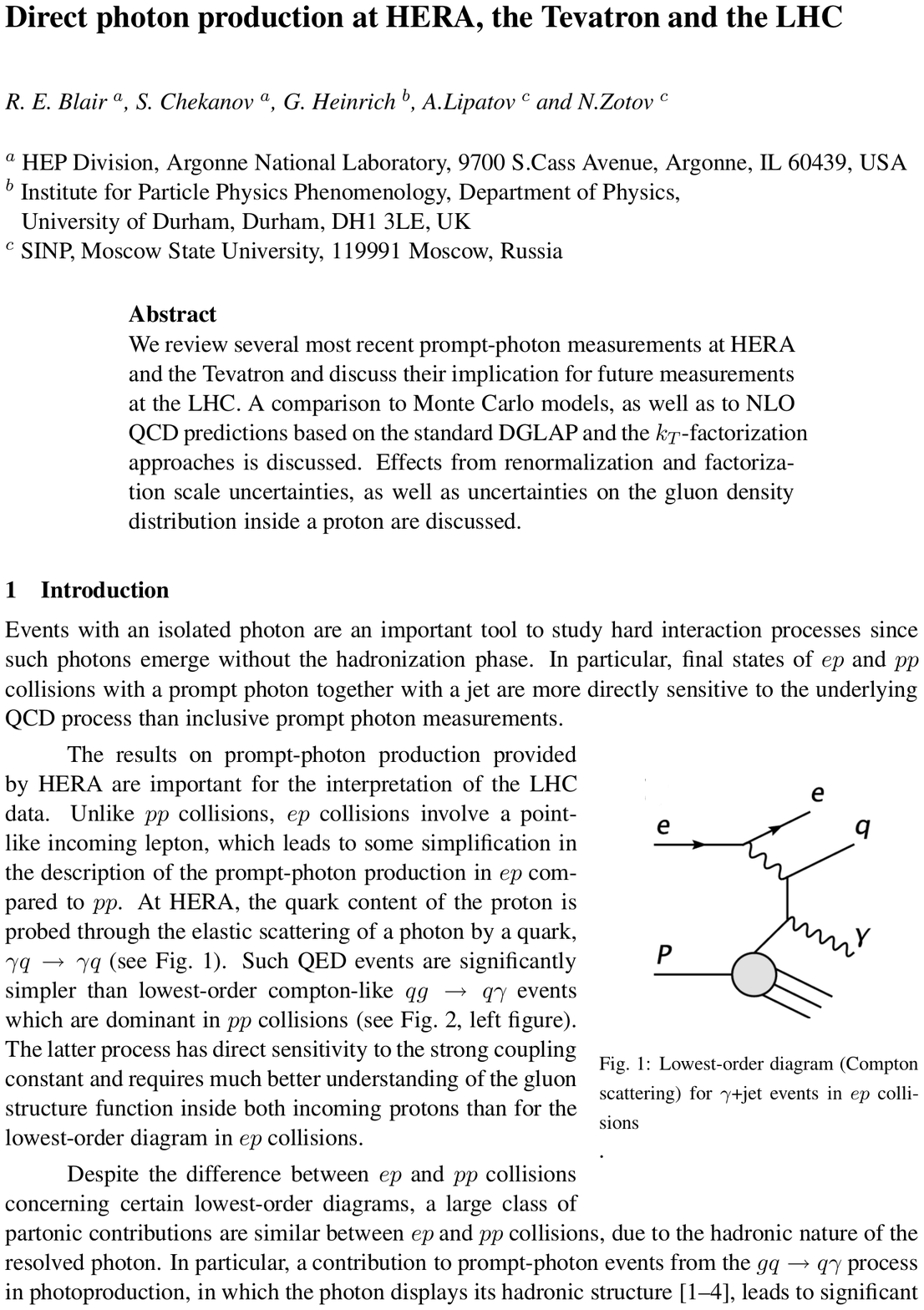}

 \coltoctitle{Propagation of Uncertainty in a Parton Shower}
 \coltocauthor{Ph.  Stephens, A.  van Hameren}
 \index{Stephens, Ph.}
 \index{van Hameren, A.}
 \Includeart{\CAUT}{\CTIT}{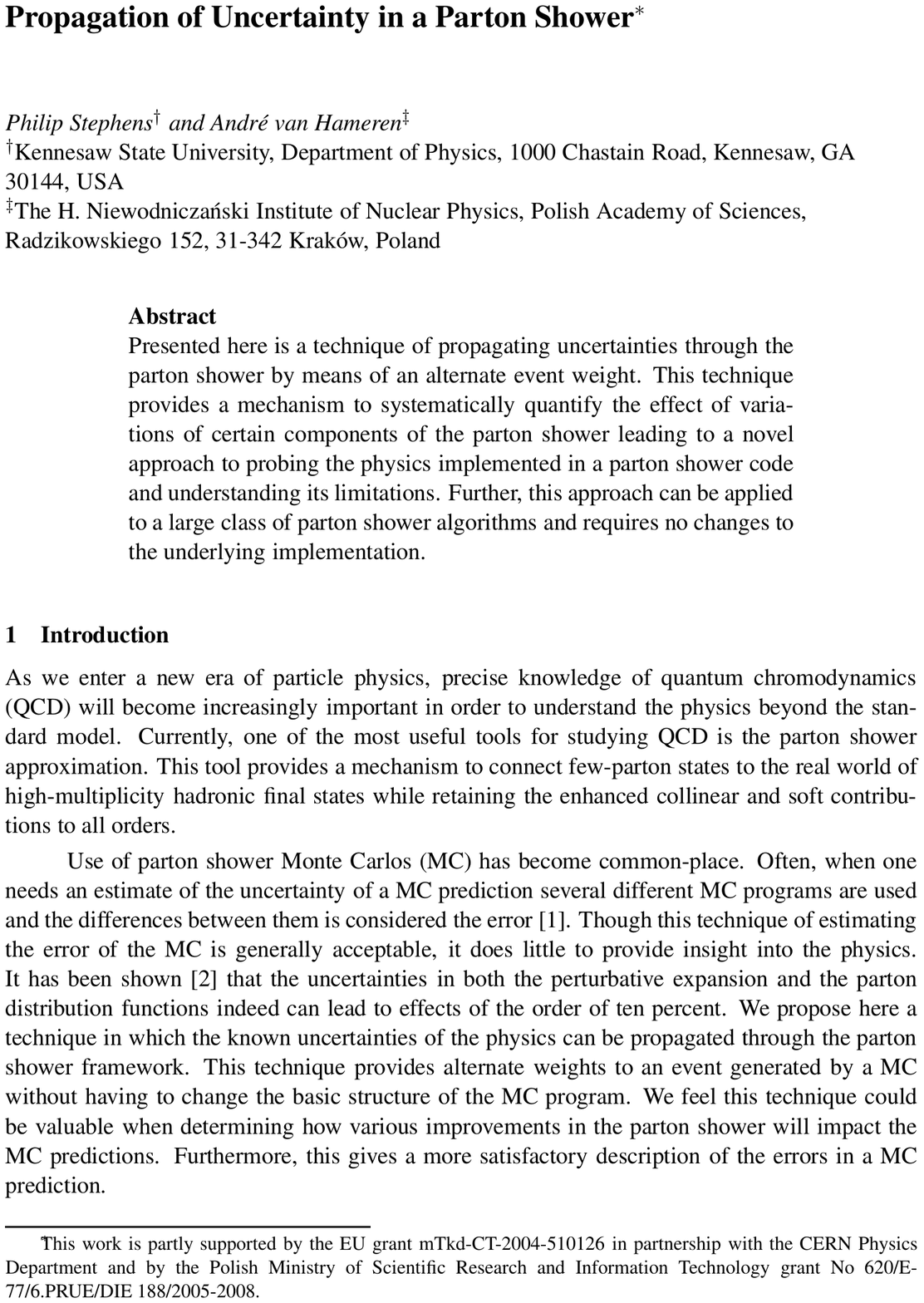}

 \coltoctitle{Perturbative description of inclusive single hadron production at HERA}
 \coltocauthor{S.~Albino}
 \index{Albino, S.}
 \Includeart{\CAUT}{\CTIT}{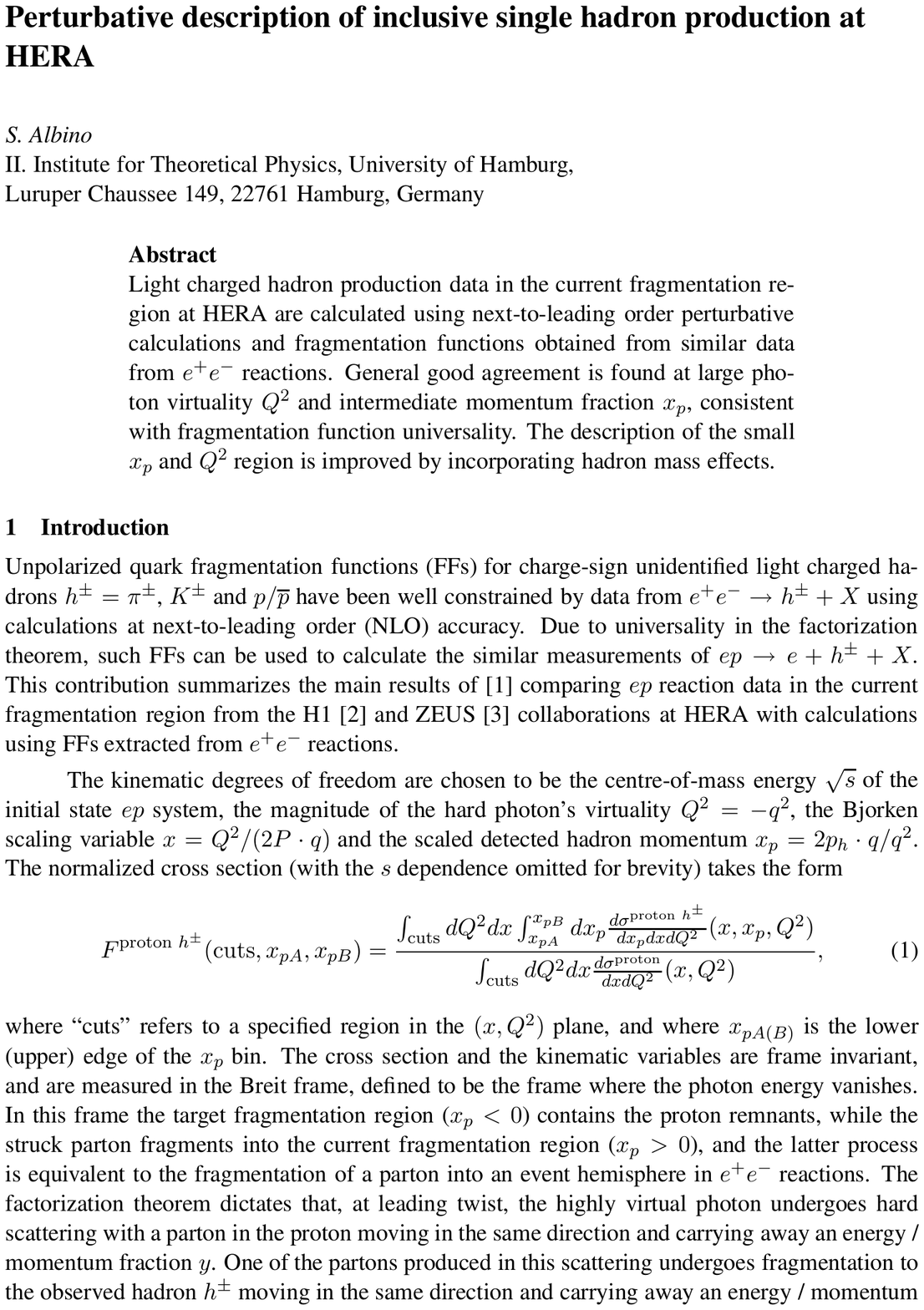}

 \coltoctitle{Nonperturbative corrections from  an s-channel  approach}
 \coltocauthor{F. ~Hautmann}
 \index{Hautmann, F.}
 \Includeart{\CAUT}{\CTIT}{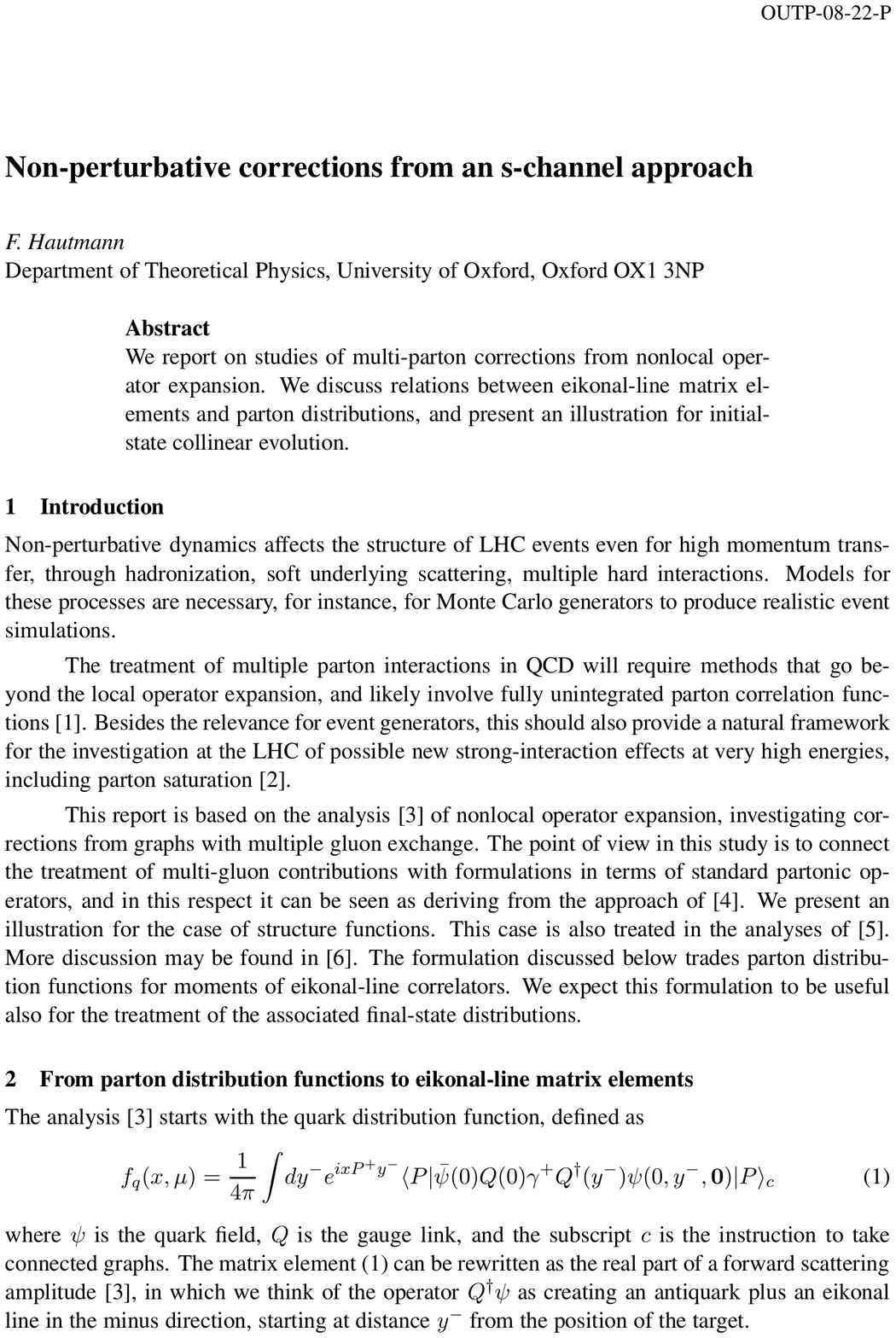}

 \coltoctitle{Single top production in the $Wt$ mode with MC@NLO}
 \coltocauthor{Ch. White}
\index{White, Ch.}
 \Includeart{\CAUT}{\CTIT}{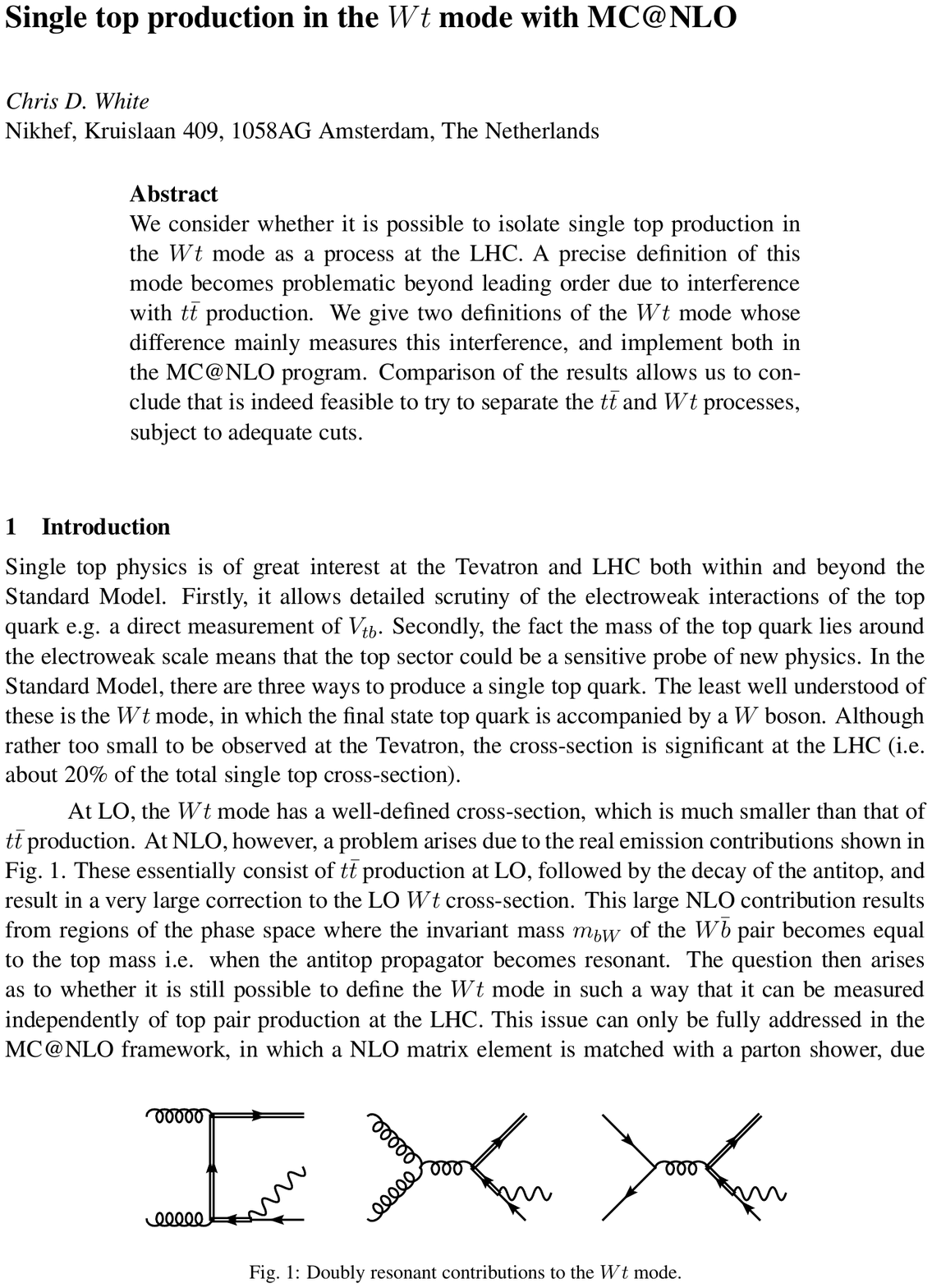}

 \coltoctitle{PYTHIA 8 Status Report}
 \coltocauthor{T. Sj\"ostrand}
 \index{Sj\"ostrand, T.}
 \Includeart{\CAUT}{\CTIT}{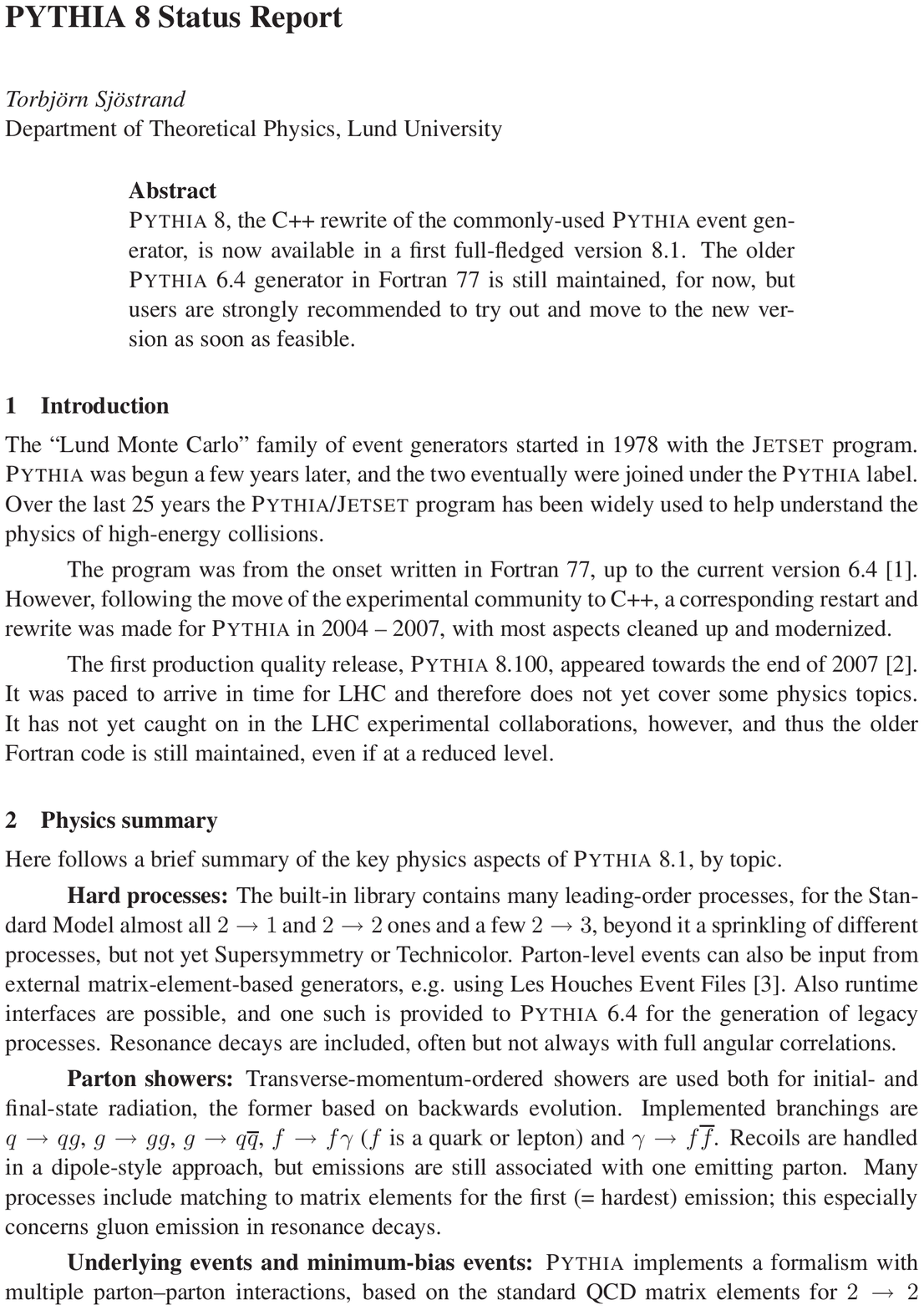}

 \coltoctitle{{\boldmath\sc ThePEG} Toolkit for High Energy Physics Event Generation}
 \coltocauthor{L. L\"onnblad}
 \index{L\"onnblad, L.}
 \Includeart{\CAUT}{\CTIT}{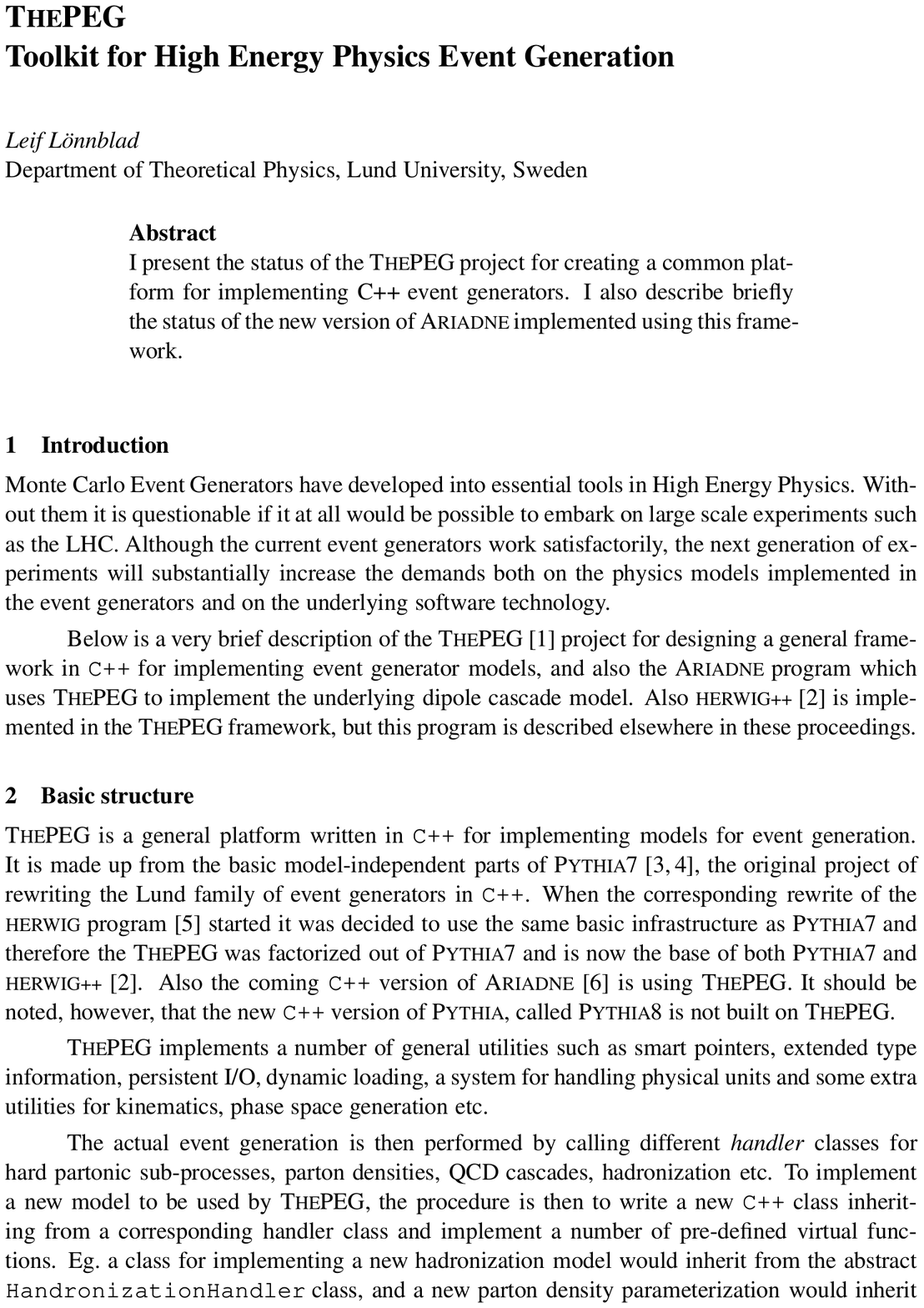}

 \coltoctitle{CASCADE}
 \coltocauthor{M. De\'ak, H. Jung, K. Kutak}
\index{De\'ak, M.}
 \index{Jung, H.}
 \index{Kutak, K.}
 \Includeart{\CAUT}{\CTIT}{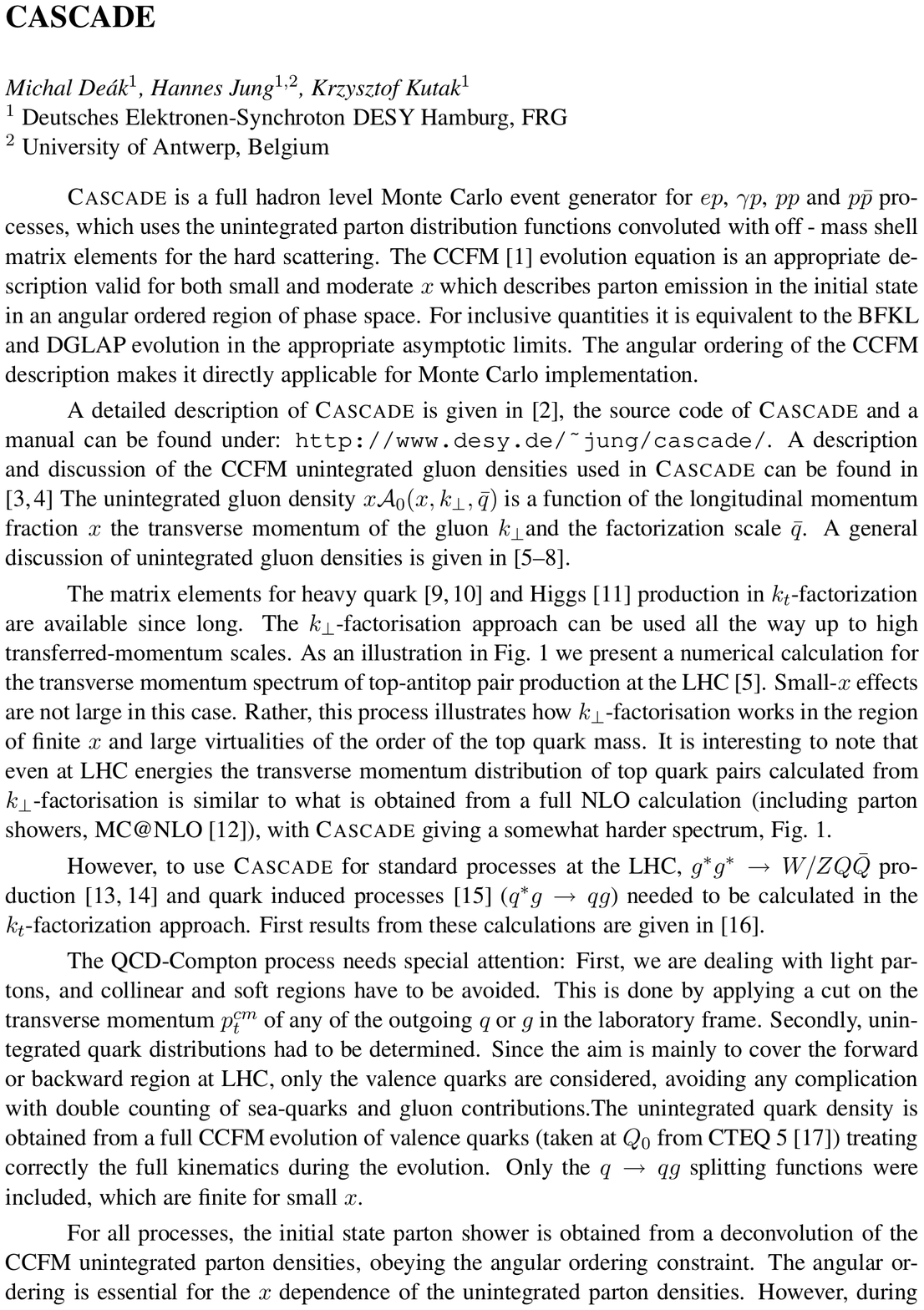}

 \coltoctitle{\genname{AlpGen} and \genname{SHERPA} in $Z/\gamma^{*}+jets$ at LHC}
 \coltocauthor{P. Lenzi}
 \index{Lenzi, P.}
 \Includeart{\CAUT}{\CTIT}{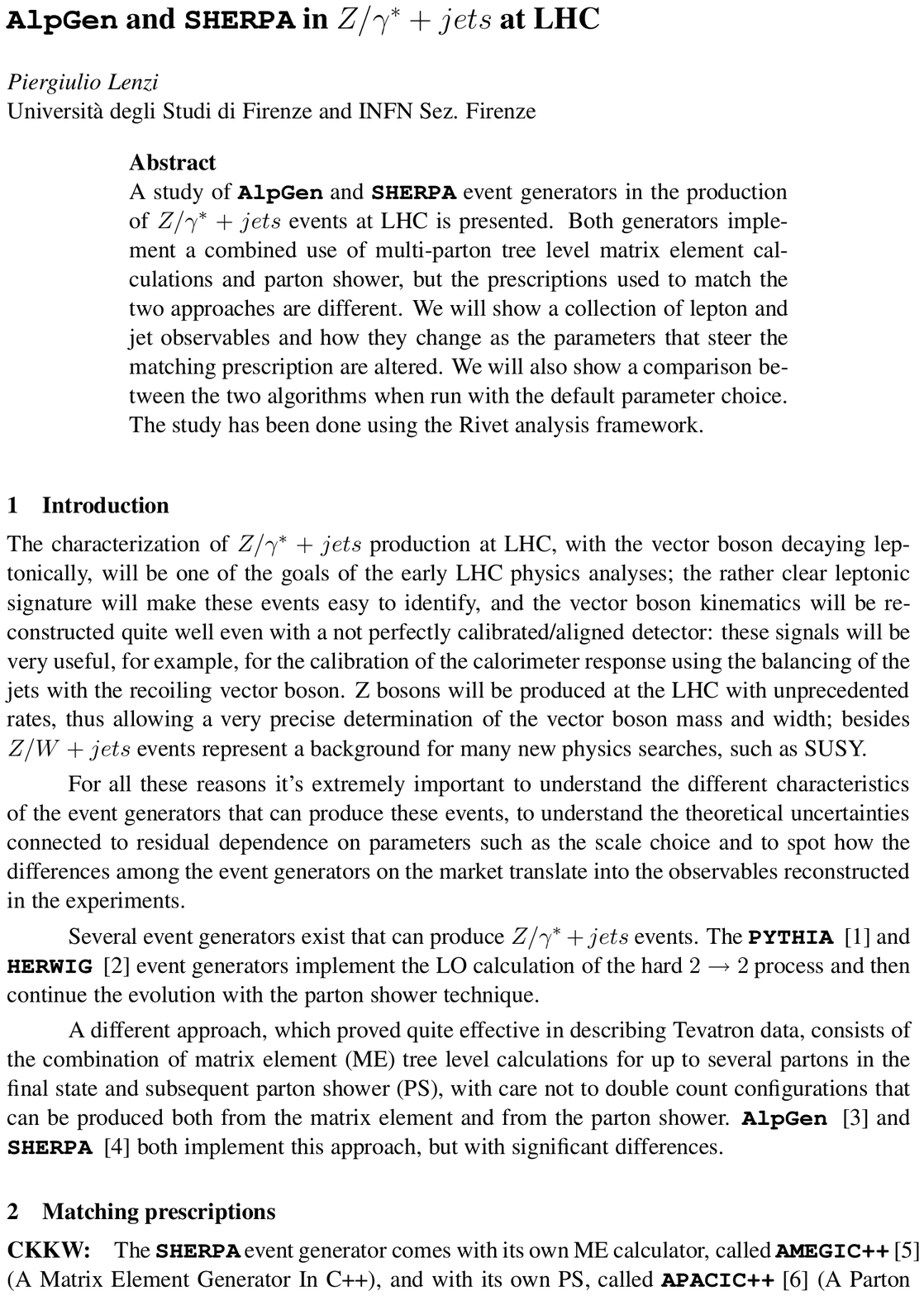}

 \coltoctitle{Generator comparison for top-pair production at CMS}
 \coltocauthor{R. Chierici}
 \index{Chierici, R.}
 \Includeart{\CAUT}{\CTIT}{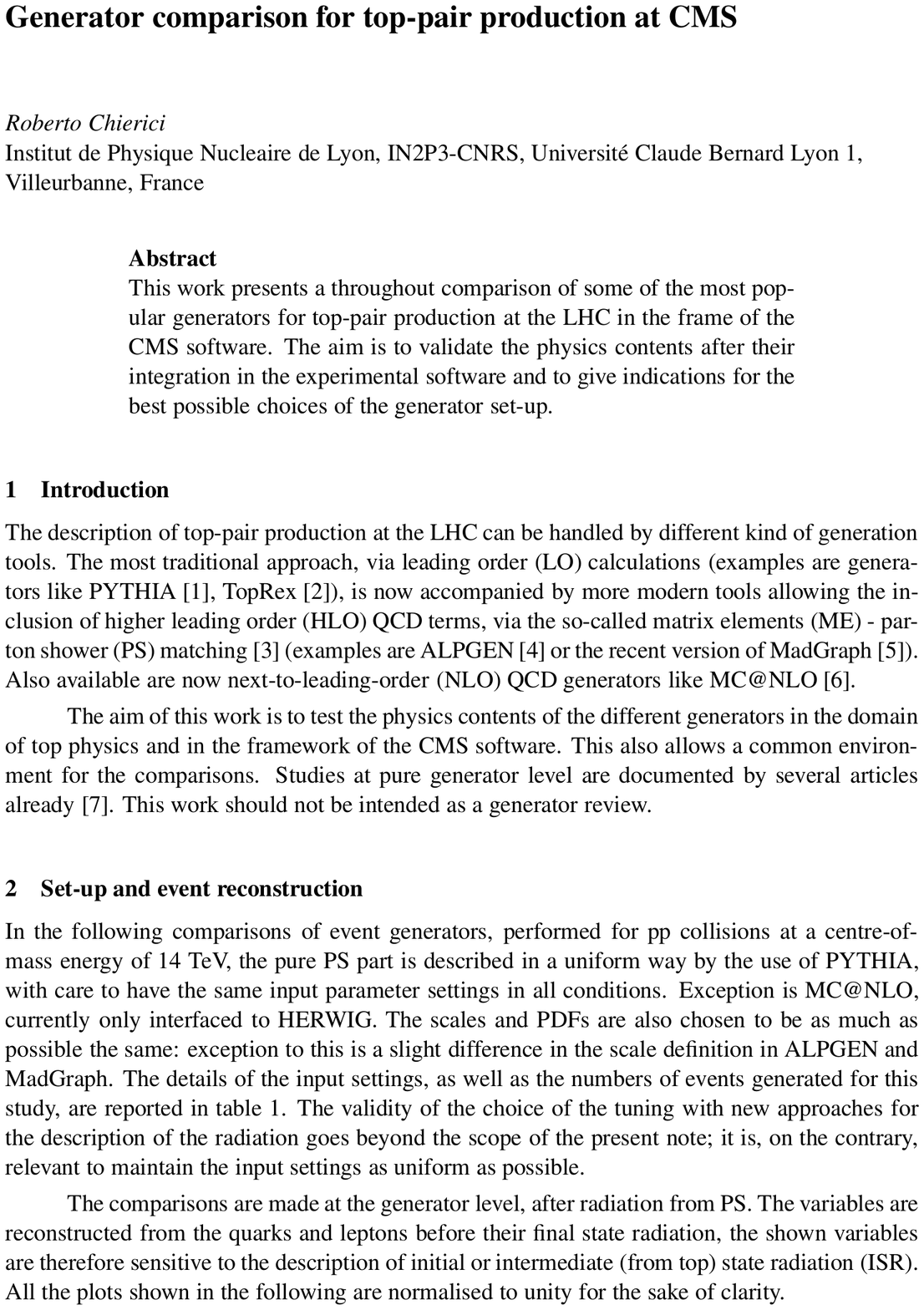}

 \coltoctitle{Herwig++ Status Report}
 \coltocauthor{M.~B\"ahr, S.~Gieseke, M.A.~Gigg, D.~Grellscheid, K.~Hamilton,
 O.~Latunde-Dada, S.~Pl\"atzer, P.~Richardson, M.H.~Seymour,
 A. Sherstnev, J.~Tully, B.R.~Webber}
 \index{B\"ahr, M.}
  \index{Gieseke, S.}
  \index{Gigg, M.A.}
 \index{Grellscheid, D.}
 \index{Hamilton, K.}
  \index{Latunde-Dada, O.} 
 \index{Pl\"atzer, S.}
  \index{Richardson, P.} 
  \index{Seymour, M.H.}
 \index{Sherstnev,  A.}
  \index{Tully, J.}
 \index{Webber, B.R.}
 \Includeart{\CAUT}{\CTIT}{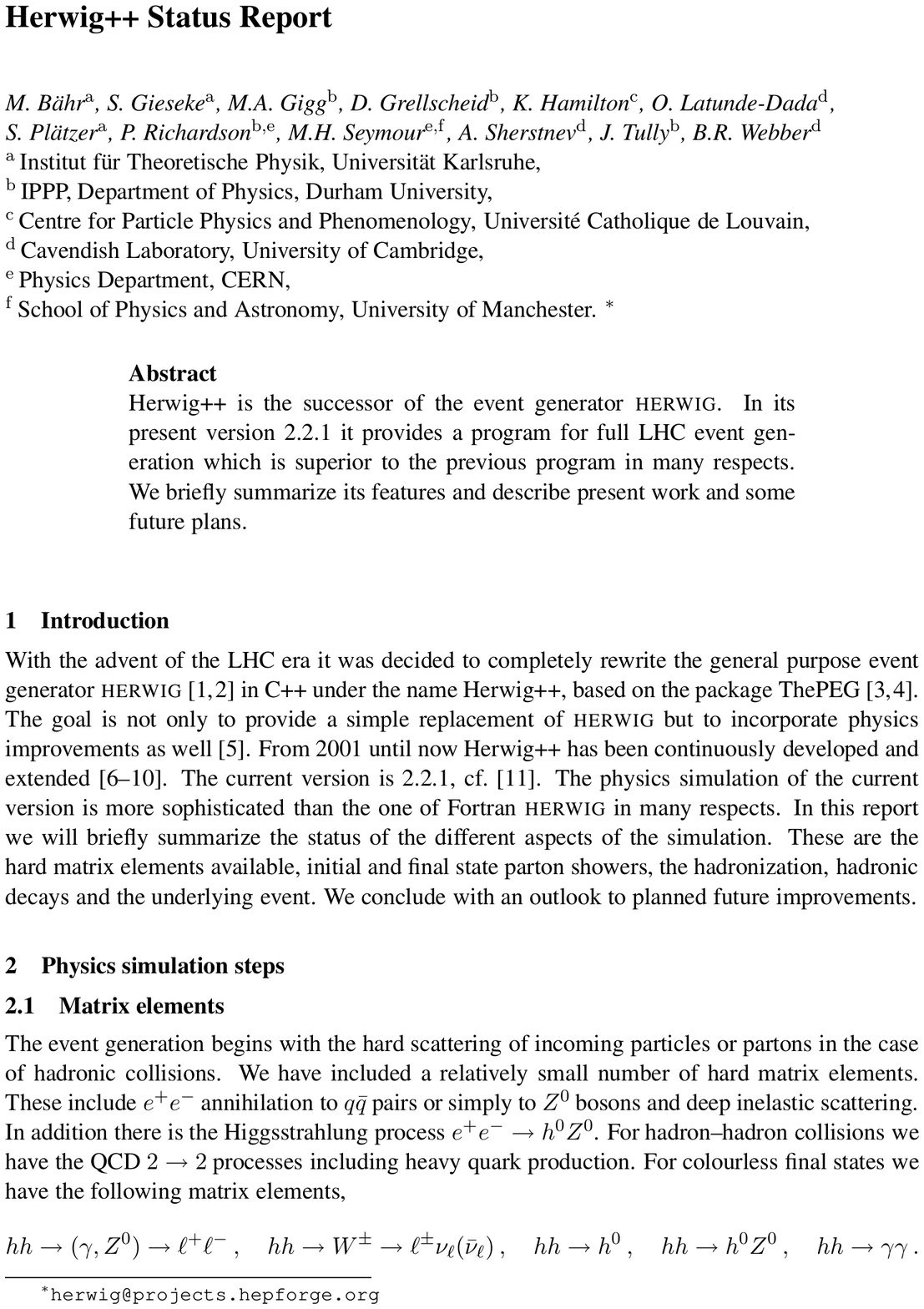}

 \coltoctitle{Forward Physics Monte Carlo (FPMC)}
 \coltocauthor{M.~Boonekamp,V.~Jur\'{a}nek, O.~Kepka, C.~Royon}
  \index{Boonekamp, M.}
    \index{Jur\'{a}nek, V.}
   \index{Kepka, O.}
   \index{Royon, Ch.}
  
 \Includeart{\CAUT}{\CTIT}{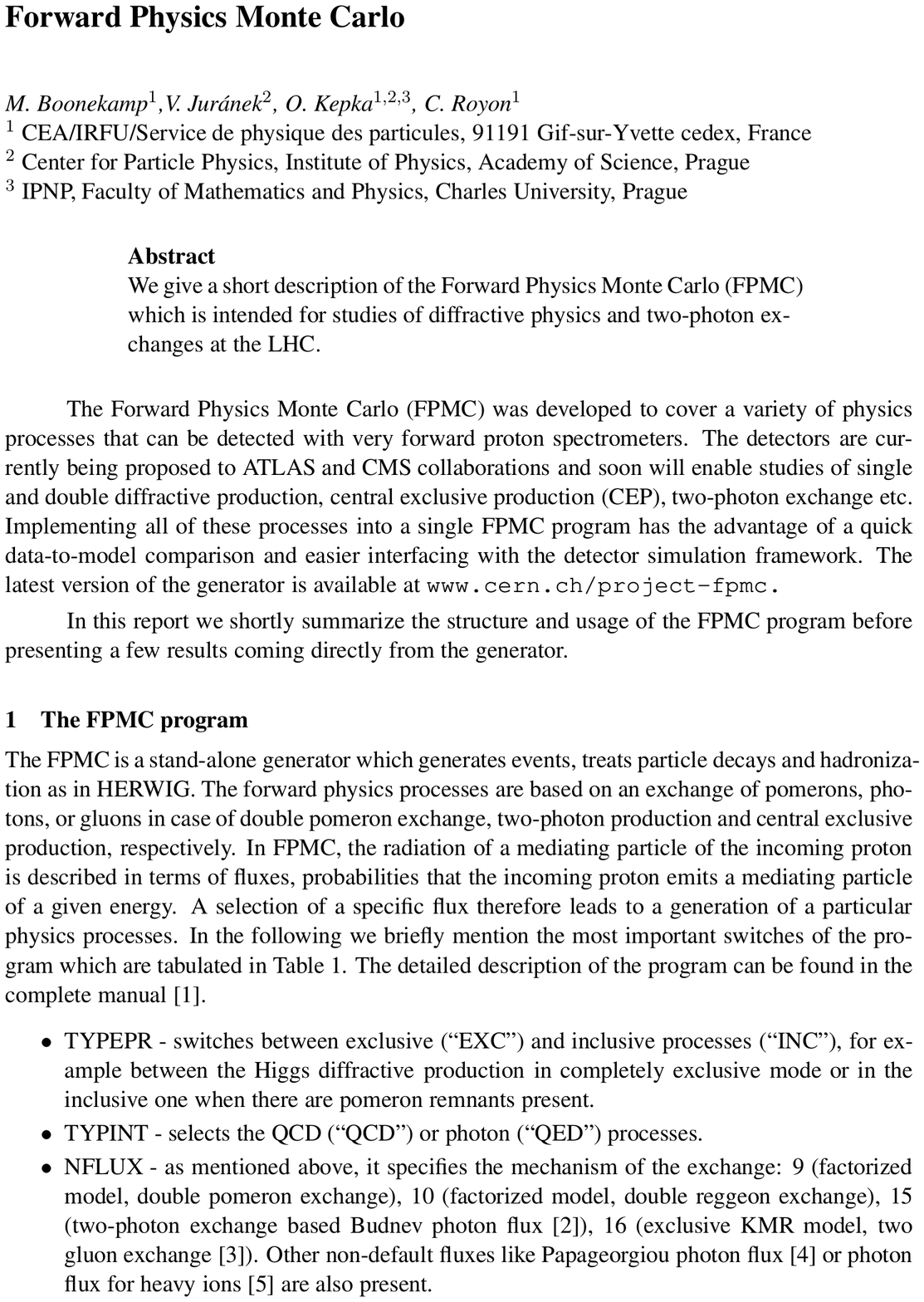}

 \coltoctitle{HEP data analysis using jHepWork and Java}
 \coltocauthor{S.~Chekanov}
   \index{Chekanov, S.}
 \Includeart{\CAUT}{\CTIT}{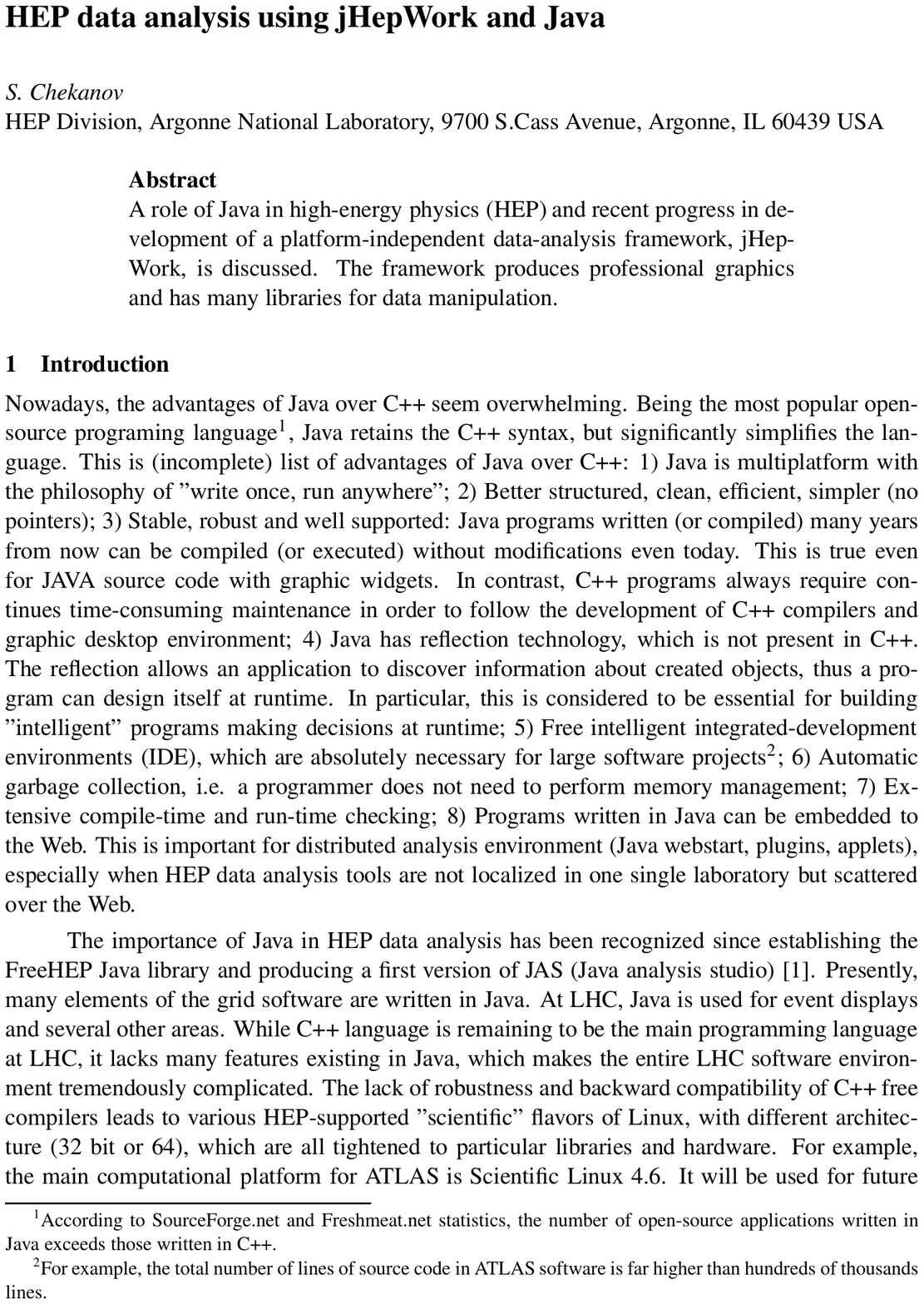}

 \coltoctitle{Tools for event generator tuning and validation}
 \coltocauthor{A. Buckley}
   \index{Buckley, A.}
 \Includeart{\CAUT}{\CTIT}{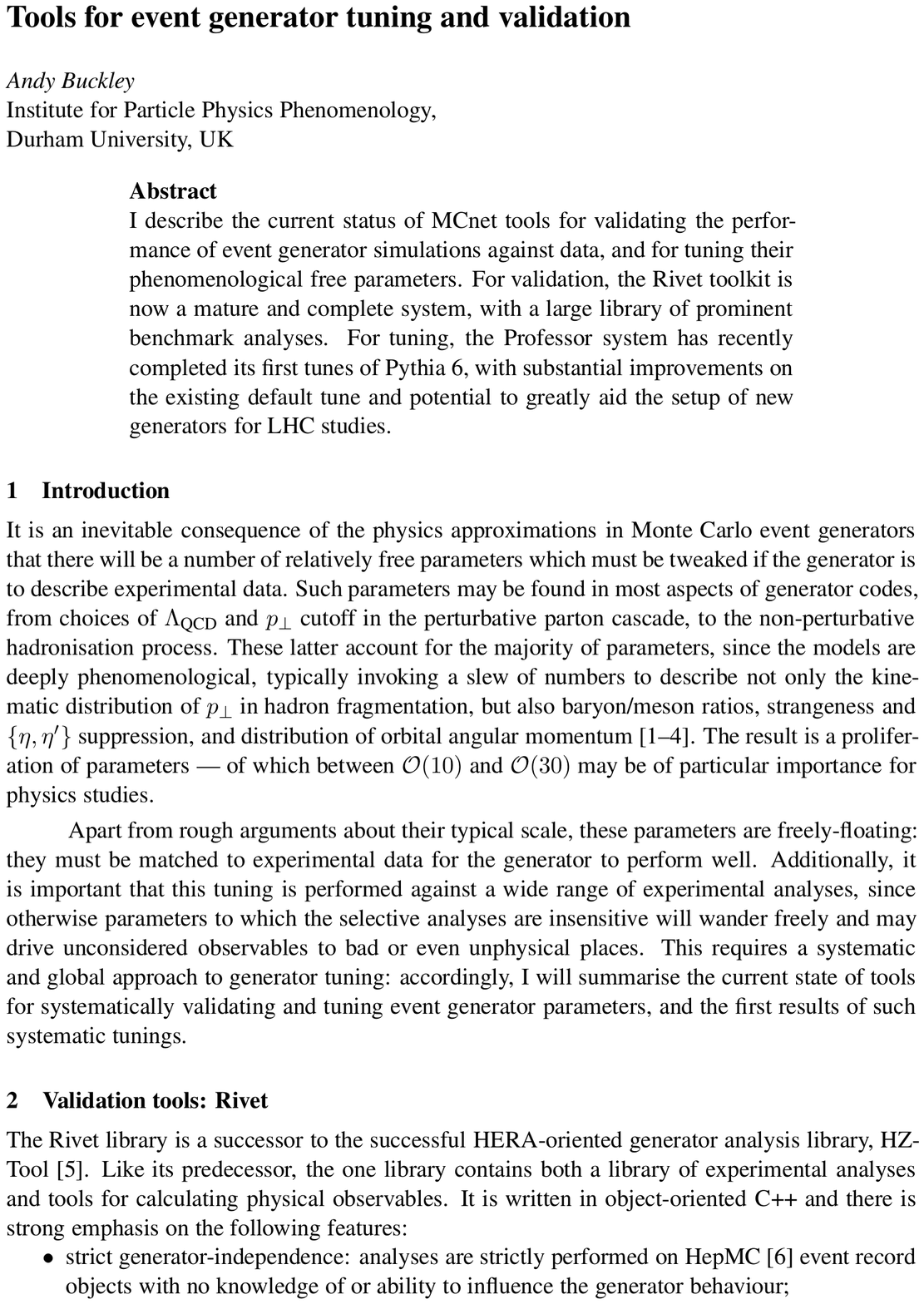}

 \coltoctitle{Prerequisites for the Validation of Experiment and Theory}
 \coltocauthor{L. Sonnenschein}
   \index{Sonnenschein, L.}
 \Includeart{\CAUT}{\CTIT}{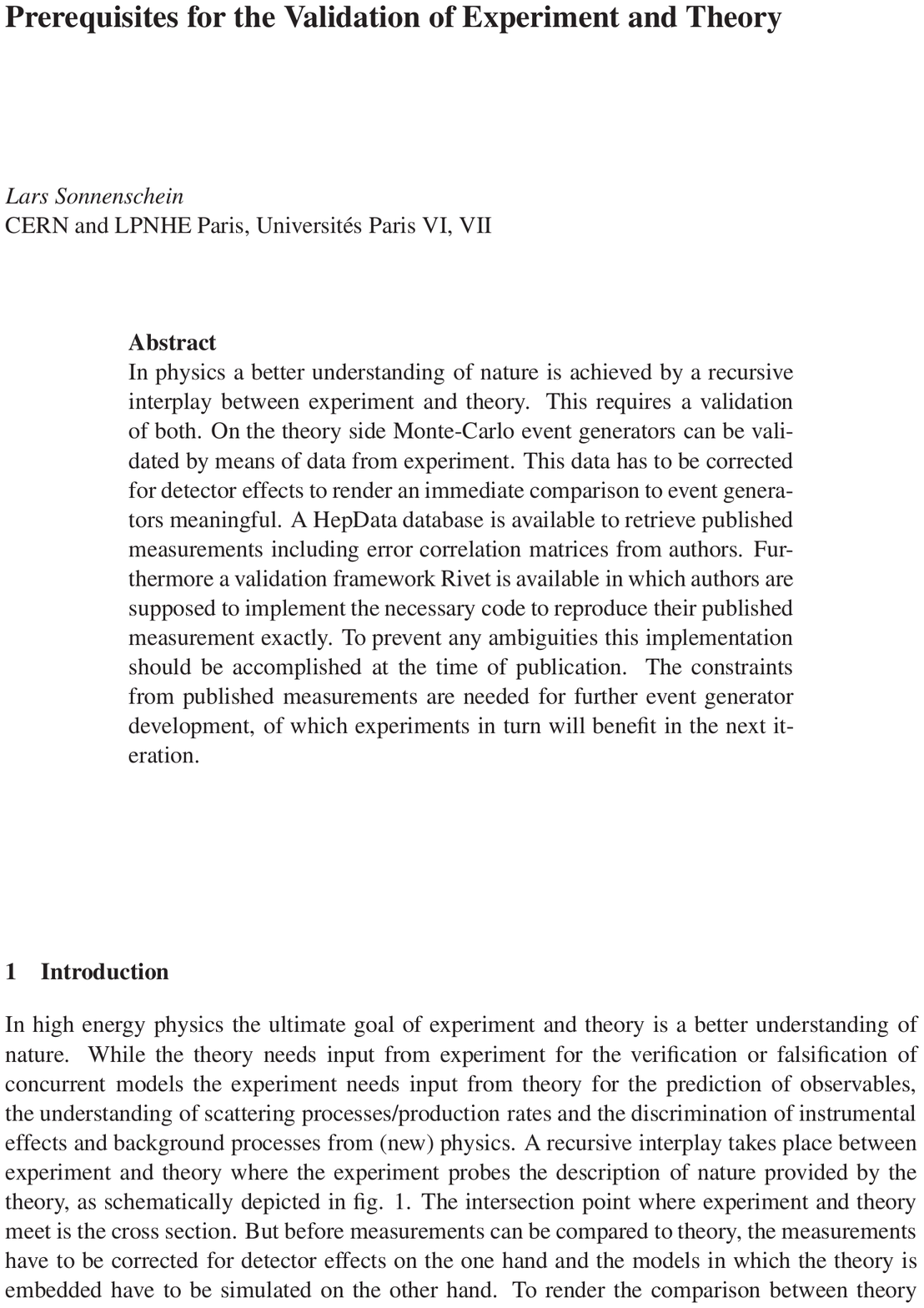}


\end{papers} 

\pagestyle{desyplain}
\CLDP 

\chapter[\large List of Authors  ]{List of Authors}
\printindex

\CLDP 
\chapter[\large List of Participants ]{List of Participants}

\include{include-parti}

\end{document}